\documentclass{aa}
\usepackage{appendix}
\usepackage{latexsym}
\usepackage{graphicx}
\usepackage{natbib}
\usepackage{lscape}
\usepackage{afterpage}
\usepackage{booktabs}
\usepackage{multirow}
\usepackage{arydshln}

\usepackage{ulem}
\usepackage[citecolor =blue]{hyperref}
\usepackage{color}

\usepackage{amstext}

\begin{document}
     
\title{QSO2 outflow characterization using data obtained with OSIRIS at the Gran Telescopio Canarias}
\author{Enrica Bellocchi\inst{1}, Montserrat Villar Mart\'in\inst{1}, Antonio Cabrera--Lavers\inst{2,3}, Bjorn Emonts\inst{4}}
\institute{$^1$ Centro de Astrobiolog\'ia, (CAB, CSIC--INTA), Departamento de Astrof\'isica, Cra. de Ajalvir Km.~4, 28850 -- Torrej\'on de Ardoz, Madrid, Spain\\  
$^2$ Instituto de Astrof\'isica de Canarias (IAC), V\'ia L\'actea s/n, E--38200 La Laguna (Tenerife, Spain)\\
$^3$ GRANTECAN, Cuesta de San Jos\'e s/n, E--38712 Bre\~na Baja (La Palma, Spain)\\
$^4$ National Radio Astronomy Observatory, 520 Edgemont Road, Charlottesville, VA 22903\\
 \email{enrica.bellocchi@gmail.com, ebellocchi@cab.inta-csic.es}\\                
 }
\date{Received 14 February 2019 / Accepted 26 April 2019}

\abstract 
{
Ionized outflows are ubiquitous in non-radio-loud obscured quasars (QSO2s) at different redshifts. However, the actual size of the outflows and their efficiency for gas ejection and star formation truncation are controversial. Large-scale ($\text{exceeding}$ several kiloparsec) extended radio structures might be necessary to identify (even to trigger) outflow signatures across such large spatial scales. }
{
We search for large-scale ionized outflows associated with six optically selected QSO2 (five non-radio-loud and one radio-loud) at $z\sim$ 0.2--0.5, targeting objects with extended radio structures. We also investigate the dynamical state of the QSO2 host galaxies.}
{
We obtained data with the optical imager and long-slit spectrograph (OSIRIS) mounted
on the 10.4m Gran Telescopio Canarias Spanish telescope (GTC) for these six QSO2 with the slit located along the radio axis. We traced the gas kinematics with the [OIII]$\lambda$4959,5007 lines to investigate ionized outflows and characterize the dynamical state of the host galaxies. 
This second study was complemented with previously published spectroscopic data obtained with the multimode focal reducer and low dispersion spectrograph (FORS2)  mounted on the Very Large Telescope (VLT) of 13 more QSO2 at similar $z$.
}
{We identify ionized outflows in four out of the six QSO2 observed with the GTC. The outflows are spatially unresolved in two QSO2 and compact in a third (radial size of $R=$0.8$\pm$0.3 kpc). Of particular interest is the radio-quiet QSO2 SDSS 0741+3020 at $z=$ 0.47. It is associated with a giant $\sim$112 kpc nebula. An ionized outflow probably induced by the radio structures has been detected along the axis defined by the central $\sim$1\arcsec\ radio structure, extending up to at least $\sim$4 kpc from the active galactic nucleus (AGN). Turbulent gas ($\sigma\sim$130 km s$^{-1}$) has also been detected across the giant gas nebula up to $\sim$40 kpc from the AGN. This turbulence may have been induced by outflows triggered by the interaction between a so-far undetected large-scale radio source and the nebula. 
Regarding the dynamical state of the host galaxies, we find that the majority of the QSO2 show v/$\sigma<$ 1, implying that they are dominated by random motions (so-called dispersion-dominated systems). Most (17 of 19) fall in the area of the E/S0 galaxies in the dynamical diagram v/$\sigma$ versus $\sigma$. None are consistent with spiral or disk galaxies. 
}
{}

\titlerunning{}
\authorrunning{Bellocchi et al.}

\maketitle

\section{Introduction}

Evidence for a tight connection between supermassive black holes (SMBHs) and their host galaxies is compelling. SMBHs have been found in many galaxies with a spheroidal component, and correlations exist between the black hole (BH) mass and some bulge properties, such as the stellar mass and velocity dispersion \citep[e.g.,][]{Ferrarese00, McConnell13}.
The origin of these relations is a topic of key importance for understanding galaxy  evolution. Quasar outflows may play a critical role. Hydrodynamical simulations show that the energy output from quasars can regulate the growth and activity of black holes and their host galaxies through these feedback mechanisms by heating and/or ejecting the gas of the interstellar medium (ISM) \citep{DiMatteo05, Bieri17, Cielo18}.

The feedback mechanisms from a growing active SMBH can be divided broadly into two types: the former is the so--called quasar (or radiative) mode feedback, which comprises wide--angle subrelativistic outflows driven by radiation from the efficient accretion of cold gas during the accretion of the BH mass, and the latter is called radio (or kinetic) mode feedback, where the bulk of the energy is ejected in a kinetic form through jets that are coupled to the galaxies gaseous environment \citep{Fabian12}. In many situations, one feedback dominates the other \citep[e.g., M87 and Mrk 231;][]{Prieto16, Rupke17}.

Quasar-mode feedback, which occurs when the AGN is very luminous \citep{Fabian12} is considered to be driven by a wind created by the luminous accretion disk. In this case, the ignition of the nucleus in a galaxy may heat up and remove gas from the interstellar medium (ISM)  of the host galaxy, potentially reducing or even stopping star formation \citep[e.g.,][]{Granato01, Croton06, Hopkins10}.

Radio-mode feedback uses the mechanical energy of the  radio structures (relativistic jets or lobes), which are more common in massive elliptical galaxies. Direct observations show that jets can influence gas at different spatial scales, up to many tens of kiloparsec from the center of the parent host galaxy \citep[e.g.,][and references therein]{Nesvadba10, Fabian12, Tadhunter14}. It has been widely discussed that AGN jets are able to exert negative feedback by effectively suppressing or even quenching star formation \citep[e.g.,][]{Best05, Croton06, Karouzos13, Gurkan15}. The jets warm up and ionize the gas they collide with, making collapse under self--gravity more difficult \citep[][]{Tadhunter14}. They may also  expel the molecular gas from the galaxy, effectively removing the ingredient required for stars to form \citep[e.g.,][]{Nesvadba06, Nesvadba11_0}.

Type 2 (obscured) quasars (QSO2) are unique laboratories for investigating  the most extreme AGN feedback. The obscuration by an optically thick structure of the highly luminous central engine, which can otherwise outshine the host galaxy, enables a detailed study of the properties of the surrounding medium. 

Because they are the most luminous AGN, the most powerful outflows with potentially the most dramatic impact are expected in quasars \citep[e.g.,][]{Page12}. However, observational evidence for this is controversial.  
Integral field spectroscopic (IFS) studies of QSO2 at z $\lesssim$ 0.7 have suggested that large-scale wide-angle AGN-driven ionized outflows are prevalent in these systems, and their action can be exerted across many kiloparsec, even the entire galaxy as shown by \citet{Liu13}, \citet{Harrison14} and \citet{McElroy15}. These works have been questioned by numerous authors, however, who claimed that the outflows are instead compact and who found no evidence for a significant impact on the host galaxies \citep{Husemann16, Karouzos16, VM16, Rose18,  Spence18, Tadhunter18}.

The role of the radio-induced feedback has not been sufficiently explored in quasars with low or moderate radio activity. Only $\sim$15\% $\pm$ 5\% of QSO2 are expected to be radio-loud \citep[][the same percentage applies to QSO1, \citealt{Kellermann94}]{Lal10}, but radio-mode feedback can also play a significant role in some non-radio-loud AGNs. It has been suggested that the most kinematically extreme nuclear outflows are often found in quasars with a modest degree of jet activity, including radio-quiet quasars  \citep[e.g.,][]{Husemann13, Mullaney13, VM14, Zakamska14}.

Only two QSO2s at $z<$0.7 have unambiguous  evidence that AGN feedback works across R$>  $ a few kiloparsec from the AGN: the so-called Teacup \citep[$z=$ 0.085,][]{Harrison15, RAlmeida17,VM18} and the Beetle \citep[$z=$0.123,][]{VM17}. Although radio-quiet, both QSO2 host large-scale extended radio structures whose morphologies are tightly correlated with that of the extended ionized gas. In the former, it is not clear whether the $\sim$10--12 kpc optical bubbles have been inflated by the radio structures or by a wide-angle large-scale AGN-driven wind \citep{Harrison15}. The Beetle indeed shows unambiguous evidence that radio-induced feedback works across a total extension of $\sim$46 kpc and up to 26 kpc from the AGN, well into the circumgalactic medium (CGM) of the system. The Beetle demonstrates that jets of modest power can be the dominant feedback mechanism acting across huge volumes in highly accreting luminous radio-quiet AGN, where radiative (or quasar) mode feedback would be expected to dominate.

Based on these results, \citet{VM17} proposed that an extended radio source capable of escaping from the dense galaxy core might be a fundamental ingredient for AGN feedback to be identified (even triggered?) across galactic or extragalactic scales. Thus, a promising strategy to find large-scale AGN-driven outflows is to search for them in QSO2s with extended radio structures ($\text{exceeding }$several kiloparsec). This is the purpose of this paper. We investigate here the possible existence of large-scale ionized outflows as traced by the forbidden [OIII]$\lambda\lambda$4959, 5007 lines along the radio axis of six optically selected (Sloan Digital Sky Survey, SDSS) QSO2 \citep[five non-radio-loud and one radio-loud; ][]{Reyes08} with extended radio structures. The study is based on long-slit spectroscopic data obtained with the optical imager and long-slit spectrograph OSIRIS mounted on the 10.4m Gran Telescopio Canarias (GTC) Spanish telescope. As an additional study, we investigate the dynamical state of the host galaxies associated with these six QSO2 at $z\sim$0.2--0.5 and 13 more SDSS QSO2 at similar $z$ observed with the multimode focal reducer spectrograph (FORS2) at the Verly Large Telescope (VLT) \citep{VM11,VM16}.

The paper is organized as follows. The sample is presented in Sect.~2. The observations and data reduction are explained in Sect.~3, while the kinematic analysis is described in Sect.~4. The results are presented and discussed in Sect.~5, including a description of the individual objects. Finally, the main results and conclusions are summarized in Sect.~6. Figures relevant to the spatially extended kinematic analysis of all objects are shown in Appendix A.

We adopt H$_0$ = 71 km s$^{-1}$ Mpc$^{-1}$, $\Omega_\Lambda$ = 0.73, and $\Omega_m$ = 0.27. At the redshifts of our sample, this gives a conversion from arcseconds to kiloparsec ranging from 3.7 to 6.5 kpc arcsec$^{-1}$ .

\section{Sample}

\begin{figure*}
\begin{center}
\includegraphics[scale=0.6]{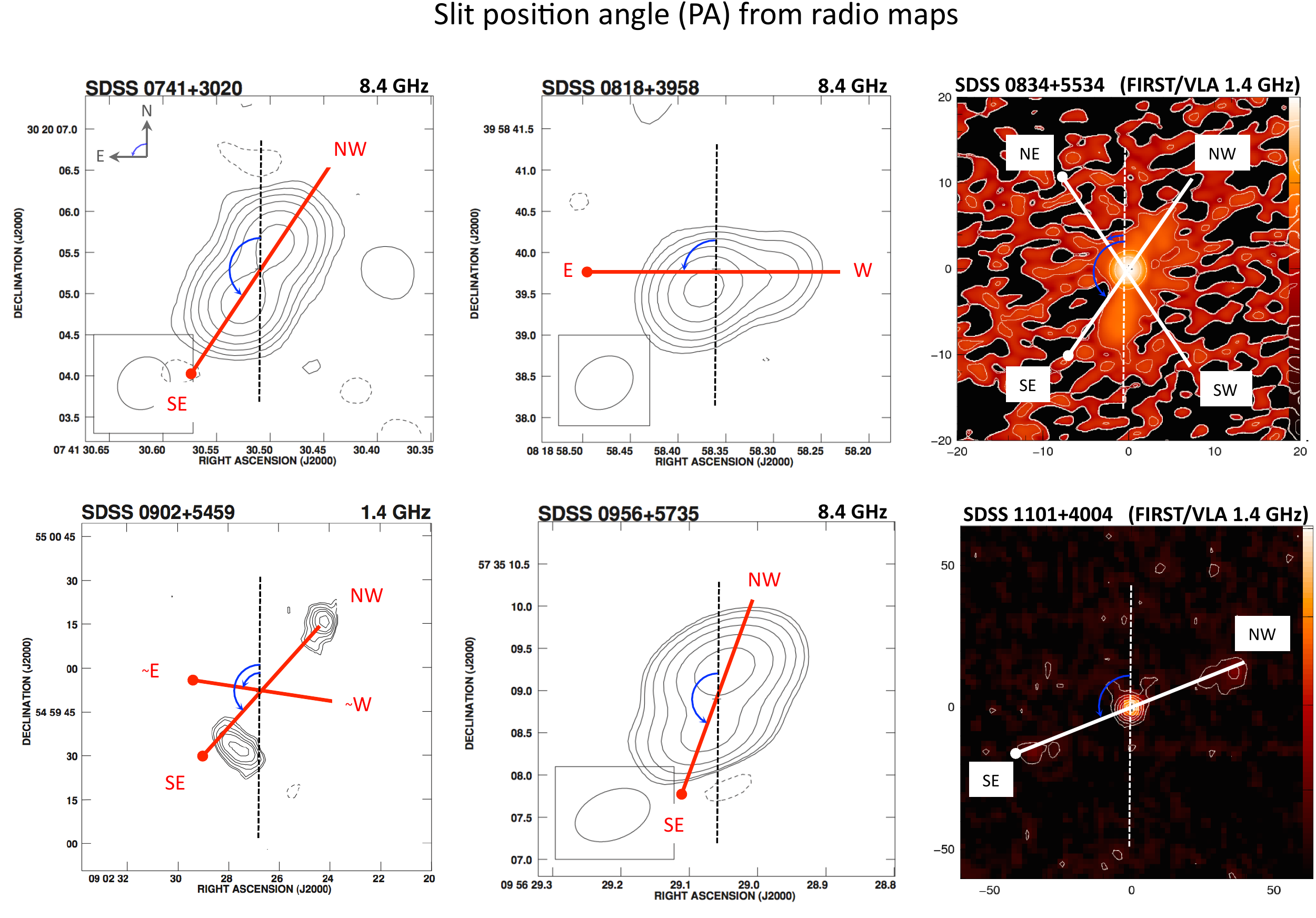}
\caption{Very Large Array radio maps (at 1.4 GHz ({\tt FIRST}) and 8.4 GHz) of the GTC sample. Four sources (SDSS 0741+3020, SDSS 0818+3958, SDSS 0902+5459, and SDSS 0956+5735) have previously been published in \citet{Lal10}. {\tt FIRST}/VLA images for SDSS 0834+5534 and SDSS1101+4004 have been obtained from the archive. All objects  show extended radio emission.
The positions of the slit used in the GTC spectroscopic observations are shown in red, except for two sources (SDSS 0834+5534 and SDSS 1101+4004), which are shown in white. The PA are shown in  Table~\ref{info1} (Col. 2). For each object, at least one PA was chosen along the main axis of the radio structures.}
\label{Lal_Ho}
\end{center}
\end{figure*}

The OSIRIS/GTC sample consists of six QSO2 at z$\sim$0.3--0.5, with log(L[OIII])$>$41.8, close to or above the threshold luminosity defined by \citet{Reyes08} for selecting type 2 AGN with luminosities in the quasar regime. All targets were selected as candidates to show potential signature of large-scale ($\text{exceeding}$~several kiloparsec) radio-induced feedback, based on the direct or indirect evidence for extended radio structures. 

\begin{figure*}
\begin{center}
\includegraphics[width=0.4\textwidth, height = 0.25\textwidth]{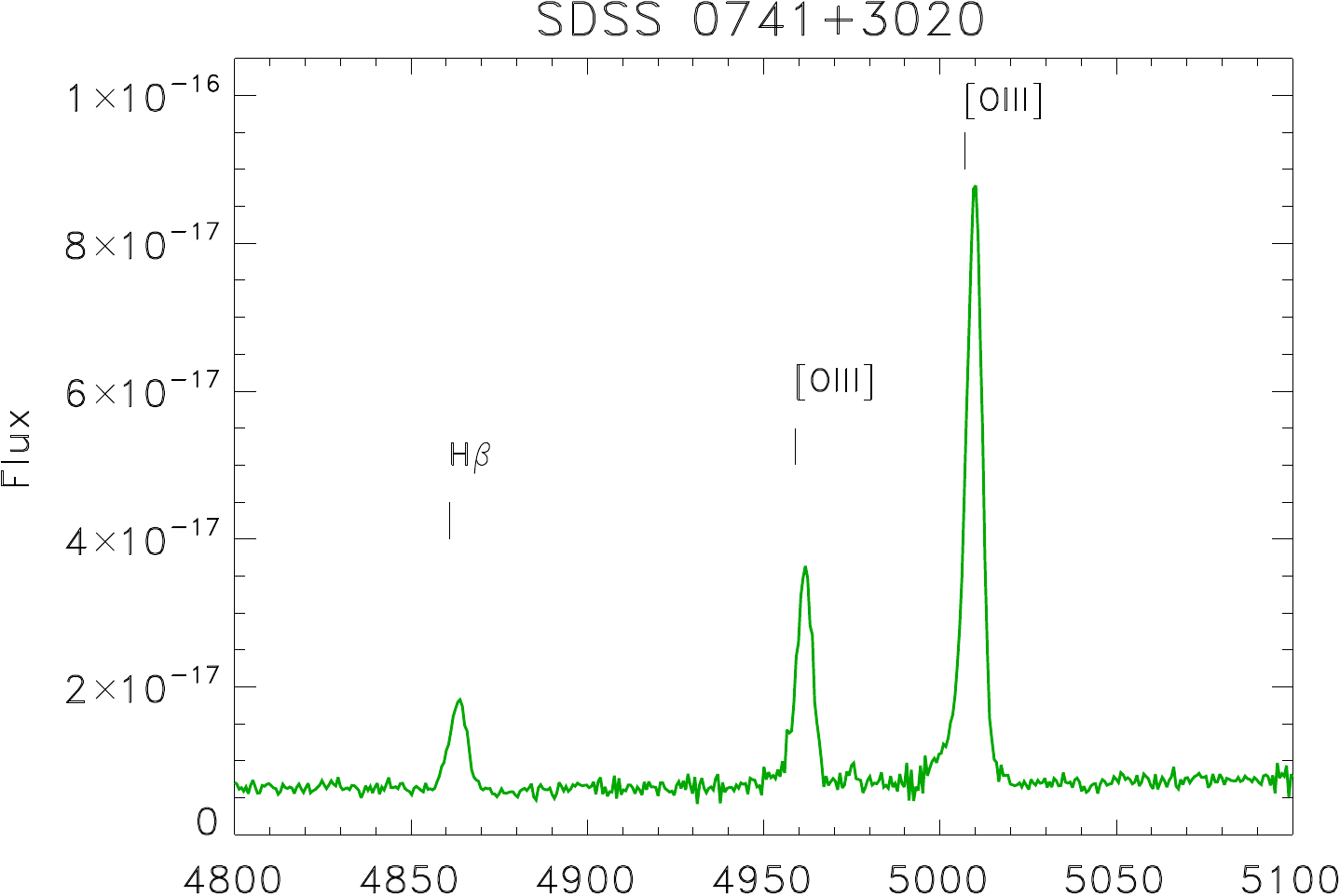}
\includegraphics[width=0.38\textwidth, height = 0.25\textwidth]{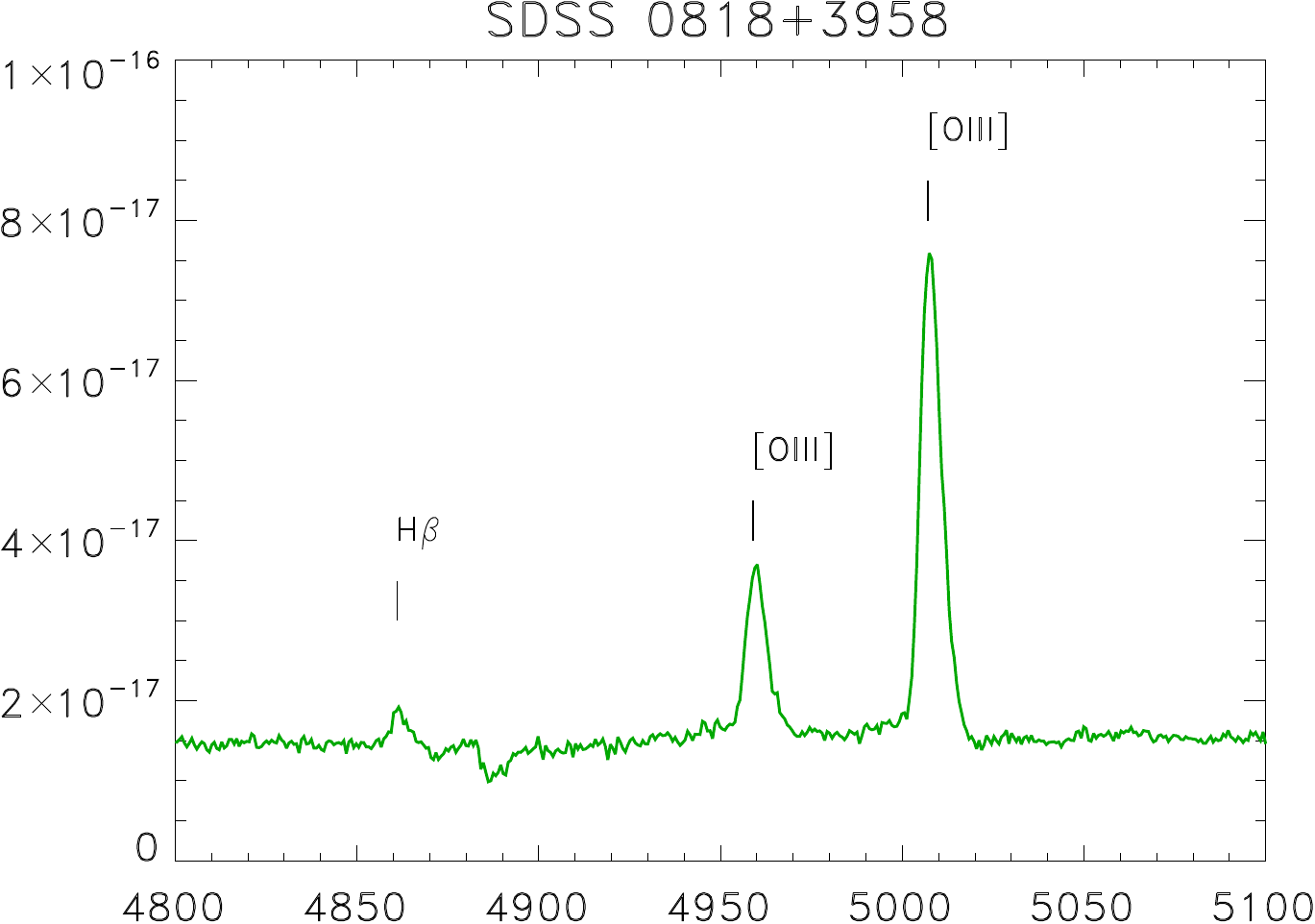}
\vskip3mm
\includegraphics[width=0.4\textwidth, height = 0.25\textwidth]{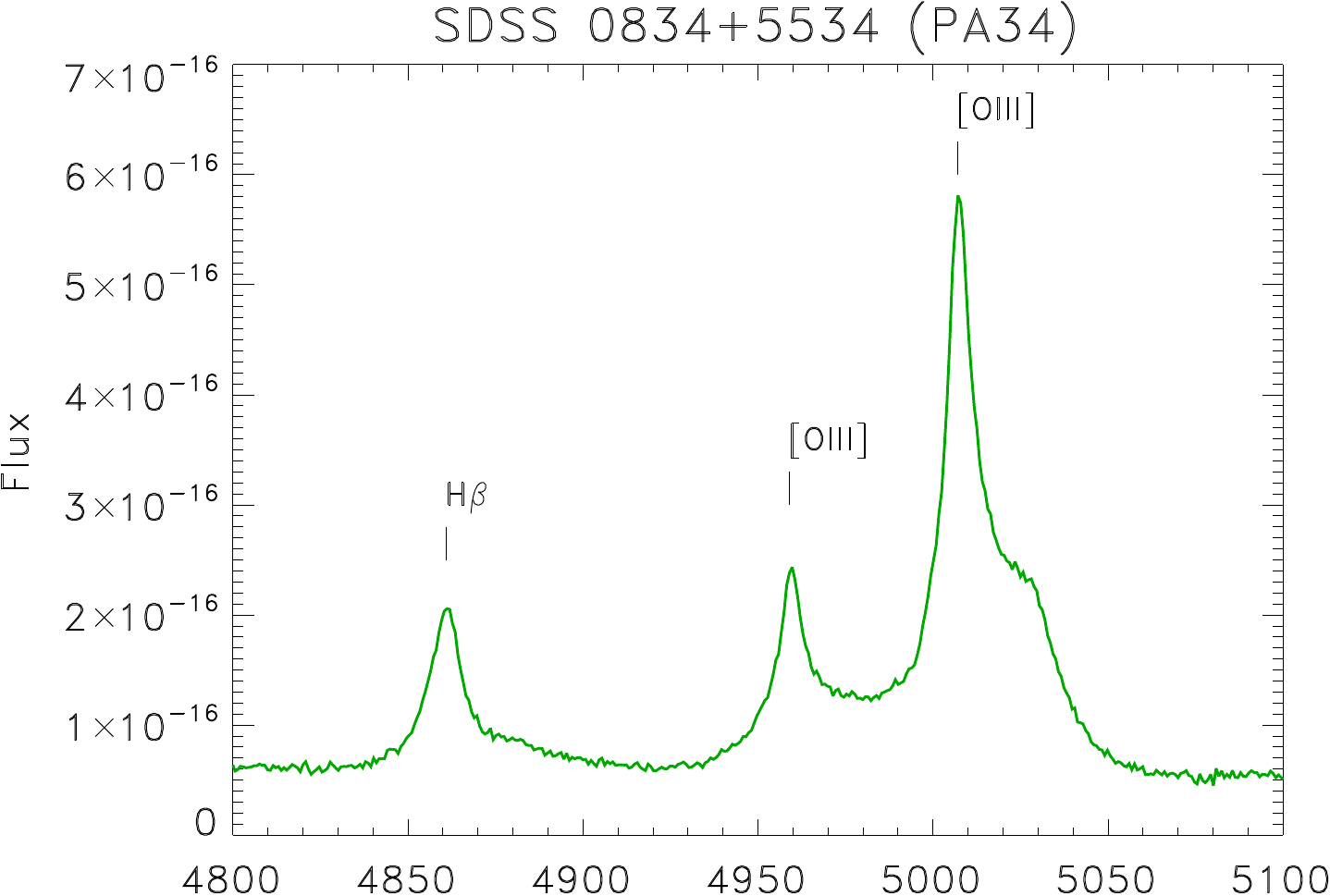}
\includegraphics[width=0.38\textwidth, height = 0.25\textwidth]{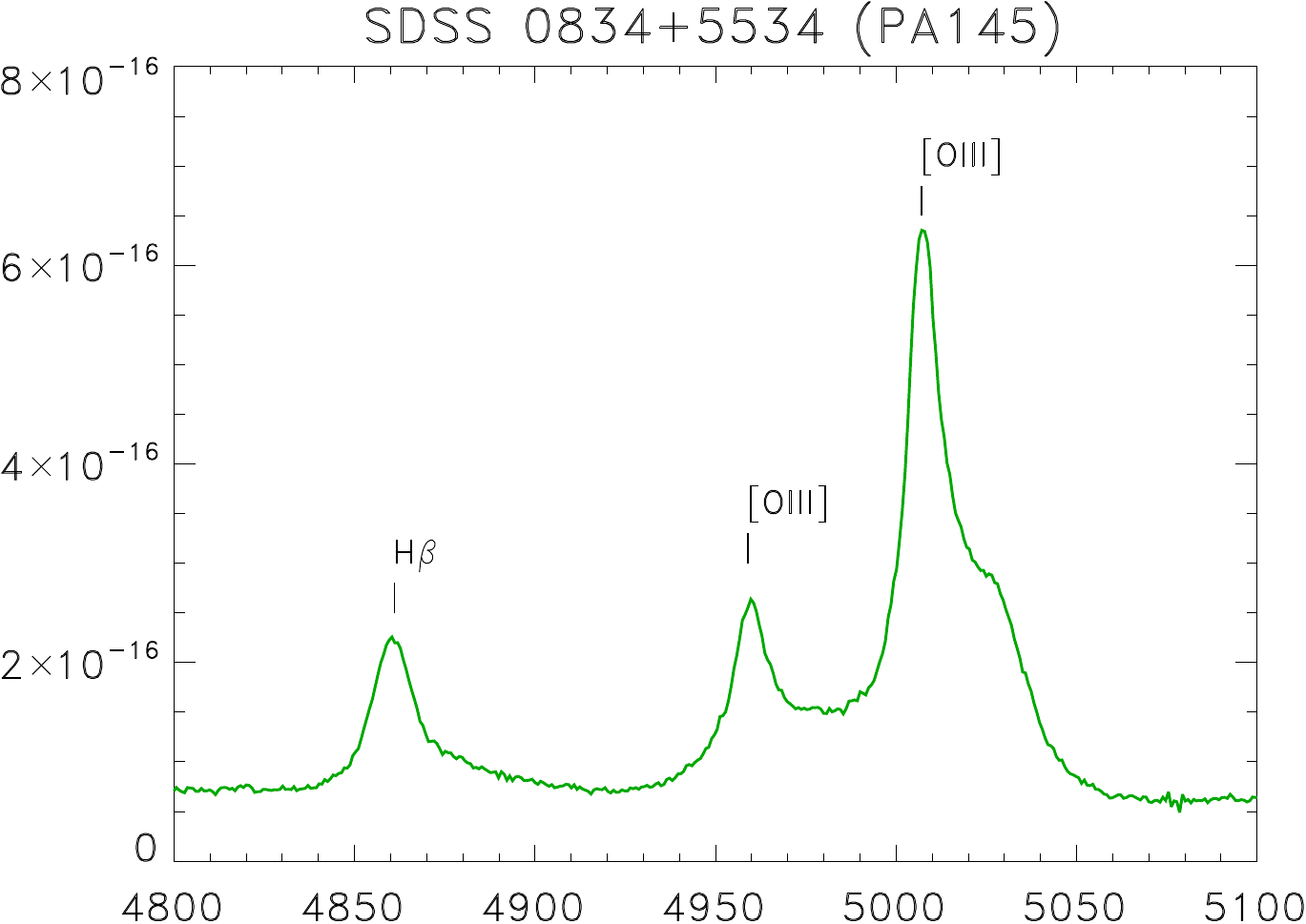}
\vskip3mm
\includegraphics[width=0.4\textwidth, height = 0.25\textwidth]{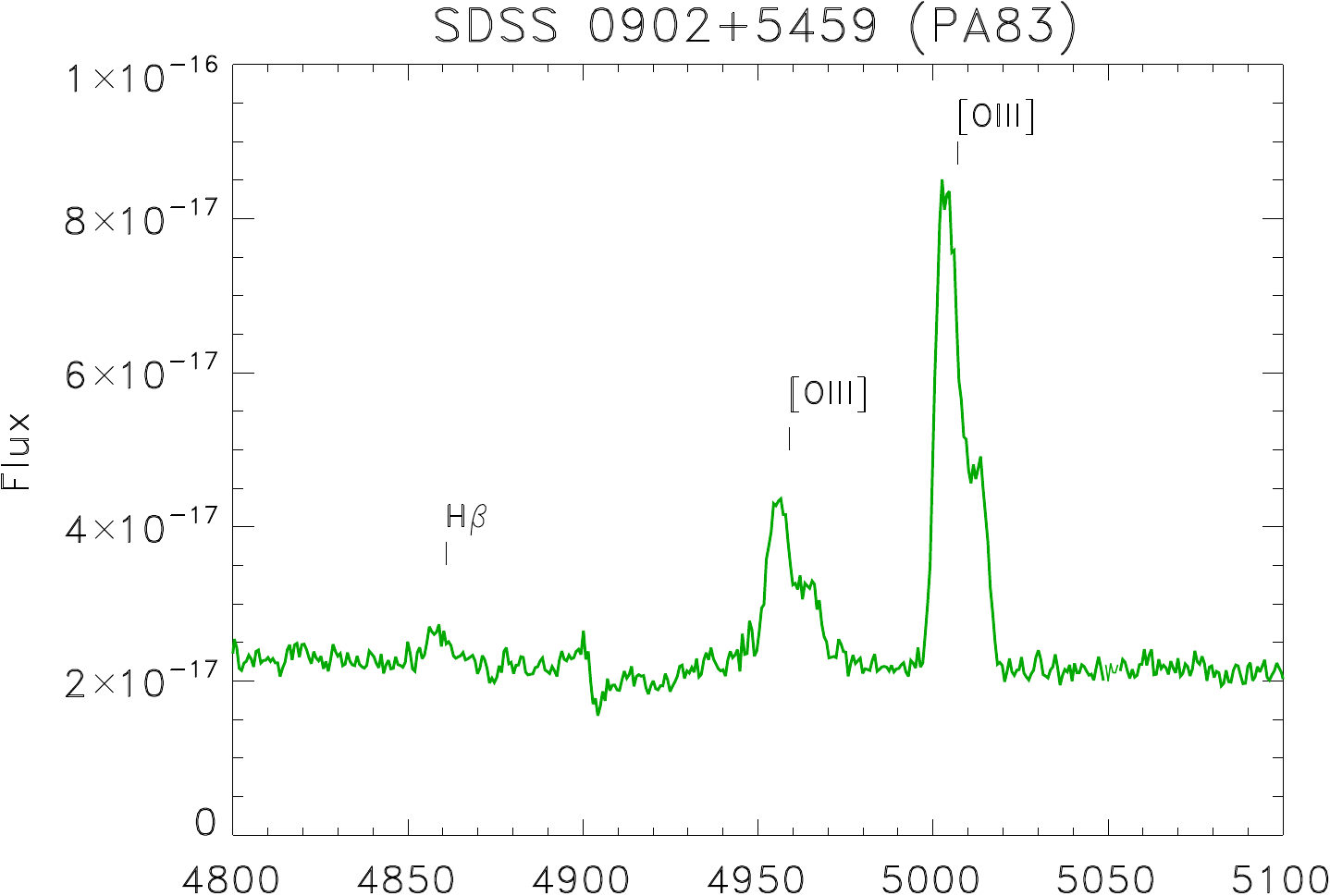}
\includegraphics[width=0.38\textwidth, height = 0.25\textwidth]{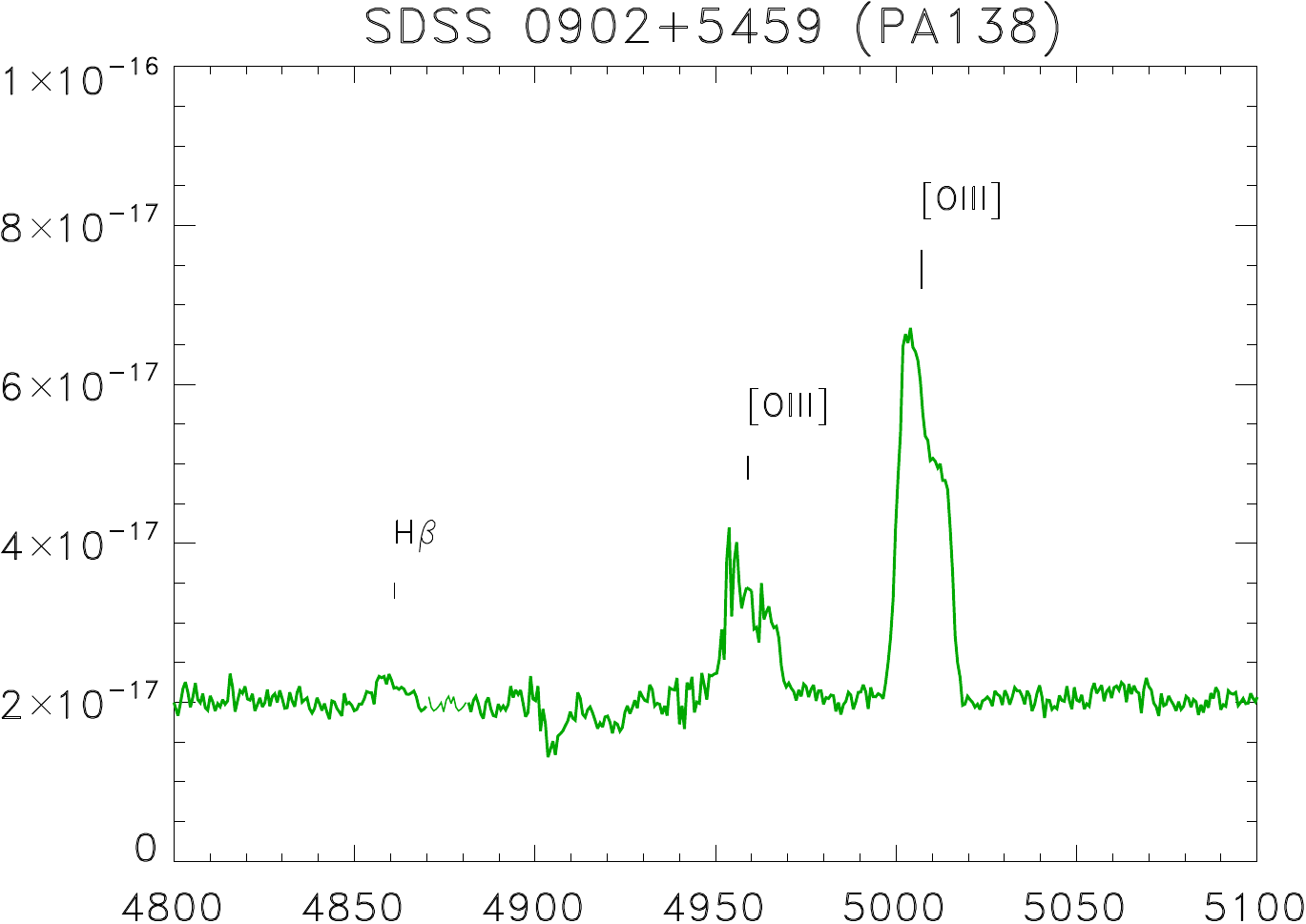}
\vskip3mm
\includegraphics[width=0.4\textwidth, height = 0.27\textwidth]{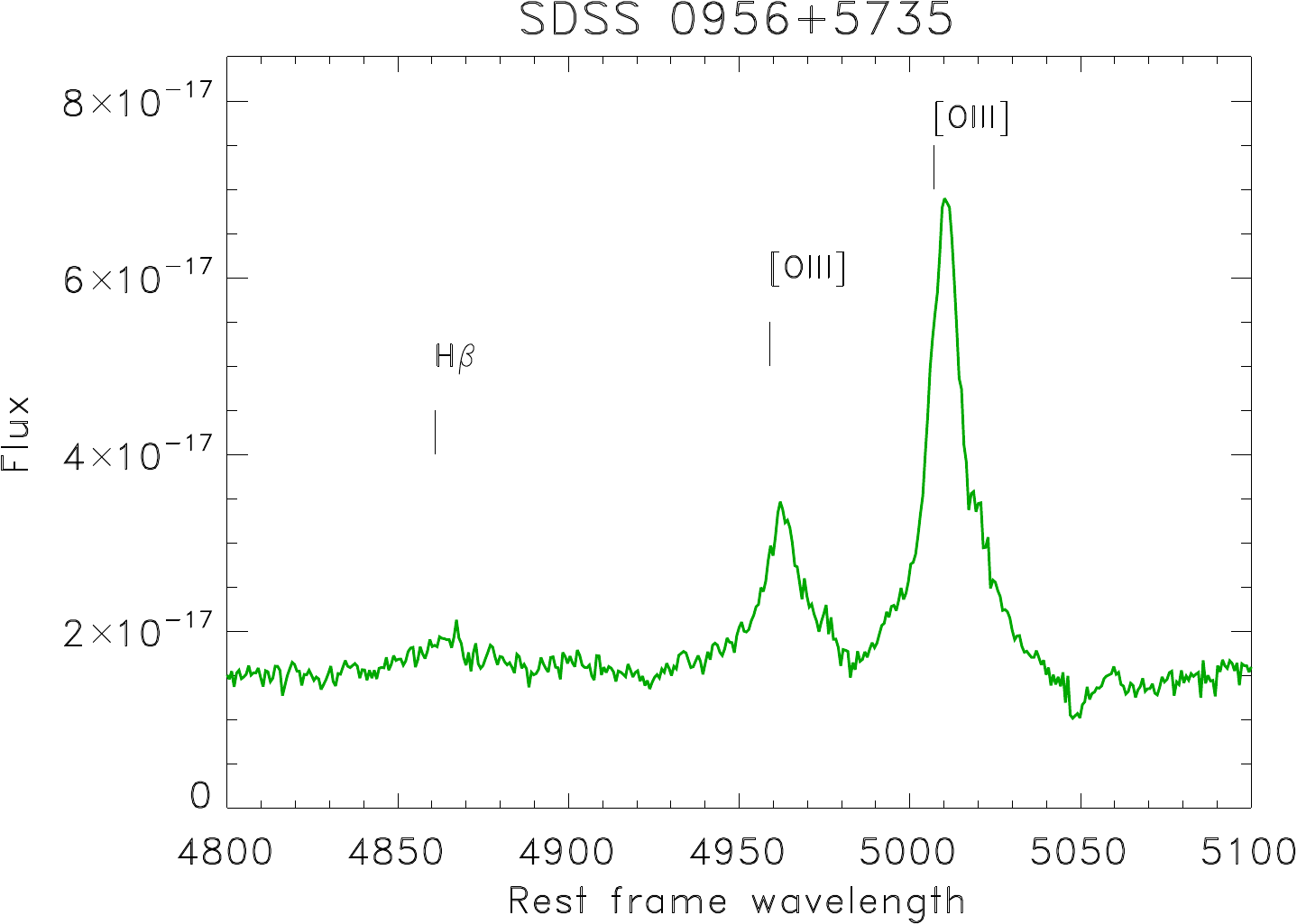}
\includegraphics[width=0.38\textwidth, height = 0.27\textwidth]{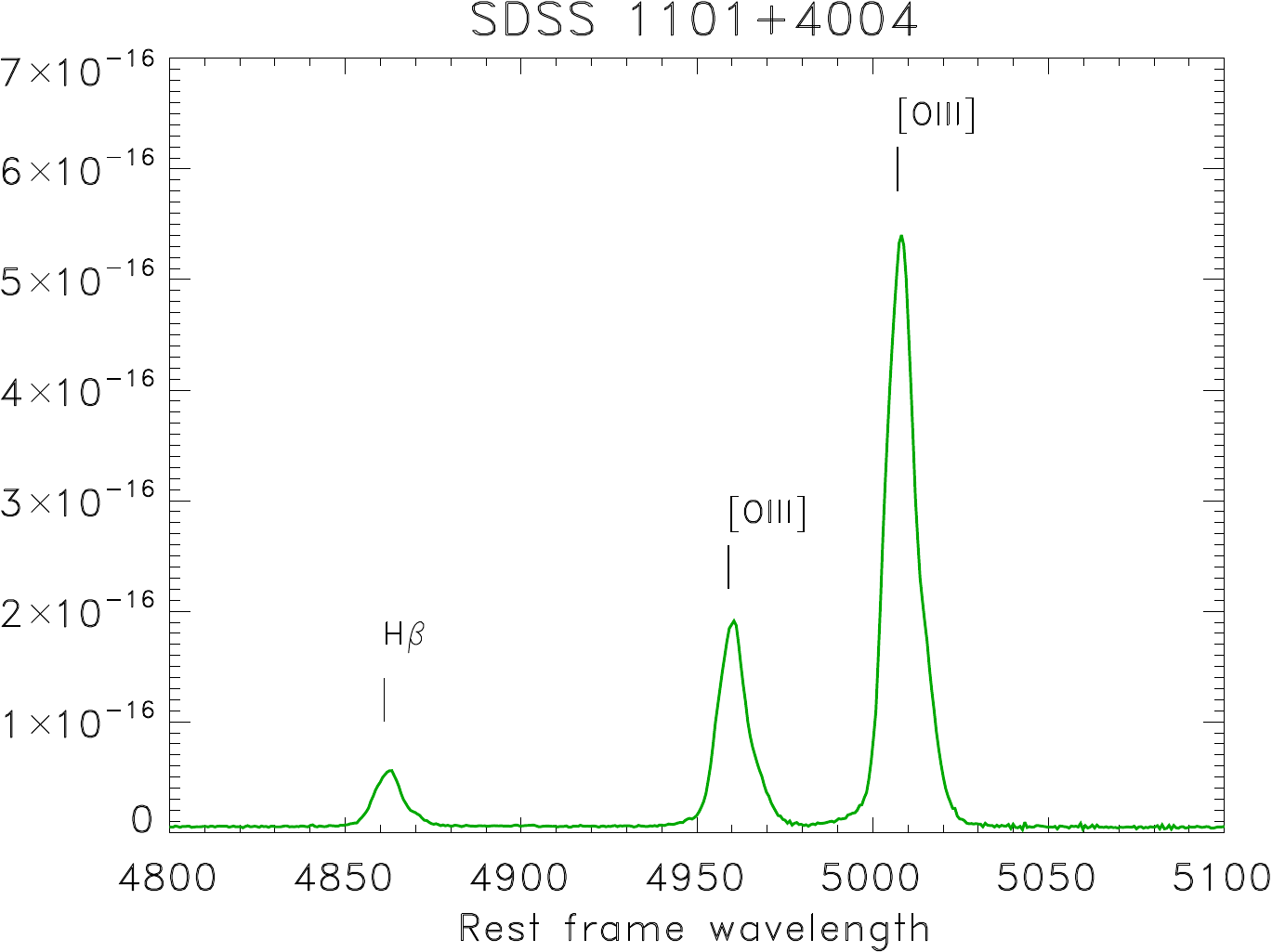}
\caption{
Nuclear OSIRIS/GTC spectra of the GTC sample. They were extracted considering apertures of 5 to 6 pixels along the slit (1.2\arcsec\ to 1.5\arcsec, respectively, depending on the slit width used). These were centered at the [OIII]$\lambda$5007 spatial centroid and selected to maximize the S/N of the line wings, where potential outflow signatures are usually identified. Fluxes are in erg~s~$^{-1}$ cm$^{-2}$ \AA$^{-1}$ and the rest--frame wavelengths are in \AA.}
\label{SDSS_spectra}
\end{center}
\end{figure*}

Four of them were part of the \citet{Lal10} sample, which was observed with the Very Large Array (VLA) at 8.4 GHz (SDSS 0902+5459 was also observed with {\tt FIRST}\footnote{Faint Images of the Radio Sky at Twenty Centimeters.}/VLA at 1.4 GHz), except for SDSS J0834+5534 and SDSS J1101+4004. 
SDSS J0834+5534 has been observed with {\tt FIRST}/VLA at 1.4 GHz, showing large-scale radio structures (up to 8.6$^{\prime\prime}$ or $\sim$32 kpc from the nucleus;  Fig.~\ref{Lal_Ho}).
It has also been selected for being radio-loud and for showing a complex and very broad nuclear [OIII] line profile (Fig.~\ref{SDSS_spectra}), suggestive of quite extreme nuclear outflows that might have been triggered by nuclear radio jets \citep{Mullaney13, VM14}.
SDSS J1101+4004 has also been observed using {\tt FIRST}/VLA at 1.4 GHz. The radio map shows a radio core and two radio hot spots at $\sim$~1\arcmin\ E and W (see Fig.~\ref{Lal_Ho}).

Table~\ref{info1} (Col.~8) lists the percentage of extended radio emission that we computed using the {\tt FIRST}/VLA and {\tt NVSS}/VLA radio flux values. The former has a resolution of 5\arcsec\ and the latter of 45\arcsec. The comparison of the two fluxes could give us an idea of the flux that is contained in the two beams (i.e., annular ring). We cannot completely rule out that any difference in flux between {\tt NVSS} and {\tt FIRST} is due to variability of the radio source. 
However, this is unlikely because this variability would occur in the core, and in Fig.~\ref{Lal_Ho} most of the radio emission comes from extended radio structures. Moreover, the fact that these sources are QSO2 makes it unlikely that a central radio jet would point along our line of sight, as is typically the case for rapidly varying blazars.

\begin{figure}
\begin{center}
\includegraphics[scale=0.6]{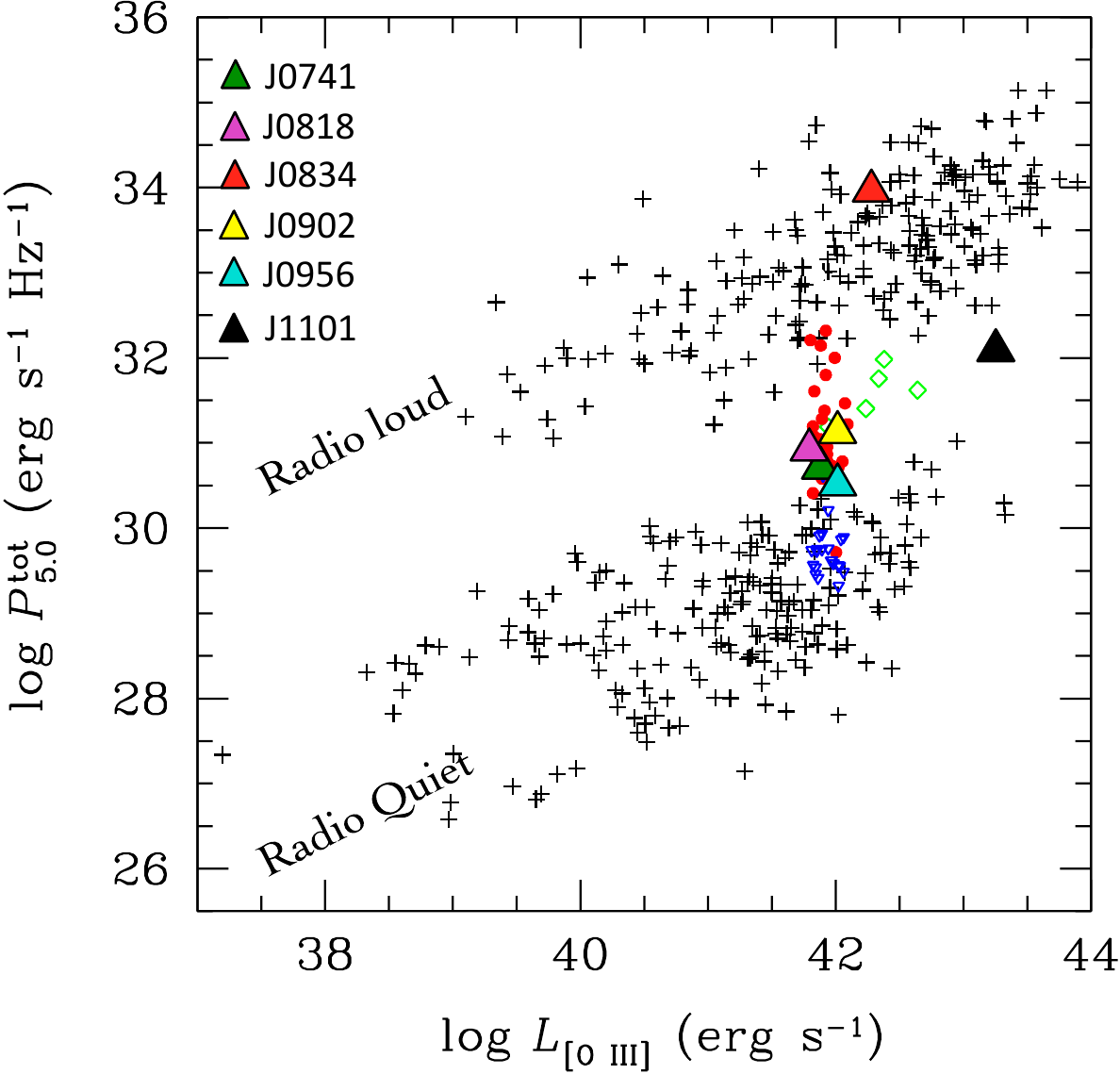}
\caption{Classification of the sample according to radio loudness. The figure is the same as Fig. 7 in \citet{Lal10}. The locations of our objects are shown with colored triangles. The cross symbols come from the AGN sample of \citet{Xu99}; the open green diamonds represent their radio-intermediate sources. The \citet{Lal10} sample of QSO2 at 0.3~$\leq$~z~$\leq$~0.7 is represented with filled red circles. Open blue triangles  are upper limits.}
\label{sample_qso}
\end{center}
\end{figure}

We classified the objects according to their radio loudness. For this, we located them in the P$_{5 GHz}$ versus L$_{[OIII]}$ plane (Fig.~\ref{sample_qso}) following \citet{Xu99}. P$_{5 GHz}$ was computed using the formula P$_{5 GHz}$ = 4$\pi$ D$_L^2$ S$_{5 GHz}$ (1+z)$^{-1-\alpha}$, where D$_L$ is the luminosity distance, S$_{5 GHz}$ is the observed flux density at 5 GHz, and $\alpha$ is the spectral index such that S$_\nu$ $\propto\nu^\alpha$. The $\alpha$ index was taken from \citet{Lal10} for the four QSO2 in their sample, computed as $\alpha = \frac{log(S_1/S_2)}{log(\nu_2/\nu_1)}$.
For the other two targets, we computed $\alpha$ from the available P$_{1.4 GHz}$ and P$_{8 GHz}$ fluxes. This index covers the range between --0.9 to 0.7.  Two of the six targets are radio-quiet (SDSS 0741+3020 and SDSS 0956+5735), three are radio-intermediate (SDSS 0818+3958, SDSS 0902+5459 and SDSS 1101+4004), and only one is classified as radio-loud (SDSS 0834+5534). Throughout the paper we refer to these targets as the GTC sample.

\begin{figure}
\begin{center}
\includegraphics[scale=0.45]{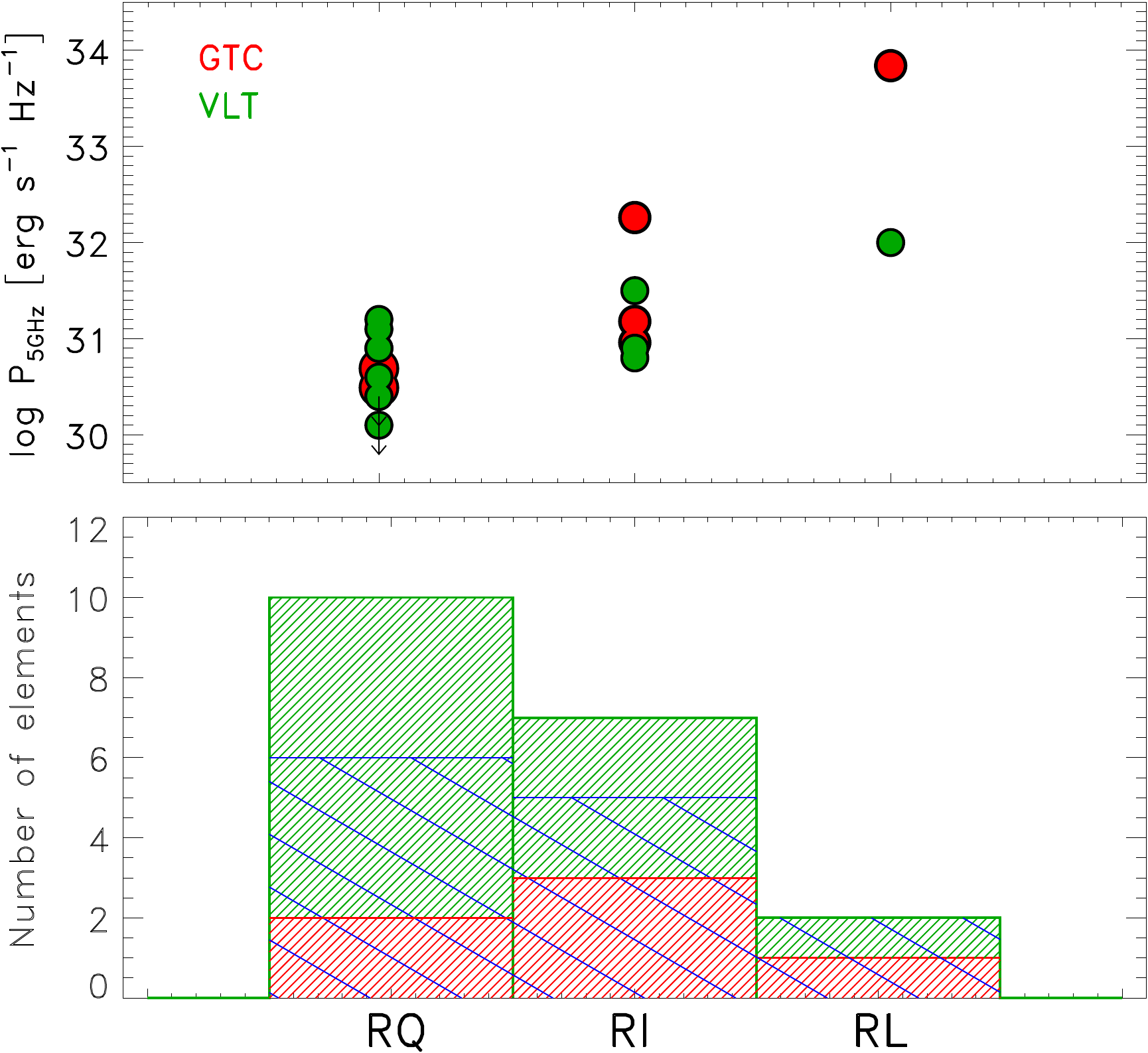}
\caption{{\it Top:} Distribution of the objects of the GTC and VLT samples in the radio power at 5 GHz (logarithmic scale) -- radio-loudness plane (i.e., radio-quiet (RQ), intermediate (RI), and loud (RL), see text). The GTC (red) and the VLT (green) objects are considered. 
{\it Bottom:} Number of elements in each radio-loudness bin, identified according to their respective color code. The number of objects with confirmed or indirect evidence for extended radio emission on scales larger than 5\arcsec \ is indicated with the dashed blue lines.}
\label{class_radio}
\end{center}
\end{figure}

We then enlarged our sample considering 13 QSO2 observed with FORS2/VLT at similar redshift that were previously analyzed in VM11 and VM16 (hereafter, the VLT sample). Eight of 13 sources are in the radio-quiet regime, 4 of 13 are radio-intermediate, and only one system is radio-loud.  Evidence for extended radio emission is found in 7 of 13 objects. This is based on the extended radio morphology in two cases (SDSS 1014+0244 and SDSS 1247+0152 , {\tt FIRST}/VLA images).  Five more objects show indirect evidence for extended radio emission, which also suggests extended radio structures. We show in Fig.~\ref{class_radio} the distribution of the GTC and VLT objects according to their radio power and radio loudness. Six radio-quiet, 5 radio-intermediate and 2 radio-loud systems show direct or indirect evidence for extended radio sources.

\begin{table*}
\centering 
  \caption{General properties of the QSO2 sample observed with OSIRIS/GTC.}
\label{info1}
\begin{tiny}
\begin{tabular}{c cccccccc}
 \hline\hline\noalign{\smallskip}
Galaxy ID        &       Short name      &              z       & Scale   & log(L$_{[OIII]}$) & log(P$_{5GHz}$)  &       Radio  & Extended radio emission\\
{\smallskip} 
SDSS code & SDSS\_PA& & &  &    &  &in the annular ring ({\tt FIRST--NVSS})\\
 &  & & (kpc/$^{\prime\prime}$) & (erg s$^{-1}$) & (erg s$^{-1}$ Hz$^{-1}$) & &  (\%)\\
 (1) & (2) &(3) & (4) & (5) & (6) & (7) & (8)\\
\hline\noalign{\smallskip}
J074130.50+302005.1 & 0741+3020\_148            &  0.476 & 5.922 &      41.93         &       30.69   &       RQ      &        38\\
\hline\noalign{\smallskip}
J081858.36+395839.7 &  0818+3958\_90    &  0.406 & 5.392 &      41.83    & 30.96 &       RI &  36\\
\hline\noalign{\smallskip}
J083454.90+553421.1 &  0834+5534\_34    &  0.241        & 3.773         &     42.27  & 33.84      &       RL & 3\\
J083454.90+553421.1 &  0834+5534\_145   &  0.241        & 3.773         &          =  &  =        &       --\\
\hline\noalign{\smallskip}
J090226.73+545952.2 &  0902+5459\_83    &  0.401 & 5.356 &    42.03   & 31.18         &       RI & 8$^{a}$\\
J090226.73+545952.2 &  0902+5459\_138   &  0.401 & 5.356 &  =    & =    &       --\\
\hline\noalign{\smallskip}
J095629.03+573508.8 &  0956+5735\_160   &  0.361 & 5.007        & 41.98    &     30.49   &       RQ      & 41$^b$        \\
\hline\noalign{\smallskip}
J110140.53+400422.9 &  1101+4004\_111   &  0.456 & 5.780        & 43.29     &    32.26   &       RI & 29\\
\hline\hline\noalign{\smallskip}
\end{tabular}
\vskip0.2cm\hskip0.0cm
\end{tiny}
\begin{minipage}[h]{18.5cm}
\footnotesize
{ {\bf Notes:} Column (1): SDSS galaxy ID.
Column (2): Short name galaxy ID with the information of the slit PA used (angle given from north to east). 
Column (3): Redshift of the SDSS sources from the NASA Extragalactic Database (NED).
Column (4): Scale. 
Column (5): Total luminosity of the [OIII]$\lambda$5007\AA\ line (logarithmic scale).
Column (6): Radio power  at 5 GHz. 
Column (7): Radio loudness (RQ: radio-quiet; RI: radio-intermediate; and RL: radio-loud). 
Column (8): Percentage of extended radio flux emission at 1.4 GHz between the {\tt FIRST} and the {\tt NVSS} beams (5$^{\prime\prime}$ vs. 45$^{\prime\prime}$ resolution, respectively).  \\ $^a$ The {\tt FIRST} image shows that almost all the emission is in the lobes with marginal evidence for a radio core (Fig. 1). $^b$ This value is quite  uncertain because the source is detected at only 3$\sigma$ level in {\tt NVSS}.}
\end{minipage}
\end{table*}

\begin{table*}
\centering   \caption{Properties of the complementary QSO2 sample observed with FORS2/VLT.}
\label{info2}
\begin{tiny}
\begin{tabular}{cccccccc}
 \hline\hline\noalign{\smallskip}
Galaxy ID        &       Short name      &              z       & Scale   & log(L$_{[OIII]}$) & log(P$_{5GHz}$)  &       Radio  & Extended radio emission\\
{\smallskip} 
SDSS code & SDSS  &  & & &  & & in the annular ring ({\tt FIRST--NVSS})\\
  &   &  & (kpc/$^{\prime\prime}$) & (erg s$^{-1}$) & (erg s$^{-1}$ Hz$^{-1}$) & &(\%) \\
 (1) & (2) &(3) & (4) & (5) & (6) & (7) & (8)\\
\hline\noalign{\smallskip}
J092318.06+010144.8&  J0923+0101 & 0.386 & 5.226  &     42.36           &       30.6         &       RQ      &       51\\
\hline\noalign{\smallskip}
J095044.69+011127.2 &  J0950+0111 & 0.404 &  5.376  &   41.80           &       30.9         &       RI      &       --\\
\hline\noalign{\smallskip}
J095514.11+034654.2 &  J0955+0346 & 0.422 & 5.520   &   42.05           &       30.9            &       RI      &       12\\
\hline\noalign{\smallskip}
J101403.49+024416.4 &  J1014+0244 & 0.573 &6.526   &    41.87           &       31.5         &       RI      &       -- \\
\hline\noalign{\smallskip}
J115314.36+032658.6 &  J1153+0326 & 0.575 & 6.537  &    43.24           &       31.2            &       RQ      &       --\\
\hline\noalign{\smallskip}
J124749.79+015212.6 &  J1247+0152 & 0.427  & 5.559   & 41.81            &       32.0         &       RL  &   28 \\
\hline\noalign{\smallskip}
J130740.55--021455.2 &  J1307--0214& 0.424 & 5.536   &  42.59           &       30.9         &       RQ      &       --\\
\hline\noalign{\smallskip}
J133633.65--003936.4 &  J1336--0039 & 0.416 & 5.473  &  42.22           &       $\lesssim$30.4         &       RQ      & 34    \\
\hline\noalign{\smallskip}
J133735.01--012815.6 &  J1337--0128 & 0.328 &  4.700   &        42.31            &       30.9    &       RQ      &       21\\
\hline\noalign{\smallskip}
J140740.06+021748.3 &  J1407+0217 & 0.309 &  4.512 & 42.58      &       30.4         &       RQ      &       27\\
\hline\noalign{\smallskip}
J141315.31--014221.0&  J1413--0142 & 0.380 &  5.175  &  42.76           &       31.1         &       RQ      &       --\\
\hline\noalign{\smallskip}
J143027.66--005614.8 &  J1430--0050 & 0.318 & 4.602  & 42.02            &       $\lesssim$30.1         &       RQ      &       --\\
\hline\noalign{\smallskip}
J154613.27--000513.5 &  J1546--0005 & 0.383 & 5.200  &  41.76           &       30.8         &       RI      &       --\\
\hline\hline\noalign{\smallskip}
\end{tabular}
\vskip0.2cm\hskip0.0cm
\end{tiny}
\begin{minipage}[h]{18.5cm}
\footnotesize
{Caption as in Table~\ref{info1}. The minus means that it has not been possible to assign a value because the source is weak and it is only detected in {\tt FIRST}.}
\end{minipage}
\end{table*}

\section{Observations and data reduction}

Long-slit spectroscopic observations were performed on February 25--26, 2017 (program GTC13--16B) in visitor mode with the optical imager and long-slit spectrograph OSIRIS mounted on the 10.4m GTC. The R2500R grating was used, which provides a spectral coverage of 5575--7685 \AA\ with dispersion 1.04 \AA\ pixel$^{-1}$ and a pixel scale of 0.254$^{\prime\prime}$ pixel$^{-1}$. A 1.5$^{\prime\prime}$ wide slit was used with the exception of one object (SDSS 1101+4004), for which a 1.2$^{\prime\prime}$ slit has been considered (see Tab.~\ref{info_observation}).
For all sources, one slit position angle (PA) was selected to coincide with the axis of the extended radio source, as measured from \citet{Lal10} radio maps. Two slit PAs were used for the SDSS 0834+5534 and SDSS 0902+5459 targets to achieve a more complete spatial coverage (see Fig.~\ref{Lal_Ho}).

For each object at least one PA was chosen along the main axis of the radio structures. In principle, this maximizes the chances of detecting extended ionized nebulae because the quasar ionization cone axis is expected to be roughly aligned with the radio axis. It also enhances the probability of detecting the effects of radio-induced feedback in the extended ionized gas.

The total exposure time per object was 6 $\times$ 900s = 5400s, except for the systems SDSS 1101+ 4004 and SDSS 0834-5534, for which total exposure times of 6 $\times$ 600s (3600s) and 8 $\times$ 300s (2400s) were considered, respectively.

The spectral resolution FWHM$_{inst}$ values measured from several prominent sky lines (IP FWHM) are listed in Table~\ref{info_observation} for each galaxy, with a typical median value of 5.3 \AA\ for the 1.5$^{\prime\prime}$ wide slit and 4.5 \AA\ for the 1.2$^{\prime\prime}$ wide slit.
The seeing FWHM size was in the range 1.0$^{\prime\prime}$--2.6$^{\prime\prime}$ during the observations, as measured from several stars in the acquisition images. Information on the observations and weather conditions is summarized in Table~\ref{info_observation}.

The data reduction was made following the same prescriptions as in \citet{VM17}. We summarize them here. The spectra were reduced using standard procedures and IRAF tasks. First the bias and flat--field corrections were applied to the images using lamp flats. Then, the 2--dim spectra were wavelength calibrated using Xe+Ne+HgAr lamps, with a resulting error consistent with the nominal spectral resolutions of the grisms. After this, sky background was subtracted and a 2--dim spectrum was obtained. The flux was calibrated using observations of spectrophotometric standard stars, white dwarfs, which were obtained in the same nights as the scientific spectra. Finally, the different individual spectra were averaged, from which we obtained a mean spectrum for each of the orientations. Atmospheric and foreground galactic extinction corrections were also applied.

\begin{table*}
\centering 
\caption{Observing information of the QSO2 sample observed with OSIRIS/GTC.}
\label{info_observation}
\begin{tiny}
\begin{tabular}{l ccccccc}
 \hline\hline\noalign{\smallskip}
 Short name      &              IP  FWHM        & Slit width    & Seeing (slit width)   & Seeing  & DIMM seeing   & Airmass \\
 SDSS/PA &      (\AA) & ($^{\prime\prime}$)     & ($^{\prime\prime}$) & ($^{\prime\prime}$) & ($^{\prime\prime}$) & ($^{\prime\prime}$)  \\
 (1) & (2) &(3) & (4) &(5) & (6) & (7) \\
\hline\noalign{\smallskip}
 0741+3020                      &  5.29 $\pm$ 0.08      &  1.5  &        1.40 $\pm$ 0.04                 &       1.47 $\pm$ 0.04 & 1.6           &1.17   \\
\hline\noalign{\smallskip}
 0818+3958                      & 5.32 $\pm$ 0.08       &  1.5  &       1.69 $\pm$ 0.06              &       1.84 $\pm$ 0.07  & 1.6  &1.04   \\
\hline\noalign{\smallskip}
 0834+5534/34                   & 5.50 $\pm$ 0.46               & 1.5           & 0.98 $\pm$      0.02            &       1.00 $\pm$ 0.02 & 1.5   & 1.23  \\
 0834+5534/145          & 5.99 $\pm$ 0.69       & 1.5           & 0.78 $\pm$ 0.02            &       0.84 $\pm$ 0.02 & 1.5           & 1.32  \\
\hline\noalign{\smallskip}
 0902+5459/83           &  5.31 $\pm$ 0.15              & 1.5   &       1.67 $\pm$ 0.08      &       1.87 $\pm$ 0.09 & 1.7           & 1.14  \\
 0902+5459/138          &  5.24 $\pm$ 0.08      &  1.5          &      1.88 $\pm$ 0.07      &        2.16 $\pm$ 0.08 & 1.7  & 1.20  \\
\hline\noalign{\smallskip}
 0956+5735                      &  5.25 $\pm$ 0.28              & 1.5   &1.72   $\pm$ 0.15    &        2.10 $\pm$ 0.18        & 1.6           &1.20   \\
\hline\noalign{\smallskip}
  1101+4004                     & 4.46 $\pm$ 0.05       & 1.2   &       2.59    $\pm$ 0.12    &       2.64 $\pm$ 0.12 & 1.5            &      1.12\\
\hline\hline\noalign{\smallskip}
\end{tabular}
\vskip0.2cm\hskip0.0cm
\end{tiny}
\begin{minipage}[h]{18.5cm}
\footnotesize
{\bf Notes:} Column (1): Short name of the source. The slit PA is also quoted, except when only one slit is used for the observations (see Table~\ref{info1}).
Column (2): FWHM of the IP as derived from several sky lines.
Column (3): Slit width in arcsec. 
Column (4): Seeing  computed using an aperture equivalent to the slit width.
Column (5): Seeing  computed from several stars in the acquisition image.
Column (6): Differential image motion monitor seeing value. It gives an approximate estimate of the real seeing during the night.
Column~(7): Airmass during the observations.
\end{minipage}
\end{table*}

\subsection{Seeing}

We investigated whether the ionized outflows identified in the GTC spectra were spatially resolved  and constrained their sizes. For this, we compared the spatial profile of the outflow emission (always identified with the broadest kinematic component)  with that of the seeing. 

As discussed in \citet{VM16}, an accurate characterization of the size (FWHM) and shape of the seeing is key to this aim. This is particularly relevant in the current work, given the strong variability of the seeing during the observing run  with FWHM$\sim$0.8\arcsec--2.6\arcsec (Table~\ref{info1}).
We followed the same method as these authors. For each object, we built the spatial profile of the seeing disk along the slit using nonsaturated stars with well-detected wings in the acquisition images that were obtained at similar time as the spectra. The seeing profiles were extracted from apertures of the same width in arcsec as the narrow slit of the spectroscopic observations \citep[see][for details]{VM16}.

\section{Kinematic analysis method}
\label{data_red}

The [OIII]$\lambda\lambda$4959,5007 emission lines were fit with Gaussian profiles using an IDL routine (i.e., MPFITEXPR, implemented by Markwardt). Both lines were fit simultaneously.

We selected pixels where both [OIII] lines were detected with a signal-to-noise ratio S/N$>$3, which automatically fit all lines to single-Gaussian profiles (i.e., one-component fit) as a first approximation. We found, however, that two or three kinematic components are generally needed to produce a good fit at different locations for most of the sources. We used in all cases the minimum number of Gaussian components required to produce a good fit.
When more than one kinematic component was required by the fits, the doublet line flux ratio and wavelengths were fixed according to atomic physics. The widths of each doublet were constrained to be the same and greater than the instrumental profile ($\sigma_{INS}$).

 For each component we measured the line flux, central wavelength ($\lambda_c$), and intrinsic width ($\sigma_{line}$ = $\sqrt{ (\sigma^2_{obs} - \sigma^2_{INS}) }$), as well as their corresponding fitting errors.

\subsection{Derivation of the main kinematic parameters}
\label{kin_pars}

To characterize the kinematics of the sample, we used an approach similar to that used in \citet{Bellocchi13} (hereafter, B13). We note that our study is limited by the incomplete spatial coverage intrinsic to the long-slit spectroscopic technique in comparison with integral field spectroscopy (IFS). The possible impact of the incomplete spatial coverage on the analysis and interpretation of the data is specifically explained when relevant. It is not expected to affect the conclusions of this work.

\begin{figure*}[htbp]
\includegraphics[scale=0.325]{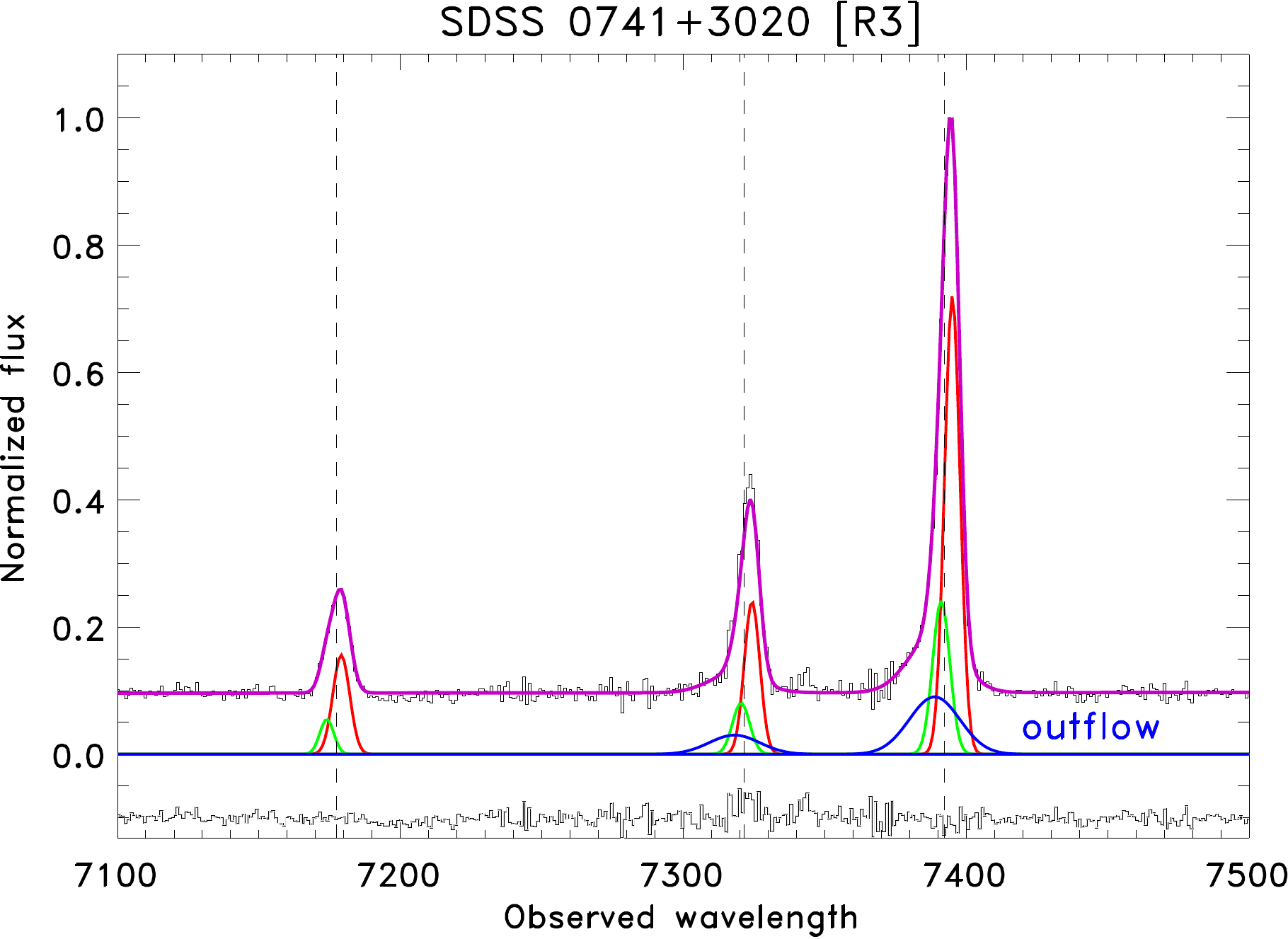}
\includegraphics[scale=0.325]{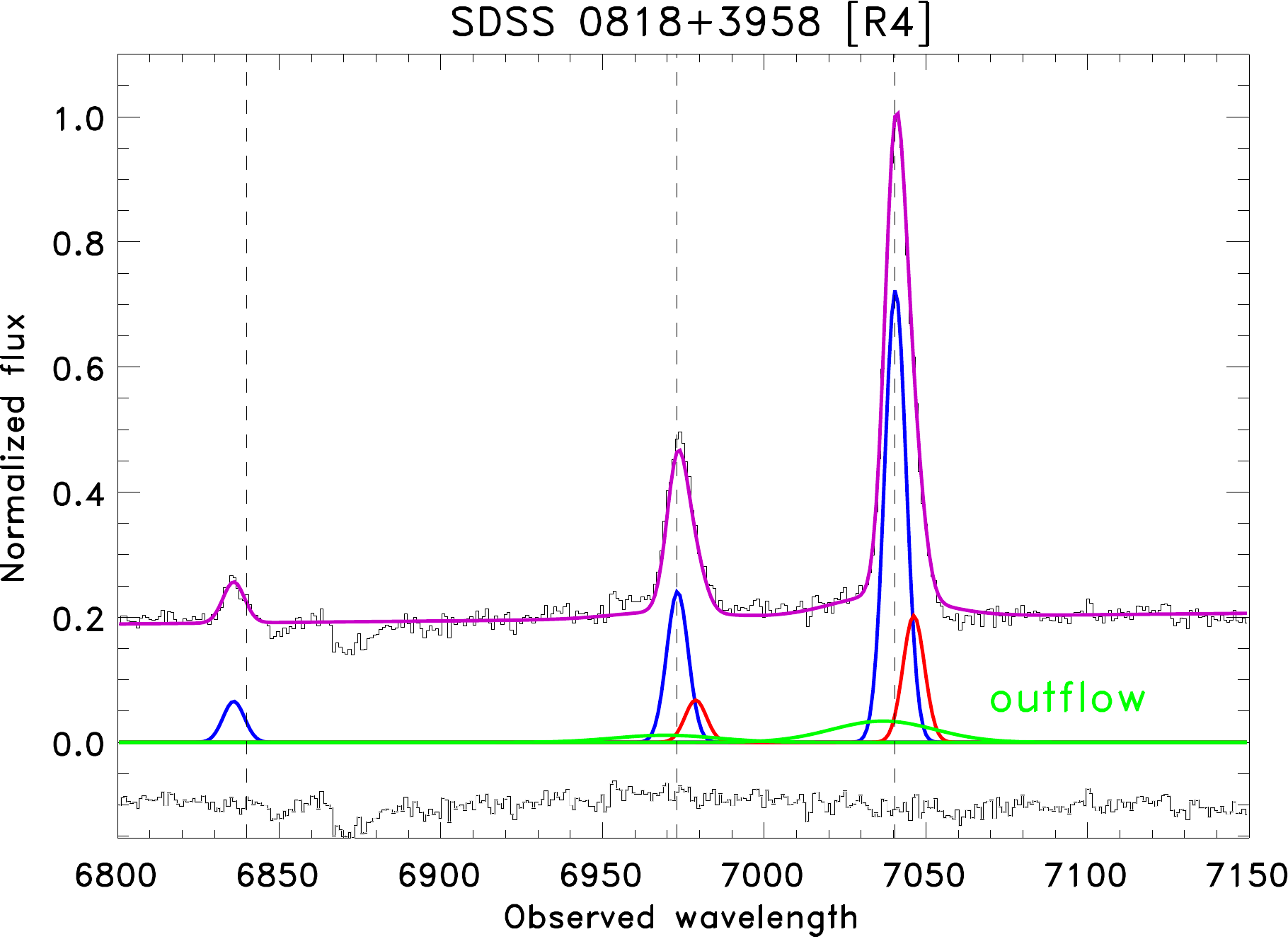}
\includegraphics[scale=0.325]{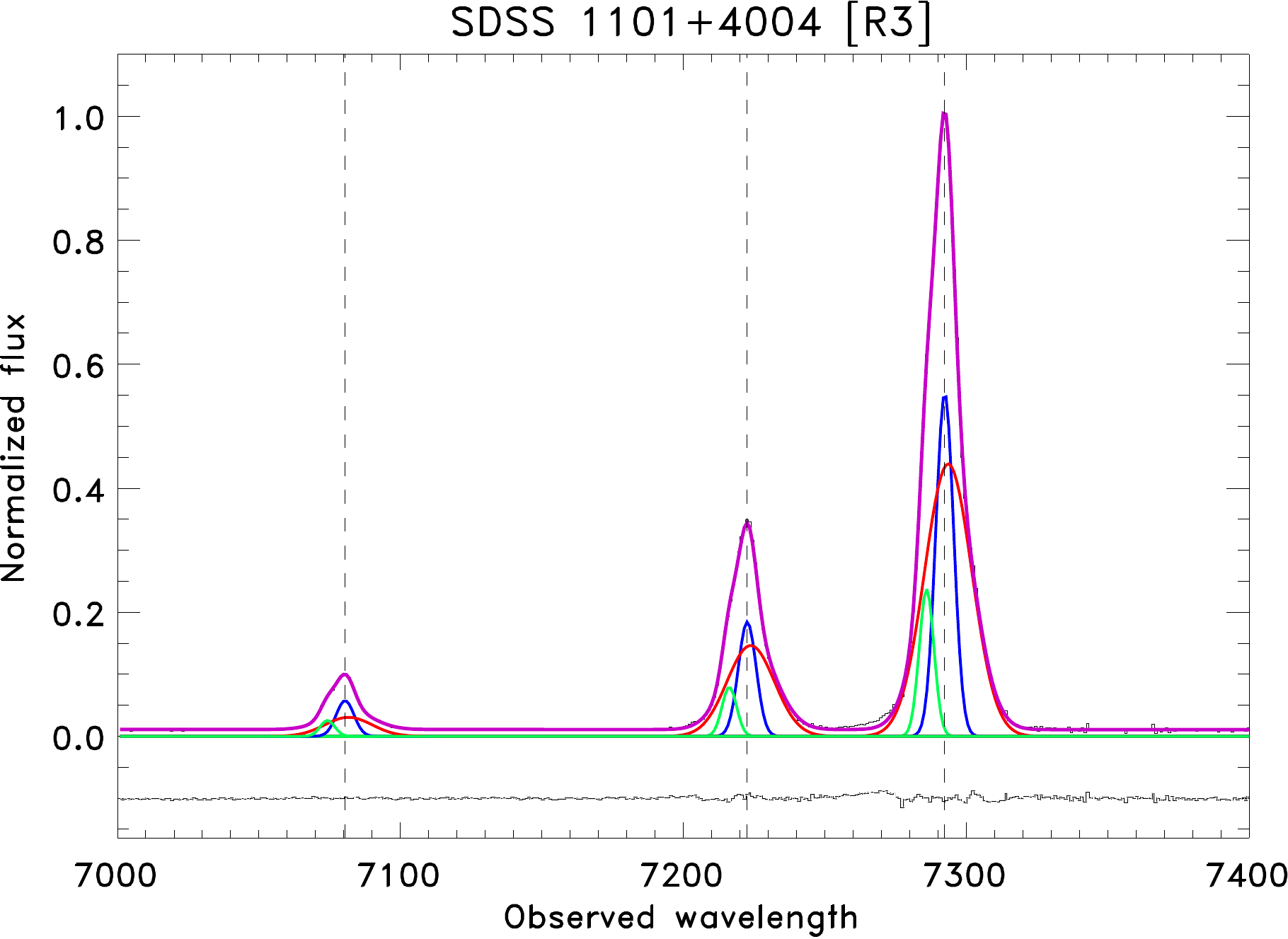}
\caption{Line-fitting results of the H$\beta$--[OIII] complex (vertical dashed lines) obtained for the integrated spectra in the inner regions of SDSS 0741+3020 (R3), SDSS 0818+3958 (R4), and SDSS 1101+4004 (R3). The purple curve shows the total H$\beta$--[OIII] components obtained from multicomponent Gaussian fits. The systemic component is in red, the secondary and third components are in blue and green, respectively. In SDSS 0741+3020, the outflow is identified with the secondary (blue) component, in SDSS 0818+3958, the outflow is identified with the third (green) component, and in SDSS 1101+4004, no outflow is found. }
\label{example_spectra}
\end{figure*}

We are especially interested in the behavior of the systemic and outflow components. Previous studies of nearby QSO2 have demonstrated that the emission line profiles are in general complex: two or three kinematic components are usually identified \citep{VM11, VM14, VM16, Rose15, Spence18}. The outflow component is in general easily recognized based on the kinematic properties because it is always the broadest and frequently shows a significant blueshift. In Fig.~\ref{example_spectra} we show the spectra of different sources (i.e., SDSS 0741+3020, SDSS 0818+3958, and SDSS 1101+4004) that were fit using three components: the secondary and third components are interpreted in different ways according to their width and velocity relative to the systemic component. It can be challenging to identify the systemic component in some cases when three kinematic components are identified, two of which have similarly narrow widths and relative fluxes. The criteria for selecting the systemic component in these ambiguous cases are explained when appropriate.

Throughout the paper we consider as `systemic velocity' of the galaxy the velocity of the systemic component at the continuum centroid (i.e., the spatial location of the peak of the continuum). It is used to determine all relative velocity shifts, v$_{shift}$.

We derived the following parameters for the systemic component. The velocity shear, v$_{shear}$, is defined as half of the difference between v$_{max}^{10\%}$ and v$_{min}^{10\%}$:

\vskip-4mm
\begin{equation}
        \hskip5mm v_{shear} = \frac{1}{2} (v_{max}^{10\%} - v_{min}^{10\%}).
\end{equation}
 
\begin{figure}[!h]
\vskip5mm\begin{center}
\includegraphics[width=0.45\textwidth, height=0.52\textwidth]{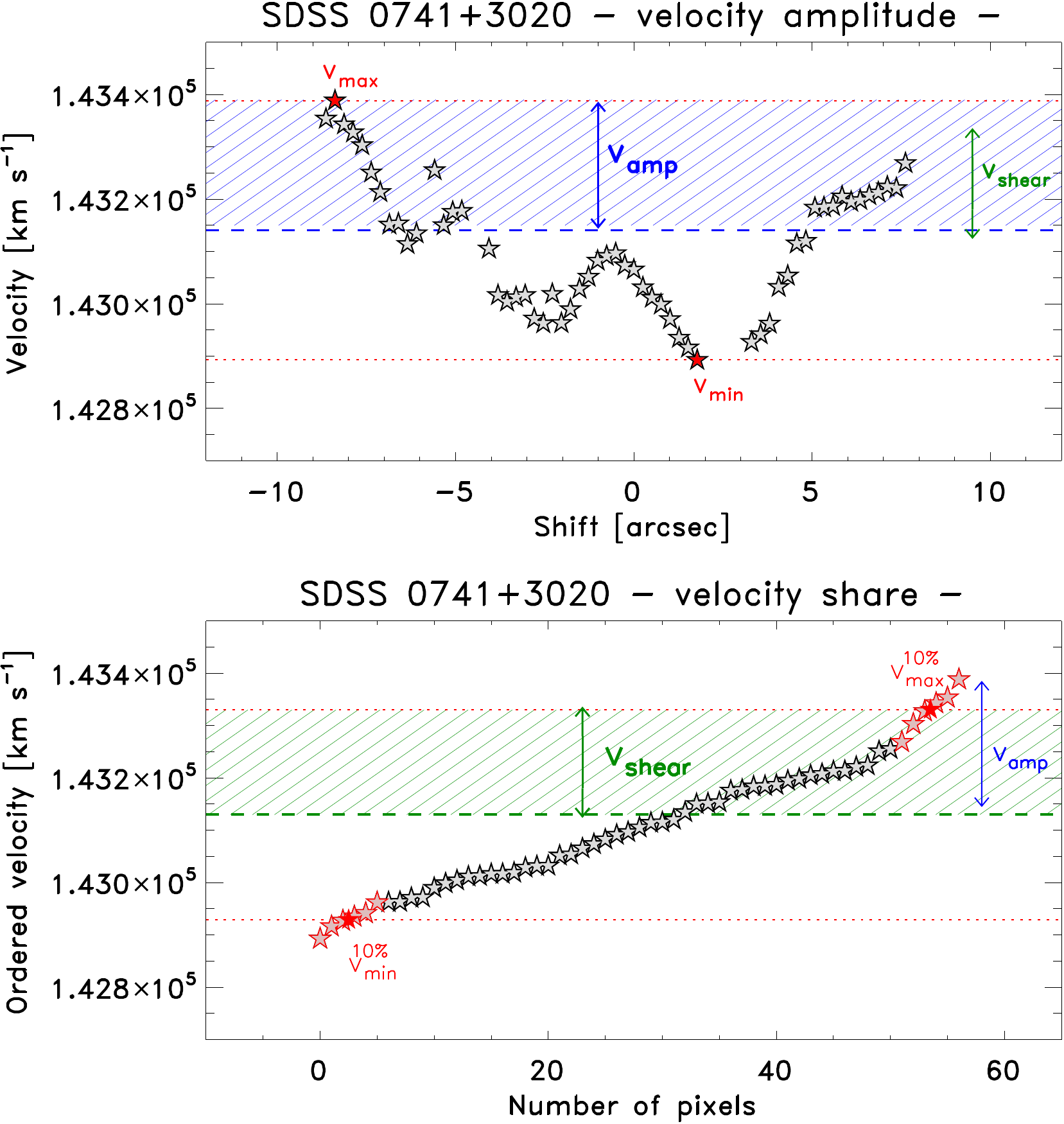}
\caption{Visual definition of the velocity amplitude (v$_{amp}$) and velocity shear (v$_{shear}$) using as an example the systemic component of SDSS 0741+3020. {\it Top:} Red dotted horizontal lines identify the maximum (v$_{max}$) and minimum (v$_{min}$) velocity values used in the v$_{amp}$ definition.
{\it Bottom:} Red dotted horizontal lines identify in this case v$^{10\%}_{max}$ and v$^{10\%}_{min}$, as defined in Sect.~\ref{kin_pars}, used in the v$_{shear}$ definition. The blue and green arrows mark v$_{amp}$ and v$_{shear}$ in both diagrams.}
\label{v_amp_v_shear}
\end{center}
\end{figure}

After ordering the velocity distribution according to increasing values, we calculated v$_{min}^{10\%}$ as the average of all velocities below the 10th percentile, while v$_{max}^{10\%}$ is the average of all velocities above the 90th percentile (see Fig.~\ref{v_amp_v_shear}, bottom). 
We used a version that was modified compared to the version in previous works \citep[e.g., ][B13]{Gon10}: the higher percentile we considered arises because fewer pixels are available when long-slit data are used to derive this parameter. V$_{shear}$ could be derived only for the GTC sample. We also defined the velocity amplitude, v$_{amp}$, for both samples as half of the observed `peak-to-peak' velocity (i.e., half the difference between the maximum v$_{max}$ and minimum v$_{min}$ values is considered; see Fig.~\ref{v_amp_v_shear}, top):

\vskip-4mm
\begin{equation}
        \hskip5mm v_{amp} = \frac{1}{2} (v_{max} - v_{min}).
\end{equation}

V$_{amp}$ and v$_{shear}$ characterize the velocity gradient of each kinematic component in a galaxy. The former just considers the two most extreme values (v$_{min}$ and v$_{max}$) of the velocity field, while the latter provides a more robust derivation of v$_{min}$ and v$_{max}$ (more data points are considered). We found that v$_{amp}$ and v$_{shear}$ deviate in general by $<$20\% (see Table~\ref{NARROW}).
The mean velocity dispersion, $\sigma_{mean}$, was measured as a simple average of the velocity dispersion $\sigma$ over all pixels.

The main kinematic parameters we measured for the broad component are $\sigma^{(b)}_{mean}$ (calculated as above) and the velocity offset, $\Delta$v. $\Delta$v is computed as the difference between the mean velocity of the broad component and the systemic velocity. For the VLT sample, they were derived from the information provided in \citet{VM11, VM12, VM16}.

\begin{table*}
 \centering
\caption{Kinematic properties of the systemic component of the whole QSO2 sample.}
\label{NARROW}
\begin{scriptsize}
   \begin{tabular}{ccccccc}
\hline\hline\noalign{\smallskip}
 Galaxy ID             &    z$_{sys}$     & v$_{shear}$ & v$_{amp}$  &  $\sigma_{mean}$  &    v$_{shear}$                  \\
{\smallskip}
         &  &&    &  &  $\overline{\sigma_{mean}}$              \\
(SDSS code)          & (km s$^{-1}$)  &(km s$^{-1}$) &(km s$^{-1}$)     & (km s$^{-1}$)&              \\
  (1) & (2) & (3) & (4) & (5) & (6)   \\
\hline\hline\noalign{\smallskip}
  \multicolumn{7}{c}{\bf OSIRIS/GTC}  \\
\cmidrule(lr){1-7}\cmidrule(lr){1-7}
    0741+3020              &    0.477    &  200.83 $\pm$ 47.74   &  247.88 $\pm$ 29.55     & 114.91 $\pm$ 42.68   &   1.75 $\pm$ 0.77             \\
\hline\noalign{\smallskip}
    0818+3958a             &    0.406    &  138.53 $\pm$ 41.59          &   156.91 $\pm$ 82.44    &  136.46 $\pm$ 17.38    &    1.02 $\pm$ 0.53             \\
    0818+3958b             &    0.406    &  60.42 $\pm$ 3.16    &   61.93 $\pm$ 14.50     & 73.30 $\pm$ 9.60        &   0.82 $\pm$ 0.65            \\
\hline\noalign{\smallskip}
    0834+5534\_34     &    0.241    &  73.90 $\pm$ 14.18        &   83.00 $\pm$ 14.18     & 146.38 $\pm$ 20.29   &    0.50 $\pm$ 0.12           &      \\
    0834+5534\_145         &    0.241    &  162.41 $\pm$ 6.82   &   165.81 $\pm$ 11.51             & 160.93 $\pm$ 68.75   &    1.01 $\pm$ 0.43            &     \\
\hline\noalign{\smallskip}
    0902+5459a\_83      &    0.401    &  83.48 $\pm$ 22.19  &   110.97 $\pm$ 17.04           & 121.30 $\pm$ 10.28   &   0.69 $\pm$ 0.19            &        \\
    
0902+5459a\_138         &    0.400    &  105.24 $\pm$ 25.32  &  127.66 $\pm$ 61.89   &168.41 $\pm$ 16.30     &   0.62 $\pm$ 0.16            &  \\
 
    0902+5459b\_83      &    0.401    &  99.60 $\pm$ 21.78 &   150.26 $\pm$ 19.43   & 163.85 $\pm$ 33.07  &   0.61 $\pm$ 0.18            & \\
 
0902+5459b\_138         &    0.400    &   69.85 $\pm$ 19.64 &   78.92 $\pm$ 12.85   & 151.15 $\pm$ 27.70    &  0.46 $\pm$ 0.16            & \\

\hline\noalign{\smallskip}
    0956+5735              &   0.362     &   27.25 $\pm$ 7.56 &   30.81 $\pm$ 9.84            & 181.21 $\pm$ 48.95    &   0.15 $\pm$ 0.06        &    \\
\hline\noalign{\smallskip}
    1101+4004a     &    0.456    & 131.46 $\pm$ 61.97 &   131.13 $\pm$ 69.19    & 309.74 $\pm$ 28.09    &   0.42 $\pm$ 0.20            &      \\
    1101+4004b     &    0.456    &  124.43 $\pm$ 17.66 &   133.01 $\pm$ 19.61       & 100.26 $\pm$ 17.25   &   1.24 $\pm$ 0.28            &       \\
\cmidrule(lr){1-7}\cmidrule(lr){1-7}
 \multicolumn{7}{c}{\bf FORS2/VLT}  \\
\cmidrule(lr){1-7}\cmidrule(lr){1-7}
J0923+0101 & 0.386 &  -- & 283.42 $\pm$ 8.75 & 285.35 $\pm$ 68.33       & 0.99 $\pm$ 0.24\\
\hline\noalign{\smallskip}
J0950+0111a & 0.404 & --  &     43.37 $\pm$ 14.29 & 94.46 $\pm$ 19.62    &  {0.46} $\pm$ 0.18    \\
J0950+0111b & 0.404 & --  &     17.36 $\pm$ 6.42 & 132.56 $\pm$ 63.57   & 0.13 $\pm$ 0.08 &        \\
\hline\noalign{\smallskip}
J0955+0346 & 0.422 & --  &  15.26 $\pm$ 9.25 & 419.29 $\pm$ 15.09       & 0.04 $\pm$ 0.02\\
\hline\noalign{\smallskip}
J1014+0244a/pa1 & 0.573 & --  &  53.30 $\pm$ 18.46 & 136.56 $\pm$ 7.01  & 0.39 $\pm$ 0.14 &                       \\
J1014+0244b/pa1 & =& --  & 66.61 $\pm$ 24.15 & $<$107.76 $\pm$ 21.49            & $>$0.62 $\pm$ 0.26      &  \\
J1014+0244a/pa2 & = & --  & 126.00 $\pm$ 33.64 & 127.86 $\pm$ 33.54     &0.99 $\pm$ 0.37                              \\
J1014+0244b/pa2 & = & --  & 22.54 $\pm$ 8.36 & $<$125.59 $\pm$ 34.43            & $>$0.18 $\pm$ 0.08      &               \\
\hline\noalign{\smallskip}
J1153+0326 & 0.575 & --  & 92.32 $\pm$ 25.93 & 314.07 $\pm$ 10.54 & 0.29 $\pm$ 0.08      \\
\hline\noalign{\smallskip}
J1247+0152 & 0.427  & --  & 99.50 $\pm$  7.25 & 328.84 $\pm$ 15.35      &0.30 $\pm$ 0.03  \\
\hline\noalign{\smallskip}
J1307--0214& 0.424 &  -- &      117.99 $\pm$ 15.30 & 165.26 $\pm$ 17.06  & 0.71 $\pm$ 0.12\\
\hline\noalign{\smallskip}
J1336--0039 & 0.416 & --  & 49.24 $\pm$ 10.93 & 163.18 $\pm$ 30.54 & 0.30 $\pm$ 0.09  \\
\hline\noalign{\smallskip}
J1337--0128 & 0.328 &  -- & 53.49 $\pm$ 15.12 & 153.50 $\pm$ 19.21      & 0.35 $\pm$ 0.11\\
\hline\noalign{\smallskip}
J1407+0217 & 0.309 & --  & 59.38 $\pm$ 14.45 & 136.82 $\pm$ 25.63   & 0.43 $\pm$ 0.13\\
\hline\noalign{\smallskip}
 J1413--0142 & 0.380 &  -- & 25.78 $\pm$ 7.22 & 237.16 $\pm$ 30.50      & 0.11 $\pm$ 0.03 &       \\
\hline\noalign{\smallskip}
J1430--0050 & 0.318 &  -- &  36.83 $\pm$ 3.15 & $<$110.92 $\pm$ 17.28   & $>$0.33 $\pm$ 0.06      & \\
\hline\noalign{\smallskip}
 J1546--0005 & 0.383 & --  & 32.92 $\pm$ 8.99 & 145.10 $\pm$ 14.08 & 0.23 $\pm$ 0.66      \\
\hline\hline\noalign{\smallskip}
\end{tabular}
\vskip0.2cm\hskip0.0cm
\end{scriptsize}
\begin{minipage}[h]{18cm}
\tablefoot{Column (1): SDSS short name. When a system is composed of two galaxies, we identify  them with $a$ and $b$. The information of the slit PA is shown, unless only one slit is used for the observations.
Column (2): Systemic redshift. 
Column (3):~[OIII] velocity shear defined as half the difference between the average of all velocities below the 10th percentile and the average of all velocity above the 90th percentile (see Sect.~\ref{kin_pars}). The minus for the VLT sample means that this value could not be measured (see text for details). 
Column (4): [OIII] `peak-to-peak' velocity amplitude.
Column (5): [OIII] mean velocity dispersion.
Column~(6): Dynamical ratio, computed as the ratio between the velocity shear (velocity amplitude) for the GTC (VLT) sample and the mean velocity dispersion values.}
\end{minipage}
\end{table*}

\begin{table*}
 \vskip1cm 
\centering
\caption{Kinematic properties of the broad (outflow) component of the whole QSO2 sample.}
\label{BROAD}
\begin{scriptsize}
   \begin{tabular}{cccccc}
\hline\hline\noalign{\smallskip}
 Galaxy ID                 &  $\sigma_{mean}^{(b)}$       & $\Delta$v   & R$_{outflow}$   \\
{\smallskip}
(SDSS\_PA code)          &   (km s$^{-1}$)  &  (km s$^{-1}$) &   (kpc) &  \\
  (1) & (2) & (3) & (4)   \\
\hline\hline\noalign{\smallskip}
  \multicolumn{5}{c}{\bf OSIRIS/GTC}  \\
\cmidrule(lr){1-5}\cmidrule(lr){1-5}
    0741+3020  Central outflow$^{(a)}$             &      291.82 $\pm$ 41.30       &  --380.54 $\pm$ 107.80 & 4.2$\pm$0.4  &       \\

   0741+3020    Giant nebula$^{(a)}$               &      $\sim$130      &  $\sim$100  & $\sim$40  &        \\
\hline\noalign{\smallskip}
    0818+3958$^{(b)}$              &    491.02 $\pm$ 61.29   &   --568.80 $\pm$ 185.75                      & $<$ 2.2&          \\
\hline\noalign{\smallskip}
    0834+5534\_34         &  948.84 $\pm$ 48.78 &    629.20 $\pm$ 63.18          & 0.7 $\pm$ 0.4 &  \\
    0834+5534\_145         &    871.51 $\pm$ 74.79     &   452.05 $\pm$ 56.84         & 0.8 $\pm$ 0.3  &      \\
\hline\noalign{\smallskip}
    0902+5459\_83a                              & --                                     &   --                  &       &                             \\
    0902+5459\_138a                     & --                                     &   --                  &     &             \\
    0902+5459\_83b                              & --                                     &   --                  &       &                                     \\
    0902+5459\_138b                     & --                                     &    --                 &       &     \\
\hline\noalign{\smallskip}
    0956+5735              &      624.09 $\pm$ 24.36  &    110.76 $\pm$ 65.54         & $<$ 3.3&   \\
\hline\noalign{\smallskip}
    1101+4004a            &--   & --    &                 &   \\
    1101+4004b                     &--       &   --     &      \\
\cmidrule(lr){1-5}\cmidrule(lr){1-4}
  \multicolumn{5}{c}{\bf FORS2/VLT}  \\
\cmidrule(lr){1-5}\cmidrule(lr){1-5}
\hline\noalign{\smallskip}
J0923+0101      &       {708.16} $\pm$ 75.19            & {--33 } $\pm$ 49      & 0.16 -- 1.0   \\
\hline\noalign{\smallskip}
J0950+0111      &       {696.69} $\pm$ 18.69            & {--506} $\pm$ 35       &       0.38 -- 1.1     \\
\hline\noalign{\smallskip}
J0955+0346 &            {1062.02} $\pm$ 84.96   &   {--810} $\pm$ 190           & 0.20 -- 3.5  \\
\hline\noalign{\smallskip}
J1014+0244a/pa1 & {652.93} $\pm$ 30.59          & {--126} $\pm$ 53              &  0.20 -- 1.0 \\
J1014+0244a/pa2   & {671.20} $\pm$ 48.85        & {--72} $\pm$ 72                &  =\\
\hline\noalign{\smallskip}
J1153+0326 & {615.97} $\pm$ 33.98               & {325} $\pm$ 80         &   0.52 -- 1.7 &\\
\hline\noalign{\smallskip}
{J1247+0152}&  {498.30} $\pm$ 84.11             & {--301} $\pm$ 194             & {0.28 -- 1.4} \\
\hline\noalign{\smallskip}
J1307--0214     & {424.80} $\pm$ 12.74          &{--246} $\pm$ 29                                & 1.3 $\pm$ 0.4 \\
\hline\noalign{\smallskip}
J1336--0039  & {1055.64} $\pm$ 73.49    & {--942} $\pm$ 112                     &       0.19 -- 1.0  \\
\hline\noalign{\smallskip}
J1337--0128 &{543.76} $\pm$ 25.49               & {--320} $\pm$ 16      & 0.19 -- 1.7     \\
\hline\noalign{\smallskip}
J1407+0217 &{709.43} $\pm$ 50.98                & {--80} $\pm$ 33       &  1.0 $\pm$ 0.2  \\
\hline\noalign{\smallskip}                      
 J1413--0142 & {505.52} $\pm$ 21.24             & {--185} $\pm$ 38              &0.21 -- 1.5   \\
\hline\noalign{\smallskip}
J1430--0050 & {679.69} $\pm$ 84.96              & {--520} $\pm$ 60               & 0.12 -- 0.80      \\
\hline\noalign{\smallskip}
 J1546--0005 &  {331.35} $\pm$ 12.74            & {--270} $\pm$ 55               & 0.20 -- 1.4   \\
\hline\hline\noalign{\smallskip}
\end{tabular}
\vskip0.2cm\hskip0.0cm
\end{scriptsize}
\begin{minipage}[h]{18cm}
\tablefoot{Column (1): SDSS short name. When a system is composed of two galaxies, we identify  them with $a$ and $b$. The information of the slit PA is shown, unless only one slit is used for the observations.  The minus means that the broad component (and thus, an outflow) has not been detected.
Column (2): Mean velocity dispersion. The FORS2/VLT values are taken from the works of \citet{VM11, VM12, VM16}.
Column (3): Velocity offset, computed as the difference between the mean velocity of the broad component and the systemic velocity. Negative values represent blueshifted values, and positive values are redshifted. The FORS2/VLT values are taken from the works of \citet{VM11, VM12, VM16}.
Column (4): Intrinsic radius of the outflow. It has been computed as half the difference between the FWHM of the observed [OIII] flux profile of the broad component and the FWHM$_{seeing}$ computed within the slit, according to the formula: R$_{out}$ [kpc] = scale [kpc/$^{\prime\prime}$]$\times$ $\frac{\sqrt{FHWM_{[OIII] obs}^2 - FHWM_{seeing}^2}}{2}$. For the FORS2/VLT data the minimum and maximum radial sizes  of the outflows are given \citep[see Tab.~5 in][for details]{VM16}. 
($a$) See Sect.~\ref{outflows}.
($b$) The kinematic values of the broad component for SDSS 0818+3958 are those obtained using the integrated values over several pixels.
}
\end{minipage}
\end{table*}

\section{Results and discussion}
\label{kinematics}

In this section we present the kinematic results of the whole sample for the systemic and broad components (Tables~\ref{NARROW} and~\ref{BROAD}). A detailed description of the individual sources is also given.

\subsection{Notes on individual objects}
\label{individual}

\begin{itemize}
\item{\bf SDSS 0741+3020}

The long-slit spectrum of this radio-quiet quasar shows very extended line emission along PA148. [OIII] is detected across $\sim$19$^{\prime\prime}$ or 112 kpc. Another galaxy at a different $z\sim$ 0.489 is detected along the slit; it is unrelated to the QSO2 (Fig.~\ref{panel_0741}, top). 

For this object, the pixel-to-pixel analysis shows that [OIII]$\lambda$5007 can be well fit using two (narrow and broad) components. Using larger apertures, a third narrow component is identified (see region R3). The broad component ($\sigma$$\sim$220--480 km s$^{-1}$) is blueshifted at all locations by $\sim$200--500 km s$^{-1}$ relative to the assumed systemic narrow component ($\sigma$$\sim$60--120~km s$^{-1}$ in the central region).
It is naturally explained by an outflow. This has a steep and compact spatial profile (Fig.~\ref{panel_0741}, bottom, flux panel), although it is clearly resolved relative to the seeing disk. The insufficient S/N at increasing distances from the center prevents us from clearly detecting and tracing the emission toward the wings. If the spatial profile follows the shape of the high surface brightness regions, it is barely resolved toward the SE (negative spatial shift) and clearly resolved toward the NW (positive spatial shift), where the spatial profile has a clear excess above the seeing wings. When we account for seeing broadening, the intrinsic FWHM of the outflow can be roughly estimated by deconvolving the seeing size \citep[see][for details]{VM16}. This results in a radial extension of the outflow of 1.8$\pm$0.4 kpc toward the SE and 4.2$\pm$0.4 kpc toward the NW.

The giant emission line nebula shows very narrow lines ($\sigma\lesssim$60 km s$^{-1}$) in the outer parts ($>$10$^{\prime\prime}$ to the SE and $>$5$^{\prime\prime}$ to the NW). The [OIII]$\lambda$5007 width reaches $\sigma\sim$130~km s$^{-1}$ at large distances from the AGN,~at~$\sim$~--6$^{\prime\prime}$-~--7$^{\prime\prime}$ or 35--40 kpc SE and $\sim$4$^{\prime\prime}$--5$^{\prime\prime}$ or 24--30 kpc NW of the spatial centroid. This emission is well in excess of the seeing wings and is dominated by the emission from the nebula. The gas must be well outside the main body of the galaxy. 
The extended non-outflowing ionized gas in mergers shows typical $\sigma<$ 100 km s$^{-1}$, even in the most dynamically disturbed systems with signs of AGN activity \citep[B13,][]{Arribas14}. The large velocity dispersion at some distant locations across the SDSS 0741+3020 nebula are suggestive of turbulent gas that might be related to an outflow \citep[see][]{VM17}. We discuss this further in Sect.~\ref{outflows}.

Gas with $\sigma\gtrsim$ 200 km s$^{-1}$ is also located at --3$^{\prime\prime}$ $\pm$ 1$^{\prime\prime}$ (to the SE) of the AGN, which is also broad for such distant locations. However, it is not clear in this case whether it is an effect of the low S/N that prevents an isolation of the central broad (blue component). The broadening at this location may thus be due to contamination by the blue central broad component, rather than due to turbulence in the giant nebula.

\vskip2mm
\item{\bf SDSS 0818+3958}

This radio-intermediate QSO2 shows clear extended emission in the 2--dim spectrum (Fig.~\ref{panel_0818}, top).
The pixel-to-pixel analysis reveals two narrow components in the central region of this system. They have $\sigma\sim$ 30--90 km s$^{-1}$ (component N1, blue, Fig.~\ref{panel_0818} bottom panels) and $\sim$100--200 km s$^{-1}$ (component N2, red, Fig.~\ref{panel_0818} bottom panels). They are shifted spatially by $\sim$0.8$^{\prime\prime}$. Although unclear, we consider N2 the systemic component in this object because its spatial centroid is at the same location as the continuum centroid and it is the more prominent in terms of relative flux. 

The seeing profile is not accurately characterized for this object. The fact that the spatial distribution of N2 is more compact than the stellar profile demonstrates that the seeing improved during the spectroscopic observations in comparison to the acquisition image.

A clear excess above this profile is visible toward the east, revealing extended emission up to $\sim$--5$^{\prime\prime}$ or 27 kpc from the continuum centroid. This extended gas, which emits very narrow lines ($\sigma \la$ 60 km s$^{-1}$), shows very distinct kinematics relative to the central regions, with a sharp redshift of $\sim$400 km s$^{-1}$. It is not certain whether this is a separate galaxy or an extended nebula.

A nuclear outflow was identified when we analyzed the spectra of larger apertures (green open squares in Fig.~\ref{panel_0818} bottom panels): a broad component with $\sigma\sim$510 km s$^{-1}$ that is blueshifted by $\sim$400 km s$^{-1}$ was detected. Nothing can be said about the spatial extension. Because it is only detected in the central region, we consider it to be spatially unresolved. This sets an upper limit on its intrinsic radial size of R $\la$ 2.2 kpc. This is a high upper limit, given the bad seeing (FWHM$\sim$1.7\arcsec).

It is spatially shifted $\sim$0.4$\arcsec$$\pm$0.2$\arcsec$ relative to the continuum centroid toward the E (Fig.~\ref{panel_0818} bottom panels). \citet{Lal10} found that the radio core of SDSSJ0818+3958 is slightly resolved and might have a core-jet morphology that is extended along PA$\sim$103 deg. A similar spatial shift is seen between the radio core and the optical position of the galaxy. If the location of the AGN coincided with the radio core instead of the optical continuum centroid, the outflow would also coincide with the AGN location. If instead the continuum centroid marked the location of the AGN, this might suggest that the outflow is shifted and may be induced by the interaction between the radio structures and the ambient gas.

\vskip2mm
\item{\bf SDSS 0834+5534}

This galaxy has been observed using two different position angles (PAs = 34, 145; see Tab.~\ref{info1}, Figs.~\ref{panel_0834} and \ref{panel_0834_2}). It is the only radio-loud QSO2 in the GTC sample.

The VLA radio map (Fig.~\ref{Lal_Ho} and overplotted contours in Fig.~\ref{0834_hst}, left) shows that this radio-loud QSO2 is associated with a bright radio core and two jets or lobes that are clearly misaligned. One extends toward the NW at PA 140 (N to E) for $\sim$6.4\arcsec. One of the slit positions (PA 145) was chosen to roughly coincide with this radio axis (Fig.~\ref{panel_0834_2}). The other jet or lobe extends roughly along the N--S direction (PA 170) and extends up to $\sim$8.4\arcsec\ from the QSO2. The broadband acquisition image shows that this QSO2 is hosted by an elliptical galaxy. The slit PA=34 was chosen to be aligned with the main axis of the galaxy.

Certain similarities between the optical (Hubble Space Telescope, HST, image) and the radio morphologies suggest that the radio structures are interacting with the QSO2 environment. In Fig.~\ref{0834_hst} (left and middle panels) the HST image shows a peculiar and narrow feature extending toward the NW for $\sim$0.9$\arcsec$ or 3.4 kpc at an angle of $\sim$138 deg N to E. A shorter extension is found in the opposite direction. These  features extend along a similar angle as the large-scale radio axis as defined by the NW radio jet (PA 140).

At smaller spatial scales, the Very Large Baseline Array (VLBA) radio map (Fig.~\ref{0834_hst}, right) reveals an inner radio jet on scales of 6.4 mas along PA 116  and a separate radio feature located at 33 mas with PA 137. This angle is basically identical to the optical collimated feature. Although this clearly extends beyond, its collimated morphology and the similarity of position angles suggest a physical link.

This HST image was obtained with the F606W filter, which covers the spectral range of 4632 --7179 \AA. The H$\beta$ and [OIII] doublet are therefore within the filter. Line emission may dominate regions of pure gas emission. It is thus possible that this collimated optical feature is gas where line emission has been enhanced through the interaction with the radio structures.

The visual inspection of the [OIII] 2--dim spectra shows that the [OIII] emission is spatially extended along both PA Figs.~\ref{panel_0834} and \ref{panel_0834_2}). The optical collimated feature is expected to be unresolved or barely resolved in the PA 145 spectrum (seeing FWHM = 0.84\arcsec\ $\pm$ 0.02\arcsec).

The kinematic analysis for the PA = 34 reveals three components, two narrow (N1, N2, $\sigma\la$ 250 km s$^{-1}$) and a broader (B, $\sigma\sim$ 900--1000 km s$^{-1}$) one. We assume that N1 is systemic, based on its dominant flux contribution compared to N2 and the very narrow $\sigma$ of this component.

The broad component is redshifted with respect to the systemic component of $\sim$500--700 km s$^{-1}$. Based on its broad intrinsic velocity dispersion, we identify this highly turbulent gas with an outflow. The systemic component N1 is clearly resolved relative to the seeing disk. The broad component B appears barely resolved. This is also confirmed by the kinematic substructure within the seeing disk. We estimate an intrinsic size of the outflow along PA=34 of 0.7$\pm$0.4 kpc. N2 is unresolved.

\begin{figure*} 
\begin{center}
\includegraphics[scale=0.7]{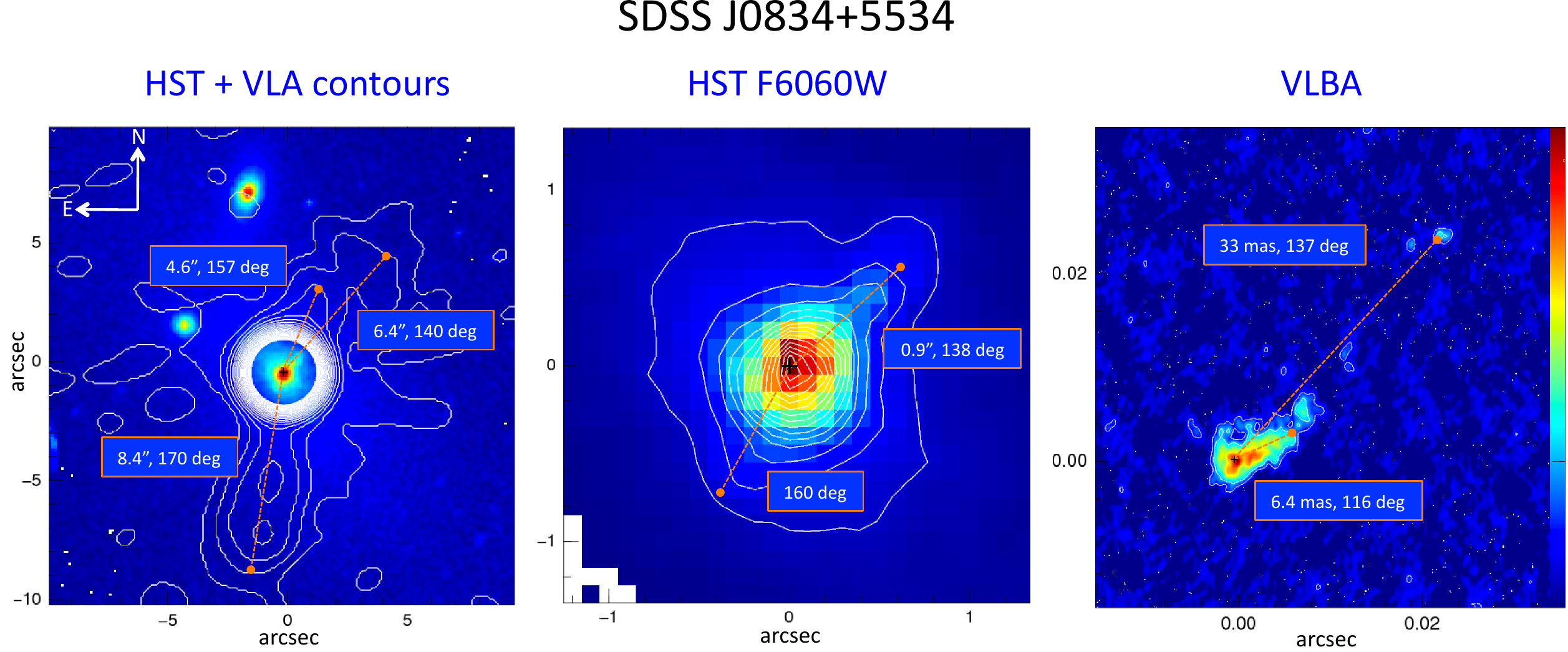}
\caption{Comparison between the optical HST and the radio VLA and VLBA images of SDSS 0834+5534. The position angles (north to east)  of the different structures (optical and radio) and their respective extensions in arcsec are shown.   The spatial centroid of the HST image  has been  aligned the VLA peak emission (black cross). {\it Left:} HST F606W image with the VLA contours overplotted (in white; also shown in Fig.~\ref{Lal_Ho}; $\sim$ 20\arcsec$\times$20\arcsec or 75 kpc$\times$75 kpc). {\it Middle:} HST image ($\sim$ 2\arcsec\ $\times$ 2\arcsec or or 7.5 kpc$\times$7.5 kpc) and its contours. {\it Right:} VLBA image of the very inner regions of the object with its contours ($\sim$ 0.04\arcsec $\times$ 0.04\arcsec or 0.15~kpc$\times$0.15~kpc). Its intensity peak overlaps with the VLA peak (black cross).}
\label{0834_hst}
\end{center}
\end{figure*}

\begin{figure*}
\centering
\includegraphics[scale=0.7]{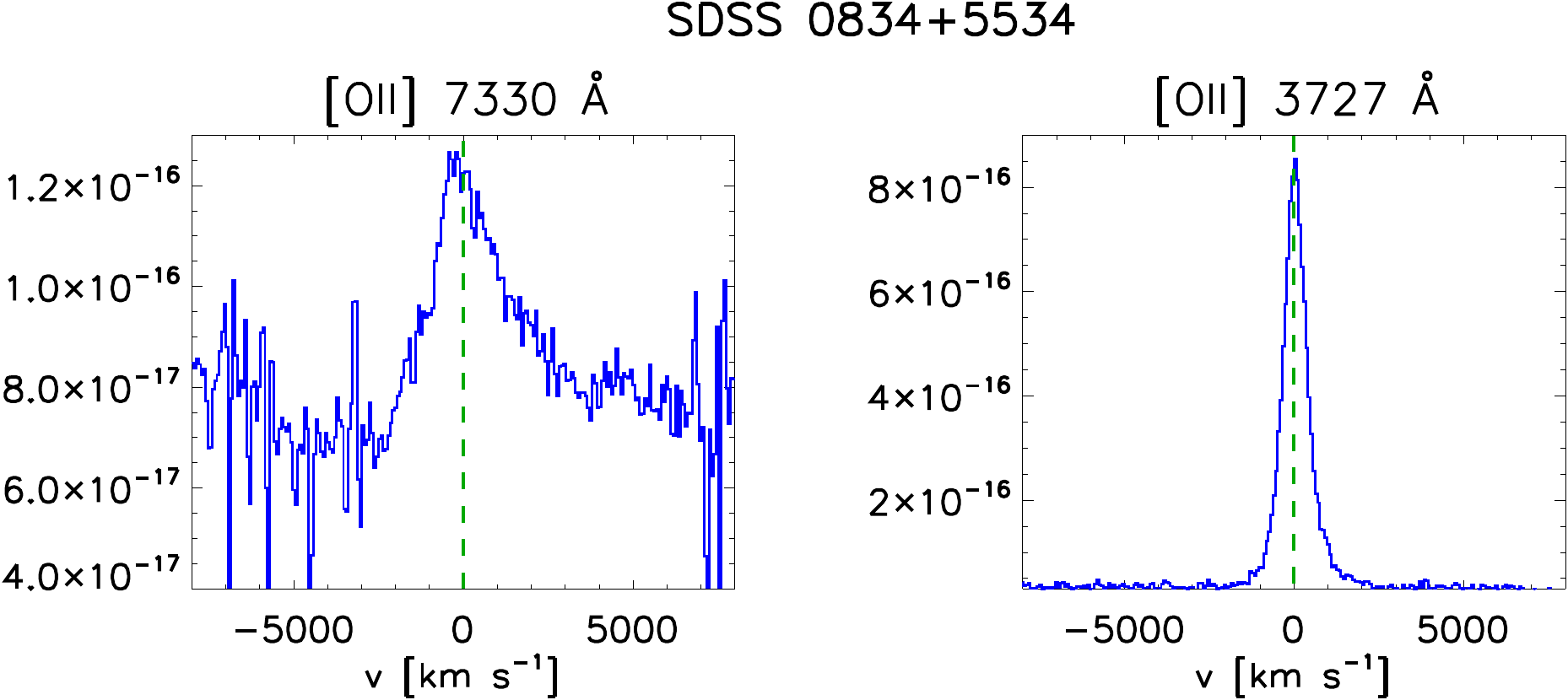}
\caption{SDSS spectrum of the galaxy 0834+5534 showing the broad [OII]7330 \AA\ and the narrow [OII]3727 \AA\ line profiles.}
\label{oxigen_0834_spectra}
\end{figure*}

The pixel-to-pixel analysis along PA=145 reveals only two components (Fig. \ref{panel_0834_2} bottom panels): the systemic component is clearly resolved. The broad component (the outflow) appears barely resolved relative to the seeing disk. 

When we integrated over several pixels a third narrow component was identified (green open squares in the v$_ {\rm shift}$ and $\sigma_{\rm int}$ plots in Fig.~\ref{panel_0834_2}).     
This is partly responsible for the kinematic spatial substructure shown by the broad component in the pixel-to-pixel analysis. Kinematic substructure is still present in the v$_{\rm shift}$ field for the broad component, however, further supporting that the outflow is resolved. We estimate an intrinsic radial size of R=0.8$\pm$0.3 kpc along PA145.

The ionized outflow in SDSS 0834+5534 has rather extreme kinematics, with $\sigma$ reaching $\sim$950 km s$^{-1}$ and a high-velocity redshift relative to the systemic component of up to $\sim$+630 km s$^{-1}$. The most extreme ionized outflows in AGN are often found in objects with some degree of radio activity \citep[e.g.,][]{Husemann13, Husemann16, Mullaney13, VM14, Zakamska14}. It is possible that the outflow is induced by the interaction with the inner radio jet in this object, as suggested by the alignment between the optical HST and the radio morphologies in the inner $\sim$1\arcsec.

The extreme kinematic properties may alternatively be favored by an intermediate orientation. Although blueshifts are much more common, redshifted ionized outflows have been identified in some systems \citep[e.g.,][]{VM11, RZ13, Bae14, Perna15}. 
\citet{VM11} found redshifted outflows in two QSO2 in their sample: one of these is SDSS J1153+03 \citep[see also Table 5 in][]{VM16}.

The authors proposed that both objects are seen at intermediate orientations. As shown by \citet{Crenshaw10}, although rare, a combination of reddening and inclination angles can be such that the blueshifted cone is entirely occulted by the disk, and the redshifted cone is not. One possible explanation for the large misalignment between the two radio jets or lobes of SDSS J0834+5534 is indeed a significant angle of the radio axis relative to the plane of the sky. As this angle increases, even a small intrinsic change in the jet angle can result in apparent large misalignments that are due to geometrical projection effects.

Another argument indicative of an intermediate orientation is the different kinematics shown by lines of the same ion and very different critical density $n_{\rm crit}$. This suggests that we are observing the inner part of the narrow-line region (NLR), which is possibly more extinguished for a pure type 2 orientation and may be better observable with an intermediate orientation. A striking feature of the nuclear spectrum of SDSS J0834+5534 is the extremely broad features at $\sim$9100 \AA\ and (using the SDSS spectrum) and at $\sim$7335 \AA. They correspond to [OII]$\lambda\lambda$7318,7339 (Fig.~\ref{oxigen_0834_spectra}, left) and [NII]$\lambda$5755. The [OII] doublet is not only very broad ($\sigma$$\sim$1520 km~s$^{-1}$), but the two components are unresolved, implying that each must have $\sigma$$\ga$800 km~s$^{-1}$. Thus, this spectral feature appears to be dominated by the outflow emission. For comparison, using the SDSS spectrum, we measure $\sigma$([OII]$\lambda$3727,3729)$\sim$330 km s$^{-1}$, which is much narrower (Fig.~\ref{oxigen_0834_spectra}, right), and shows no clear evidence for a broad underlying component. The fact that lines emitted by the same ion show such different kinematics is naturally explained by a density effect \citep{DeRob84, Espey94, VM15}. [OII]$\lambda$3727,3729 have very low $n_ {\rm crit}$=1.3--4.5$\times$10$^{3}$ cm$^{-3}$ , while [OII]$\lambda\lambda$7318,7339 have both $n_ {\rm crit}$=1.8$\times$10$^7$ cm$^{-3}$. 

The different kinematics can be naturally explained if most of the outflow is preferentially emitted in the inner denser part of the NLR, while [OII]$\lambda$3727,3729 is preferentially emitted by lower density and more distant non-outflowing gas \citep{VM15}.

Something similar occurs with [NII]$\lambda$5755 and [NII]$\lambda\lambda$6548,6583. The red [NII] lines show a prominent narrow core and faint red wings due to the outflow, while [NII]$\lambda$5755~is dominated by a very broad underlying feature ($\sigma\sim$1020 km s$^{-1}$). This suggests that the flux of this line is preferentially emitted by the outflow. Although the interpretation of this feature is not trivial because it may also be contaminated by an also broad coronal [FeVII]$\lambda$5721 line, [NII]$\lambda$5755 must in any case be very broad. If broad [FeVII]$\lambda$5721 is confirmed, it would be one additional argument in favor of an intermediate orientation, if coronal lines are emitted in the far wall of the torus, as proposed by \citet{Rose15}. [NII]$\lambda\lambda$6548,6583 have $n_ {\rm crit}$=1.2$\times$10$^{5}$ cm$^{-3}$, while [NII]$\lambda$5755 has 1.8$\times$10$^7$ cm$^{-3}$. This further supports that the outflowing gas is preferentially emitted in a high-density compact region $n\sim$10$^{7}$ cm$^{-3}$, where lines of low $n_ {\rm crit}$ are quenched. As a consequence, the relative contribution of the outflow to the total line flux increases with $n_ {\rm crit}$.

In summary, we propose that SDSS 0834+5534 is seen at an intermediate orientation and is associated with an ionized outflow of radial size R$\sim$0.8 kpc. Given the dominant outflow emission in high critical density lines, a significant fraction of the outflow emission is expected to be concentrated in a high-density ($n\ga$10$^6$ cm$^{-3}$) more compact region closer to the AGN.

\vskip2mm
\item{\bf SDSS 0902+5459}

This radio-intermediate QSO2 shows a double morphology at 1.4 GHz ({\tt FIRST} image) with two lobes separated by $\sim$1\arcmin\ \citep{Lal10}. The integrated fluxes in the {\tt NVSS} and {\tt FIRST} data are roughly consistent (Table~\ref{info1}). However, the {\tt FIRST} image shows that almost all the emission is in the lobes, and there is marginal evidence at best for a radio core, with on the order of only $\sim$3\% of the total 1.4 GHz flux. As also mentioned by Lal and\ Ho, the integrated spectral index $\alpha^{8.4}_{1.4}$=--0.89 is typical for extended radio sources.

In the optical it has been observed using two position angles (PAs = 83, 138): the second angle has been chosen to be roughly aligned with the radio axis (Fig.~\ref{Lal_Ho}).

The [OIII] 2--dim spectra shows double-peaked lines (see also the integrated spectrum in Fig.~\ref{SDSS_spectra}) with a slight spatial shift (Figs.~\ref{panel_0902} and \ref{panel_0902_2}). The kinematic analysis isolates the two spatial components along the two slit PAs. They are similarly narrow (N1, N2, $\sigma\sim$ 150 km s$^{-1}$) and show a velocity shift of 800 km s$^{-1}$. They are spatially shifted by $\sim$0.8$^{\prime\prime}$. Although uncertain, we consider that the brightest of the two components, which has $\sim$1.7 times more flux, is the systemic component.

The spatial distributions of both N1 and N2 are dominated by steep and rather compact profiles. N2 is unresolved or barely resolved along PA 83 and 138, respectively. The systemic component N1 shows a clear excess above the seeing disk toward the E along PA~=~83. Extended emission is detected up to --3$^{\prime\prime}$ or 16 kpc from the continuum centroid. 
The most distant gas from the AGN emits narrower lines ($\sigma\sim$50--100 km s$^{-1}$) than in inner regions ($\sigma\sim$ 150 km s$^{-1}$). 

Most probably, this system consists of two emission line galaxies. We discard a scenario in which they are two spatially shifted emission line regions associated with the same galaxy based on the rather compact morphologies of both components and their high relative velocity. Based on the large velocity shift, it is also unlikely that these are two interacting galactic nuclei. For comparison, the typical relative velocity in interacting nuclei with projected nuclear separation $\la$ 5 kpc is $\la$ 350 km s$^{-1}$ \citep{Tecza00, Genzel01, Dasyra06b}.
No evidence is found for an ionized outflow  in this object.

\vskip2mm
\item{\bf SDSS 0956+5735}

This radio-intermediate quasar shows a slightly resolved radio structure along PA = 160, possibly with a core-jet morphology \citep{Lal10}, which was the slit position angle we chose. On the other hand, even though the radio source appears very weak in both {\tt NVSS} and {\tt FIRST}, the comparison between the {\tt FIRST} and {\tt NVSS} fluxes suggests that almost half of the 1.4 GHz flux extends across spatial scales between 5\arcsec -- 25\arcsec (Tab.~\ref{info1}).

For this object the 2--dim spectrum shows a quite compact [OIII] profile, with no obvious evidence for extended emission (Fig.~\ref{panel_0956}).

The pixel-to-pixel kinematic analysis reveals two components, one narrow (N, $\sigma\sim$ 190 km s$^{-1}$) and one broad (B, $\sigma\sim$ 700 km s$^{-1}$). The broad component is redshifted  $\sim$150 km s$^{-1}$ with respect to the systemic velocity. According to its high intrinsic velocity dispersion, we identify this highly turbulent gas with an outflow. 
The GTC and especially the SDSS spectra of this object are rather noisy. It is not possible to apply a similar analysis as for SDSS 0834+55 to investigate whether faint nuclear emission lines provide an indication that the redshifted outflow is due to the high inclination angle relative to the sky.

The systemic component (N) appears to show a narrower spatial profile than the seeing. If confirmed, this would imply that the seeing improved during the spectroscopic observations. However, taking into account the error bars and the somewhat peculiar shape of its central spatial profile, this is not clear. We consider that the star indeed traces the seeing accurately.
Considering the broad component, if we take into account the error bars and because there is no obvious kinematic substructure within the seeing size, we conclude that the outflow is not spatially resolved. 
We estimate an upper limit to the intrinsic outflow radial size of R $<$ 3.3 kpc. This is a high upper limit because the seeing was very poor (FWHM$\sim$2.1\arcsec). The systemic component is not resolved either.

\vskip2mm
\item{\bf SDSS 1101+4004}

This is a radio-intermediate QSO2 (Tab.~\ref{info1}). The {\tt FIRST} image clearly shows a bright radio core and a pair of radio lobes oriented at 112$^\circ$ north to east and symmetrically located at $\sim$40$^{\prime\prime}$ (230 kpc) away from the nucleus (Fig.~\ref{Lal_Ho}). The slit was located along this radio axis.
The comparison between the {\tt FIRST} and {\tt NVSS} fluxes suggests that 29\% of the 1.4 GHz flux extends across spatial scales between 5\arcsec and 45\arcsec (see Table~\ref{info1}).

The pixel-to-pixel analysis of this system reveals three components (Fig.~\ref{panel_1101}). 
Two are rather narrow (narrow N2, N3, $\sigma\sim$ 80 km s$^{-1}$). The third is broader (NB, $\sigma\sim$ 350 km s$^{-1}$), and we identify it with the systemic component, associated with the AGN host galaxy. Following the M$_{\rm BH}$ versus $\sigma_{*}$ correlation \citep{McConnell13}, the $\sigma$ values of N2 and N3 would imply an unrealistically low BH mass for a quasar of $\sim$10$^6$ M$_{\odot}$. This assumption is also supported by the spatial coincidence between the centroid of the [OIII] and the spatial continuum profiles. The high $\sigma$ observed for the systemic component can be explained by the presence of a massive galaxy. For comparison, the typical mean stellar $\sigma$ in massive elliptical galaxies is in the range 230 -- 370 km s$^{-1}$ \citep{Veale17}. The $\sigma$ of the broad component would imply M$_{\rm BH}\sim 10^{9.6}$ M$_{\odot}$.

 In the v$_{shift}$ panel, the systemic (NB) and secondary narrow (N2) components show quite similar kinematics, while N3 is blueshifted by 500 km s$^{-1}$ from the systemic component. This might be a different galaxy, as shown by  IFS Gemini data by \citet{Liu13, Liu13_2}.

The seeing properties are uncertain in this case. The fact that the spatial distribution of NB is more compact than the stellar profile (Fig.~\ref{panel_1101}) demonstrates that the seeing improved during the spectroscopic observations in comparison to the acquisition image. In any case, all three kinematic components are unresolved or are barely resolved at best.
We find no evidence for an outflow in this system.

\end{itemize}

\subsection{Dynamical support of QSO2 hosts}
\label{dynam}

In this section we focus on quantifying the dynamical support that characterizes the QSO2 we analyzed. In order to kinematically characterize a system, the dynamical ratio (v/$\sigma$) between the stellar velocity amplitude and the mean stellar velocity dispersion of a system is a key parameter. This information is not available for the stellar component of our sample. \citet{Greene05} compared the stellar and gas kinematics in a large sample of type 2 AGN selected from the SDSS. They found that if the asymmetric wings [O III]$\lambda$5007 are removed, the width of the line core $\sigma_{gas}$ component can also be used to trace the stellar component $\sigma_{*}$. Because the scatter is large, they advocate the use of this relation only in statistical studies. Because our objective in this section is to analyze the general behavior of all objects in our sample rather than interpreting individual systems, we consider it a reasonable assumption.

Following their work, after applying the kinematic decomposition (Sect.~\ref{data_red}), we used v$_{shear}$ and $\sigma_{mean}$ of the systemic component (Sect.~\ref{kin_pars}) to trace the stellar kinematics and measure the dynamical ratio v/$\sigma$ as v$_{shear}$/$\sigma_{mean}$ for all galaxies (see Tab.~\ref{NARROW}). \citet{Greene05} found $\frac{\sigma_{gas}}{\sigma_{*}}$=1.22$\pm$0.78. We ignored this correction factor because it is unknown how v$_{gas}$ and v$_{*}$ are related, and accordingly, whether a correction should also be applied to v$_{gas}$. On the other hand, we compared with results for other samples that in some cases were classified based on the gas kinematics (spiral, ultra-luminous infrared galaxies (U)LIRG, and Lyman-break analog (LBA) systems) and on stellar kinematics (ellipticals). We confirm that using $\sigma_{*}\sim\sigma_{gas}/1.22$ instead of $\sigma_{gas}$ produces moderate changes in the kinematic values and would not affect our conclusions regarding the dynamical state of the systems.

\begin{figure*}
\begin{center}
\hskip0cm\includegraphics[scale=0.63]{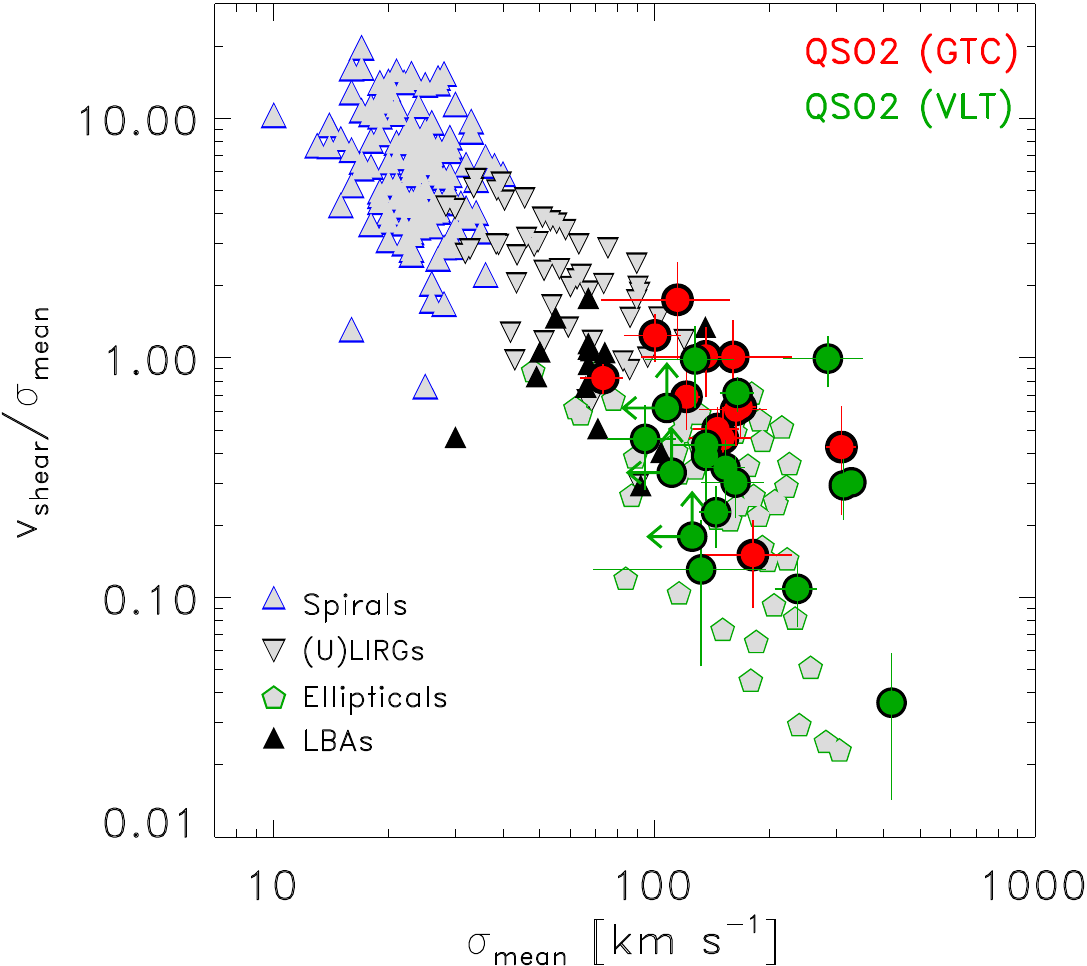}
\hskip0cm\includegraphics[scale=0.63]{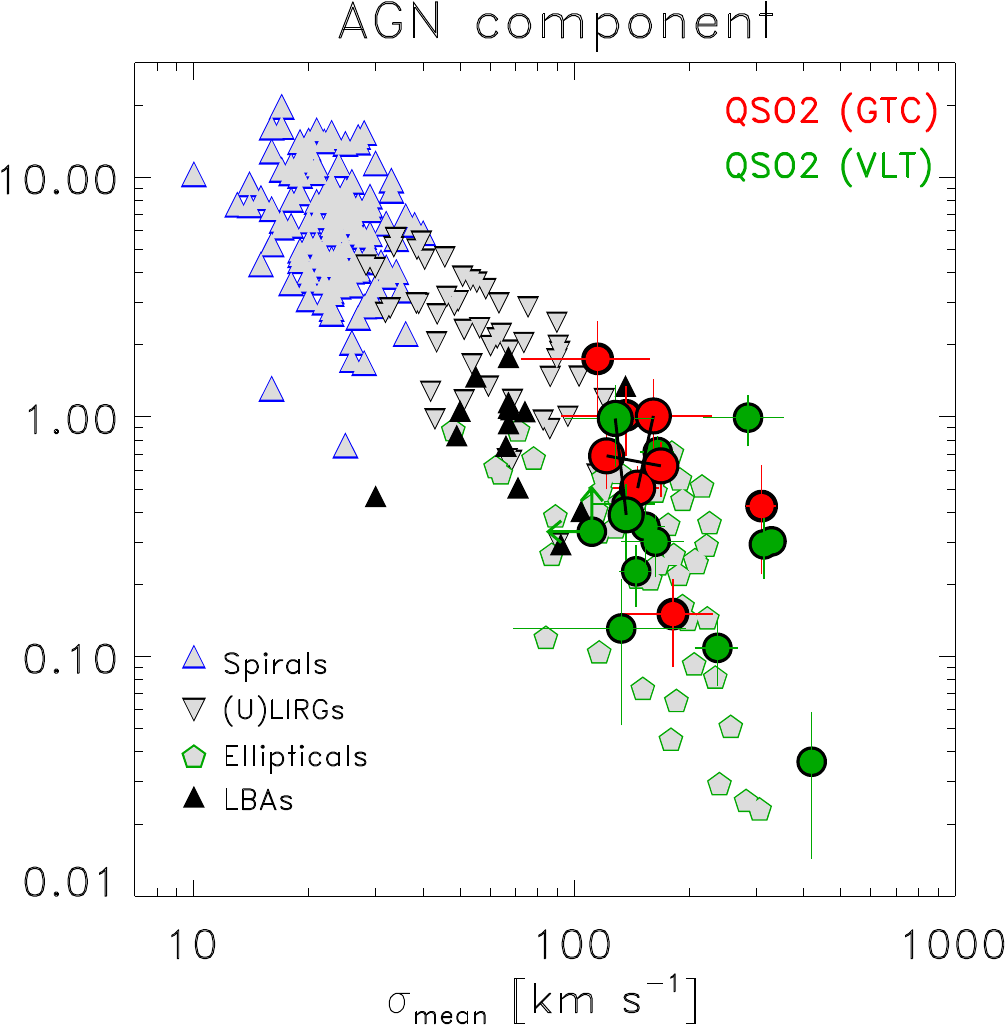}
\caption{Dynamical state of the galaxies. Relationship between the observed dynamical ratio, v$_{shear}$/$\sigma_{mean}$, and the mean velocity dispersion, $\sigma_{mean}$ for all galaxies isolated in the GTC (red) and VLT (green) samples (left). Only the AGN host galaxies are shown in the right panel. When two locations are available for the same AGN host (derived from the two spectra obtained with different slit PAs), these are joined by a solid black line. 
The blue--contoured triangles represent spiral GHASP galaxies \citep{Epi10}, green--contoured pentagons show E/SO objects \citep{Cap07}, black--contoured top-down empty triangles show (U)LIRGs (B13), and black filled triangles indicate LBAs \citep{Gon10}. For the spirals the v parameter is defined as the maximum amplitude of the rotational curve within the extent of the velocity field along the major axis, and $\sigma$ is the average of the velocity dispersion map. For E/S0, v and $\sigma$ are luminosity--weighted square quantities that we derived from the stellar velocity field and velocity dispersion maps, respectively \citep[see details in][]{Cap07}. For the (U)LIRGs, the velocity shear and mean velocity dispersion were computed as in this work (a slightly different percentile was used in B13 to derive the v$_{shear}$). For the LBAs, the velocity shear was defined as in B13, while the velocity dispersion is the flux--weighted mean value.}
\label{dyn_ratio}
\end{center}
\end{figure*}

In Fig.~\ref{dyn_ratio} v/$\sigma$ is presented with respect to $\sigma$. The panel on the left includes all galaxies (AGN hosts and companions). The panel on the right shows only the AGN hosts. Objects with v/$\sigma$~$>$~1 are classified as rotation dominated and objects with v/$\sigma$ $<$ 1 as random-motion dominated systems \citep[e.g.,][and references therein]{Epi12}.  

We compare our kinematical results with those of other samples of non-active galaxies based on IFS data: elliptical/lenticular galaxies (E/S0) are drawn from the SAURON project, studying the stellar kinematics \citep[e.g.,][]{Cap07}, local spiral galaxies were selected from the Gassendi H$\alpha$ survey of SPirals \citep[GHASP,][]{Epi10}, LBAs at z$\sim$0.2 observed in H$\alpha$ with OSIRIS/Keck \citep[i.e.,][]{Gon10}, and local (U)LIRGs analyzed in B13. We note that these works use slightly different definitions for v and $\sigma$, but this has no significant impact on the interpretation of the v/$\sigma$ over $\sigma$ relationship (see caption of Fig.~\ref{dyn_ratio}).

The majority of the QSO2s show v/$\sigma<$ 1, implying that they are dominated by random motions (dispersion-dominated systems). A few objects fall in the transition region between (U)LIRGs and elliptical galaxies in the v/$\sigma$--$\sigma$ plane. Most fall in the area of the E/S0 galaxies.

We used long slit data that are affected by an incomplete 2--dim spatial coverage. Integral field spectroscopy is clearly a more reliable technique to trace the kinematics. In spite of this limitation, we do not expect a significant impact on our results. If a scale factor of v$/\sigma_{\rm IFS}\sim$ 0.57 $\times$ v$/\sigma_{\rm LS}$ were applied to our v$/\sigma$ long-slit values following \citet{Cap07}, in general, the moderate change of location in the v$/\sigma$ vs. $\sigma$ diagram would not change the dynamical classification of the systems. This is further supported by the data on the four systems in our sample that were observed with two slit positions (see Fig.~\ref{dyn_ratio}, right). The same galaxy can change its location in the diagram slightly depending on the slit position angle, as we show in Fig.~\ref{dyn_ratio} (right), but the dynamical classification does not vary.

We list in Table~\ref{info_morpho} the classification of the host galaxies of the sample based on the dynamical state (Col. 2) and the morphological classification (Col. 3). When two galaxies are identified, only the AGN host is included in the dynamical classification. The morphological classification is based on the visual inspection of the SDSS or, when available, HST images. When a classification based on the parametric decomposition of the structural components was available \citep{UM19}, this is mentioned instead because it is more reliable. `Uncertain' means that  an unambiguous classification was not possible.

We are confident about the morphological classification of 15 of the 19 objects. Most are ellipticals (9  of 15) or complex systems (double, multiple, or highly disturbed; 6 of 15). None are classified as spiral or disks. 
The dynamical classification reveals that the vast majority of objects (17 of 19) are consistent with E/S0, including those that have highly disturbed morphologies that indicate mergers. None are consistent with spiral or disk galaxies.

A larger sample of QSO2 is required, as well as a more complete spatial coverage than that given by the long-slit data using integral field spectroscopy to study the kinematics in these objects and their dynamical state in a  more comprehensive way. In spite of these limitations, we can claim that the dynamical state of most QSO2 hosts is consistent with that of E/S0, with only one object (SDSS 0741+3020) placed in the transition region with ULIRGs. This result is consistent with morphological and parametrical studies of host galaxies of luminous type 2 AGN \citep[e.g.,][]{Bessiere12, VM12, UM19}, which show that the majority of QSO2 hosts ($\sim$80\%) are elliptical and highly disturbed systems. Disks are identified only in $\sim$24\% of QSO2 \citep{UM19}.

\begin{figure*}
\begin{center}
\hskip-5mm\includegraphics[scale=0.75]{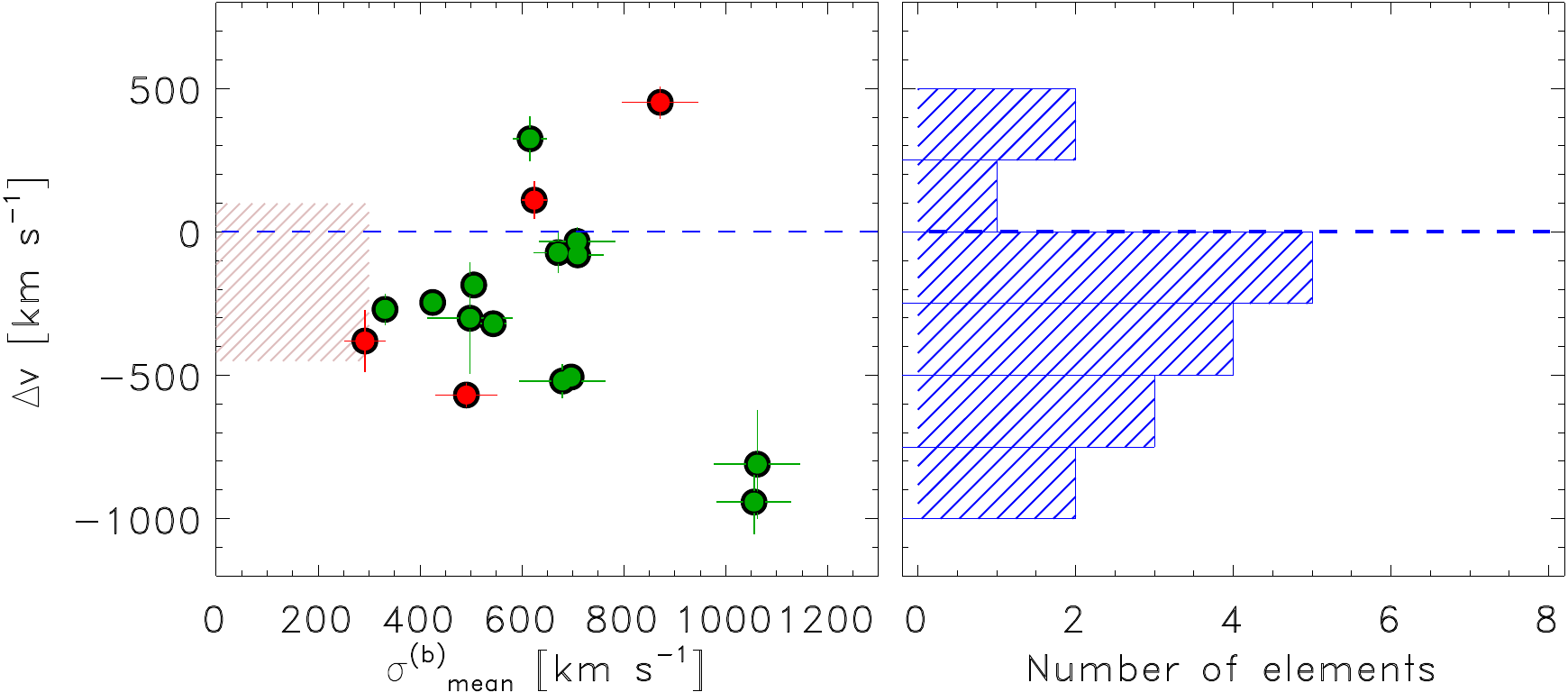}
\caption{Velocity shift, $\Delta$v, between the broad and narrow components vs. the mean velocity dispersion of the broad component, $\sigma^{(b)}_{mean}$. The horizontal dashed line indicates the zero-velocity shift. Most of the objects show a blueshifted (negative values) broad component, most of them clustered around $\Delta$v = 0 - --500 km s$^{-1}$. The dashed red region represents the typical values covered by a local sample of (U)LIRGs studied by B13. The color code is the same as in Fig.~\ref{dyn_ratio}.}
\label{broad_kin}
\end{center}
\end{figure*}

\subsection{Properties of the ionized outflows}
\label{outflows}

We confirm ionized outflows in four of the six QSO2 in the GTC sample: SDSS 0741+3020, SDSS 0818+3958, SDSS 0834+5534, and SDSS 0956+5735. No outflows were detected in SDSS 0902+5459 and SDSS 1101+4004. Outflows in these two objects are of course not discarded. This is not surprising because of the complexity of these systems, which consist of two galaxies or nuclei and the poor seeing that affected the observations. This runs against the efficiency of detecting the outflows because they are expected to be quite compact ($R\sim$1--2 kpc) in general \citep[][]{VM14, Husemann16, VM16, Fischer18, Rose18, Spence18, StorchiB18, Tadhunter18}.

Of the four detected outflows, three are indeed quite compact. Two outflows are spatially unresolved, and one has a radial size of $\sim$0.8 kpc (SDSS 0834+5534).
These three outflows have kinematic properties within the range observed in similar objects (Table~\ref{BROAD}), as shown in Fig.~\ref{broad_kin}. This figure shows that they are more extreme than ionized outflows studied in (U)LIRGs (B13), in the sense that the line widths are significantly broader ($\sigma>$300 km~s$^{-1}$, while such broad lines are not found in (U)LIRGs unless they contain a QSO). Higher velocity shifts can also be reached ($\sigma\ga$ 500 km s$^{-1}$ in some cases). That the kinematics of QSO-induced outflows are more extreme than those of (U)LIRGs is well known \citep[e.g.,][]{Greene11, Rupke11}.

The ionized outflow in SDSS 0834+5534 has rather extreme kinematics in comparison with other QSO2, with $\sigma$ reaching $\sim$950 km s$^{-1}$. It also shows a very high velocity redshift relative to the systemic component of up to $\sim$+630 km s$^{-1}$. We proposed in Sect.~\ref{individual} that this object is seen at an intermediate (type 1 vs. type 2) orientation, so that the high inclination angle allows us to see the inner regions of the outflow, deep into the inner part of the NLR where the outflow kinematics may be more extreme \citep{VM14,VM15}. It is also possible that the outflow is induced by the interaction with the inner radio jet, as suggested by the alignment between the optical HST and the radio morphologies in the inner $\sim$1\arcsec\ (Sect.~\ref{individual}). 

\begin{figure}
\vskip5mm
\begin{center}
\includegraphics[scale=0.48]{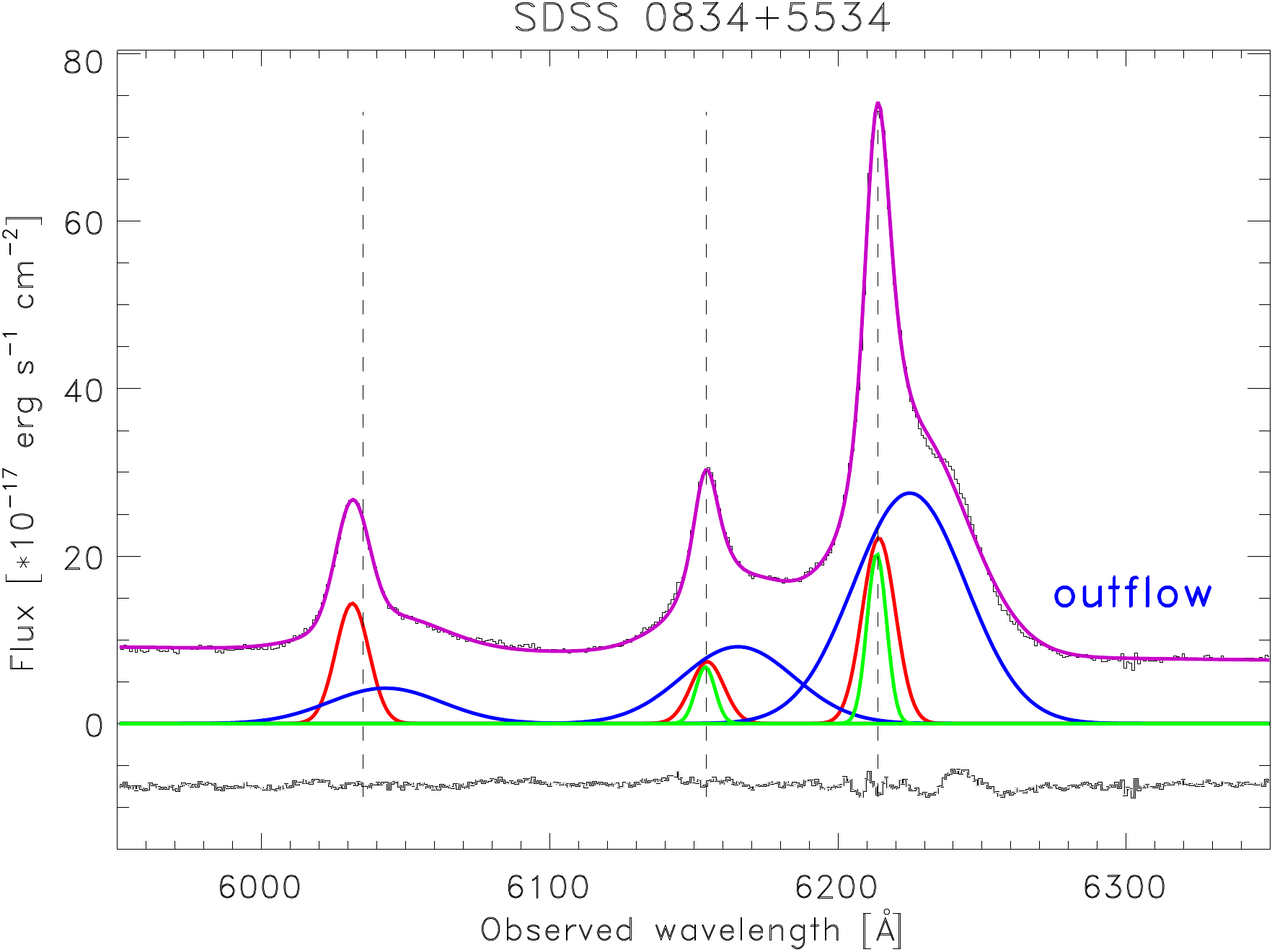}
\caption{H$\beta$--[OIII] observed spectrum of SDSS 0834+5534 where the main (systemic) and broad (outflow) components coexist. The purple curve shows the total H$\beta$--[OIII] components obtained from multicomponent Gaussian fits. The red and blue curves represent the systemic and broad components, respectively. In green we show the third narrow component (see text for details). }
\label{outflow_fig}
\end{center}
\end{figure}

We estimated the outflow mass ($M_{outflow}$), mass, and energy injections rates ($\dot M_{outflow}$, $\dot E_{outflow}$) for SDSS 0834+5534, the only object in the OSIRIS sample for which $L_{\rm H\beta}^{outflow}$=3.5$\times$10$^{41}$ erg s$^{-1}$ (blue component in Fig.~\ref{outflow_fig}) and $R_{outflow}=$0.7$\pm$0.4 kpc (instead of an upper limit) are both available.

We estimated the mass outflow \citep{Baldwin03} as
\begin{equation}
 \hskip3mm M_{\rm outflow}=\frac{\mu_{\rm H}  \hskip2mm m_{\rm H}}{\alpha^{\rm H\beta}_{\rm eff} \hskip2mm E_{\rm H\beta}} \times \frac{L_{\rm H\beta}^{\rm outflow}}{n},
\end{equation}
  
~which, normalized to $M_{\rm \odot}$ results in 
  
\begin{equation} 
 \hskip3mm M_{\rm {outflow}}  = 9.4 \times 10^{-33} \times ~\frac{L_{\rm H\beta}^{outflow}}{n} ~ [\rm{M_{\rm \odot}}]. 
\end{equation}
 
$\mu_{\rm H}$=1.42 is the mean mass per hydrogen atom assuming solar metallicity, $m_{\rm H}$  is the proton mass, $\alpha^{\rm H\beta}_{\rm eff}$ is the H$\beta$ case B effective recombination coefficient for an electron temperature $T_{\rm}\sim$10$^4$ K \citep[i.e., $\sim$3$\times$10$^{-14}$ cm$^3$ s$^{-1}$; ][]{Oster89}, $E_{\rm H\beta}$ erg is the energy of one H$\beta$ photon, and $L_{\rm H\beta}^{outflow}$ and $n$ are the H$\beta$ luminosity and density of the outflow. Moreover, 
\begin{equation}
\hskip4mm \dot M_{outflow} = \frac{M_{outflow} ~  V_{\rm max}}{R_{outflow}}   
\end{equation}

~and
\begin{equation}
\hskip4mm \dot E_{outflow} =\frac{1}{2}~\dot M_{outflow} ~ V_{max}^2. 
\end{equation}

Following \citet{Rupke05}, $$V_{\rm max}= \frac{FWHM^{(b)}}{2}+ |\Delta {\rm v}| = 2.35 \frac{\sigma_{\rm mean}^{(b)}}{2}+ |\Delta {\rm v}|$$ which gives V$_{max}$ = 1744 km s$^{-1}$  for the broad component (Table~\ref{BROAD}).    
When we assume $n$ = 1000 cm$^{-3}$ \citep{Holt11, VM14, VM16}, $M_{outflow}$ = 3.3$\times$10$^{6}$ M$_{\odot}$, $\dot M_{outflow}$ = 8.5 M$_{\odot}$ yr$^{-1}$ and $\dot E_{outflow}$ = 7.1 $\times$10$^{42}$ erg s$^{-1}$. These moderate values are within the range inferred for the VLT sample by \citet{VM16} \citep[see also][]{VM14}. As the authors discuss, these values are affected by huge uncertainties that is mainly due to the unknown $n$ because a broad range of densities (maybe up to $n\ga$10$^6$ cm$^{-3}$) may exist in the outflowing gas. 

The uncertainties on SDSS 0818+3958 and SDSS J0956+5735, for which only $L_{[OIII]}^{outflow}$ and upper limits on $R_{outflow}$ could be measured, are clearly also very high, but the outflow parameters are probably similarly modest. These objects have $R_{outflow}<$2.2 and $<$3.3 kpc, respectively. At $\sim$2--3 kpc from the AGN, low $n\la$few$\times$100 cm$^{-3}$ are possible. However, even if such large sizes were confirmed, the outflow emission is expected to be preferentially emitted in a central compact source where a broad range of $n$ is present. 

Thus, as already argued in \citet{VM16} for the VLT sample, it is not clear that the ionized outflows in SDSS 0834+5534, SDSS 0818+3958 and SDSS 0956+5735 have a significant impact on their host galaxies, neither by injecting energy and mass beyond the inner $\sim$ 1--2 kpc of the narrow-line region, well into the ISM, nor by triggering high mass outflow rates \citep[see also][]{VM14, Husemann16, VM16, Rose18, Spence18, Tadhunter18, Fischer18, StorchiB18}.

\subsubsection{Giant radio-induced outflow in SDSS 0741+3020?}

Of special interest is SDSS 0741+3020. This object is associated with an ionized central outflow that extends 1.8$\pm$0.4 kpc toward the SE and 4.2$\pm$0.4 kpc toward the NW from the AGN in the direction of the radio axis. It has $\sigma\sim$220--480 km s$^{-1}$ and is blueshifted relative to the systemic component by $\sim$200--500 km s$^{-1}$. This is an unusually large size in a radio-quiet QSO2, where, as mentioned above, the outflows are in general compact. 
Low $n$ ($\la$ few$\times$100 cm$^{-3}$) are expected at such large radial distances from the AGN, while higher $n$ exist in the central narrow-line region (see previous section). This and the lack of morphological information in two spatial dimensions prevents meaningful limits on  $M_{outflow}$, $\dot M_{outflow}$ and $\dot E_{outflow}$.

The radio source associated with this object is slightly resolved at 8.4 GHz and 1.4 GHz and appears to show a double structure located at $\sim$1$^{\prime\prime}$ or $\sim$6 kpc from the optical position of the galaxy \citep{Lal10}. The similar maximum size of the central outflow suggests that it may have been induced by the interaction between the radio structures and the gas.

SDSS 0741+3020 is also associated with a giant nebula of maximum total extension $\sim$112 kpc. While the nebula is very quiescent in the outer parts ($\sigma\la$~few~$\times$10 km s$^{-1}$), turbulent gas ($\sigma\sim$130 km s$^{-1}$) has also been detected across this gas reservoir up to $\sim$40 kpc from the AGN, {well outside the main body of the galaxy (see Sect.~\ref{individual}).

SDSS 0741+3020 is strongly reminiscent of the radio-quiet QSO2 nicknamed the Beetle at $z=$ 0.10 \citep{VM17}. This object is also associated with a giant ionized nebula (maximum extension 71 kpc).  The gas motions are turbulent ($\sigma\sim$160--200 km s$^{-1}$) within the radio structures, up to $\sim$26 kpc from the AGN, and are more quiescent beyond. We proposed a scenario in which the giant nebula consists of in situ gas, made of tidal remnants that were redistributed during galactic interactions  across tens of kiloparsec in the circumgalactic medium (CGM). The interaction between the large-scale radio structures ($\sim$46 kpc) and the giant nebula has induced outflows that are responsible for the kinematic turbulence. 
As discussed in that work, the coexistence in the Beetle of a large-scale radio source and a rich widely spread gaseous environment illuminated by the quasar facilitates the detection and occurrence of radio-induced feedback across huge volumes.

All low $z$ AGN with reported giant nebulae ($\ga$60 kpc), including the Beetle, show clear signs of past or ongoing galactic interactions \citep[see][for a comparison]{VM18}, with rich tidal features of different morphologies (shells, tails, bridges, etc). This is possibly the case of SDSS 0741+3020 as well. Although the available SDSS images and our acquisition GTC images are not deep enough to detect low surface brightness tidal features, a merger or interaction event is suggested by the complex morphology, which shows a multiple system (Table~\ref{info_morpho}). This event may be the origin of the giant nebula. 

The existence of extended radio emission on scales larger than the 5\arcsec ($>$30 kpc) in this object is suggested by the fact that $\sim$38\% of the 1.4 GHz radio flux is outside the {\tt FIRST} 5\arcsec beam (Tab.~\ref{info1}). Therefore, SDSS 0741+3020 shares several properties with the Beetle: 1) it shows preliminary evidence for a merger or interaction event, 2) it is associated with giant turbulent nebula, and 3) there is indirect evidence for a large-scale extended radio structure. 

All this suggests a similar scenario: a radio source that has so far been invisible has escaped the galaxy boundaries that extend up to at least $\sim$40 kpc from the AGN, well into the CGM. The interaction with the giant nebula may have produced outflows that are responsible for the turbulent motions. Deeper radio observations, sensitive to low surface brightness structures, are necessary to investigate whether this large  structure indeed exists. If confirmed, SDSS 0741+3020 would be the second radio-quiet quasar with confirmed unambiguous evidence for AGN radio-induced feedback acting across tens of kiloparsec.

\begin{table*}
\centering 
  \caption{Classification of the host galaxies of the GTC and VLT samples based on their dynamical state and morphology. }
  \label{info_morpho}
\begin{tiny}
\begin{tabular}{c ccc}
 \hline\hline\noalign{\smallskip}
Short galaxy ID          &      Dynamical       &       {Morphological class}   & Reference\\
         &      class   &        {(visual or parametric)}       & \\
\cmidrule(lr){3-3}
{\smallskip} 
 (1) & (2) &(3) & (4)\\
\cmidrule(lr){1-4}\cmidrule(lr){1-4}
  \multicolumn{4}{c}{\bf OSIRIS/GTC}  \\
\cmidrule(lr){1-4}\cmidrule(lr){1-4}
0741+3020  & Intermediate (U)LIRG--E/S0         &        Multiple system        &       This work\\
\hline\noalign{\smallskip}
0818+3958       &  E/S0         &        Uncertain &This work \\
\hline\noalign{\smallskip}
0834+5534   &  E/S0  &  E  & This work\\
\hline\noalign{\smallskip}
0902+5459a  &  E/S0     &       Uncertain & This work\\
\hline\noalign{\smallskip}
0956+5735 &   E/S0      & {\it E}       & UM19 \\
\hline\noalign{\smallskip}
1101+4004a  &  E/S0     &       Double system  &  This work \\
\cmidrule(lr){1-4}\cmidrule(lr){1-4}
  \multicolumn{4}{c}{\bf FORS2/VLT}  \\
\cmidrule(lr){1-4}\cmidrule(lr){1-4}
J0923+0101 &  E/S0      &       {\it E} & UM19 \\
\hline\noalign{\smallskip}
J0950+0111 &     E/S0   &       Possibly 2 nuclei       & VM16, This work\\
\hline\noalign{\smallskip}
J0955+0346 &  E/S0 &    E & This work\\
\hline\noalign{\smallskip}
J1014+0244 &  E/S0      &       Uncertain  &This work\\
\hline\noalign{\smallskip}
J1153+0326 &  E/S0      &       E?      &This work      \\
\hline\noalign{\smallskip}
J1247+0152 &  E/S0      & E&    This work\\
\hline\noalign{\smallskip}
J1307--0214&   E/S0     & Highly disturbed      &This work      \\
\hline\noalign{\smallskip}
J1336--0039 &    E/S0 &         E       &This work      \\
\hline\noalign{\smallskip}
J1337--0128 &  E/S0 & Highly disturbed  & UM19  \\
\hline\noalign{\smallskip}
J1407+0217 &  E/S0 & {\it E}    & UM19  \\
\hline\noalign{\smallskip}
J1413--0142 &  E/S0 &   {\it E}& G09 \\
\hline\noalign{\smallskip}
J1430--0050 & Uncertain & Highly disturbed      & VM12\\
\hline\noalign{\smallskip}
J1546--0005 &   E/S0 & {\it     E}      & UM19  \\
\hline\hline\noalign{\smallskip}
\end{tabular}
\vskip0.2cm\hskip0.0cm
\end{tiny}
\begin{minipage}[h]{18.5cm}
\footnotesize
{ {\bf Notes:} Dynamical and morphological classifications of the AGN host galaxies in the GTC and VLT samples. Column (1): Short SDSS galaxy ID.
Column~(2): Dynamical classification according to the values derived in Tables~\ref{NARROW} and 3 in B13.
Column~(3): Morphological classification according to the visual inspection of SDSS and/or HST images (this work) or parametrical analyses \citep[G09, VM12, VM16, UM19:][respectively]{Greene09, VM12, VM16, UM19}.
The results based on the parametric classification are differentiated from the visual classification results with italic font. Column~(4): Reference for the morphological classification in Col.~(3). Both classification methods show that none of the galaxies are classified as spirals or disks, neither dynamically nor morphologically. }
\end{minipage}
\end{table*}

\section{Summary and conclusions}

We have studied the ionized gas kinematics in six optically selected SDSS type 2 QSO (QSO2) at z$\sim$0.3--0.5, which have evidence for extended radio sources. One QSO2 is radio-loud, three are radio-intermediate, and two are radio-quiet. Our goal was to investigate the potential role of large-scale radio-induced feedback in these systems. We analyzed the spatially extended gas kinematics using [OIII]$\lambda\lambda$4959, 5007, based on Osiris/GTC long-slit spectroscopy. One or two slit positions were used in all cases, one of which was always aligned with the main radio axis to maximize the chance of detecting extended ionized nebulae and large-scale radio-induced feedback.

Ionized outflows have been detected in four of the six QSO2. The most extreme outflow in terms of kinematics is detected in the only radio-loud object in the sample, SDSS 0834+5534. It has $\sim$950 km s$^{-1}$ and is redshifted by $\sim$630~km s$^{-1}$. An intermediate orientation and/or radio-induced feedback may be responsible for the extreme kinematics.

Three of the four outflows are rather compact. Two are spatially unresolved, and a third (the outflow associated with SDSS 0834+5534) has a radial extension R = 0.8$\pm$0.3 kpc. We find no evidence that these outflows are capable of a significant impact on their host galaxies, neither by injecting energy and mass beyond the inner $\sim$1--2 kpc of the narrow-line region, well into the ISM, nor by triggering high mass outflow rates. These small sizes are consistent with the growing evidence that ionized outflows are compact in most optically selected QSO2 (R$\la$1--2 kpc). Clearly, the existence of an extended radio source does not guarantee that radio-induced feedback will occur at any spatial scale. It is necessary that it interacts with the ambient gas.

Of special interest is the radio-quiet QSO2 SDSS 0741+3020 at $z=$0.47. It is associated with a giant nebula of  total extension $\sim$112 kpc. A large-scale ionized outflow, probably induced by the radio structures, has been detected along the axis that is defined by the central $\sim$1\arcsec\ radio structure. Up to R~=~4.2~$\pm$~0.4 kpc, the outflowing gas shows $\sigma\sim$220--480 km s$^{-1}$ and is blueshifted by $\sim$200--500 km s$^{-1}$ relative to the systemic component. Moreover, turbulent gas ($\sigma\sim$130 km s$^{-1}$) has also been detected across the giant gas reservoir up to $\sim$40 kpc from the AGN. This turbulence may have been induced by the interaction between a so-far undetected large-scale ($\ga$40 kpc from the AGN) radio source and the nebula. Deeper radio observations sensitive to large-scale low surface brightness structures are necessary to test this scenario.

We have also investigated the dynamical state of the QSO2 host galaxies in the 6 QSO2 observed with GTC and 13 more QSO2 that were previously observed with FORS2/VLT \citep{VM11, VM16}. The majority of the QSO2 show v/$\sigma<$ 1, implying that they are dominated by random motions (dispersion-dominated systems). Most (17 of 19) fall in the area of the E/S0 galaxies in the dynamical diagram v/$\sigma$ versus $\sigma$. None are consistent with spiral or disk galaxies. This result is consistent with morphological and parametrical studies of host galaxies of luminous type 2 AGN, which show that the majority of QSO2 hosts ($\sim$80\%) are elliptical and highly disturbed systems.

\appendix
\section{Spatially extended kinematic analysis }

In this appendix we show the 2--dim spectra and the results of the spatially extended kinematic analysis of the GTC sample.

\begin{figure*}
\vskip-3mm
\begin{center}
\includegraphics[width=16cm]{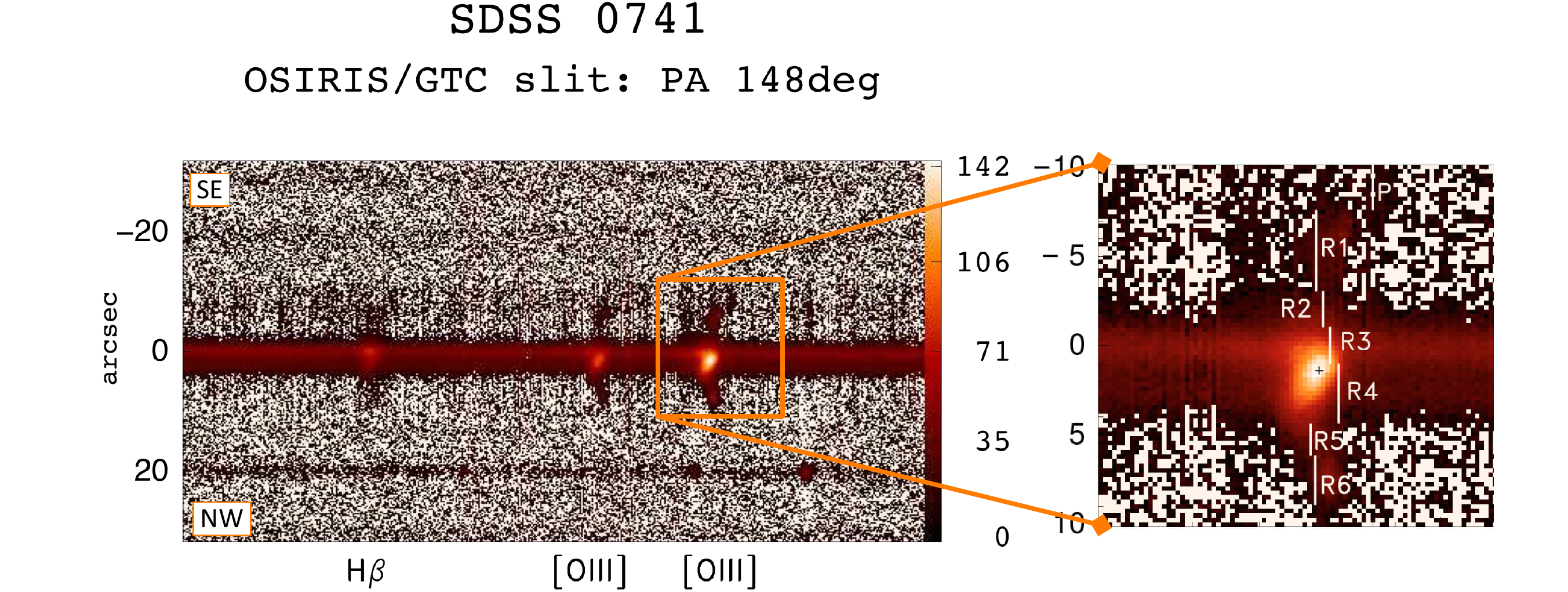}
\vskip0mm
\includegraphics[scale=0.54]{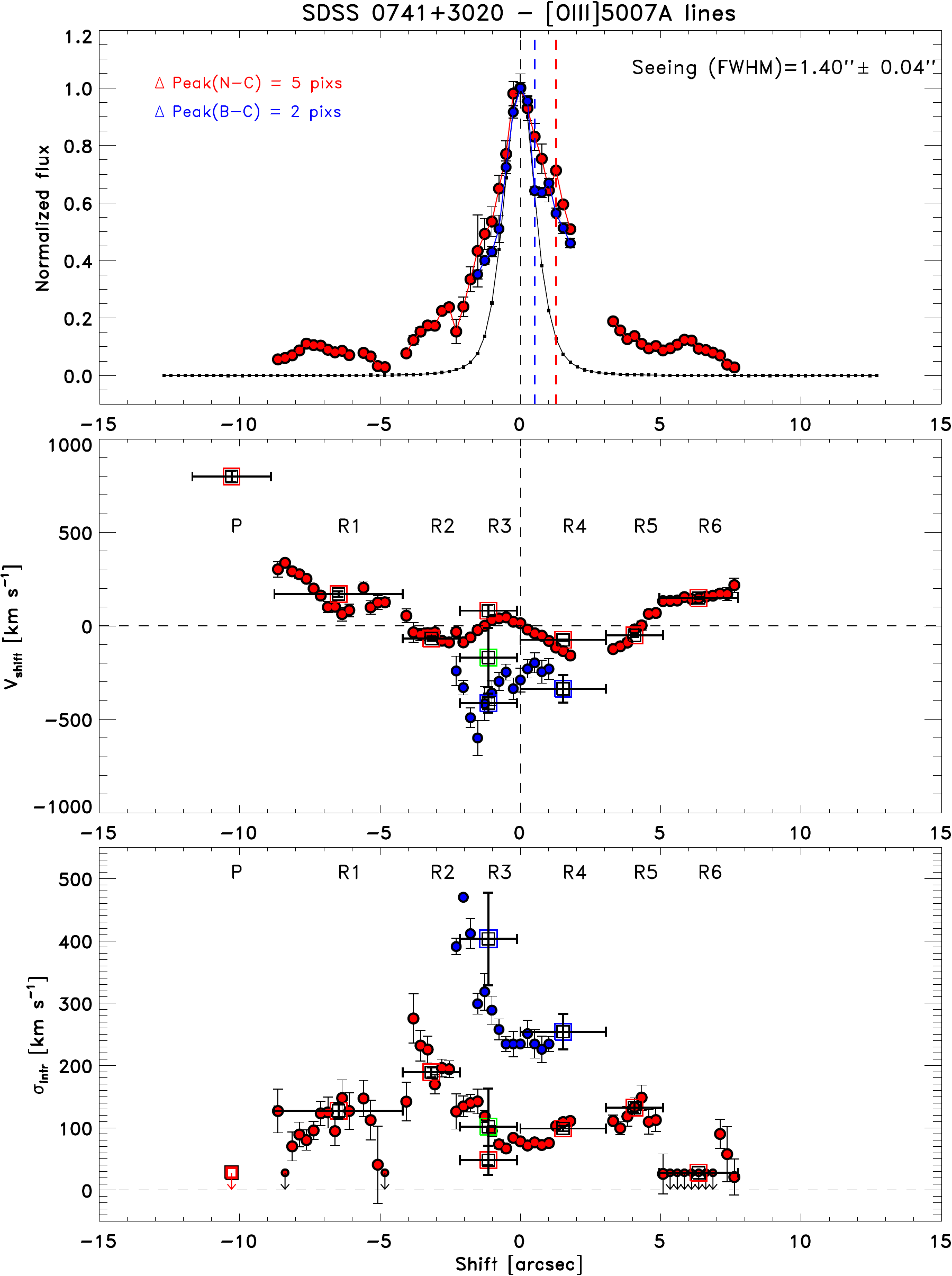}
\caption{SDSS 0741+3020: {\it Top panel. Left:}  
2--dim spectrum covering the H$\beta$ and [OIII]$\lambda\lambda$4959,5007 lines  along the slit PA. The colored bar spans the scale between the minimum and maximum fluxes in units of 10$^{-19}$ erg s$^{-1}$ cm$^{-2}$ pix$^{-1}$. {\it Right:} Zoom-in around [OIII]$\lambda$5007 highlighting all spatial regions considered in the kinematic analysis (R1, R2, etc). The spatial location of the narrow component flux peak is shown with a black cross. The spatial zero in all panels is assumed to coincide with the continuum spatial centroid (i.e., the location of the peak of the continuum). {\it Bottom panel.} Spatial distribution and kinematic properties of the individual kinematic components  identified in the [OIII] doublet. {\it Top:} Spatial distribution of the continuum-subtracted line flux. The spatial profile of each kinematic component has been shifted and centered with the seeing profile to facilitate visual comparison.
The spatial zero marks the location of the continuum peak. The line spatial centroid is assumed to coincide with the location of the line peak. The true shift between the line spatial centroid of a given kinematic component and the spatial zero (e.g., $\Delta$ Peak(N--C)) is shown with dashed vertical lines maintaining the same color--code.
Each kinematic component is identified in this panel with N (narrow), B (broad), or other, as explained in the main text for each object. {\it Middle:} Velocity field (i.e., velocity shift computed relative to the systemic velocity; see Sect.~\ref{kin_pars}). {\it Bottom:} Intrinsic velocity dispersion. Upper limits on the velocity dispersion are shown for spectrally unresolved lines. The results of the pixel-to-pixel analysis are shown with solid circles. No spatial shift has been applied to the velocity field and velocity dispersion. Red and blue are used for the systemic (narrow) and for the broad components, respectively. When a third component is isolated, green is used. The results of the analysis based on the spectra integrated within larger apertures (P, R1, R2, etc.) are shown with large open square symbols, using the same color-code as in the pixel-to-pixel analysis.}
\label{panel_0741}
\end{center}
\end{figure*}

\begin{figure*}
\begin{center}
\includegraphics[scale=0.6]{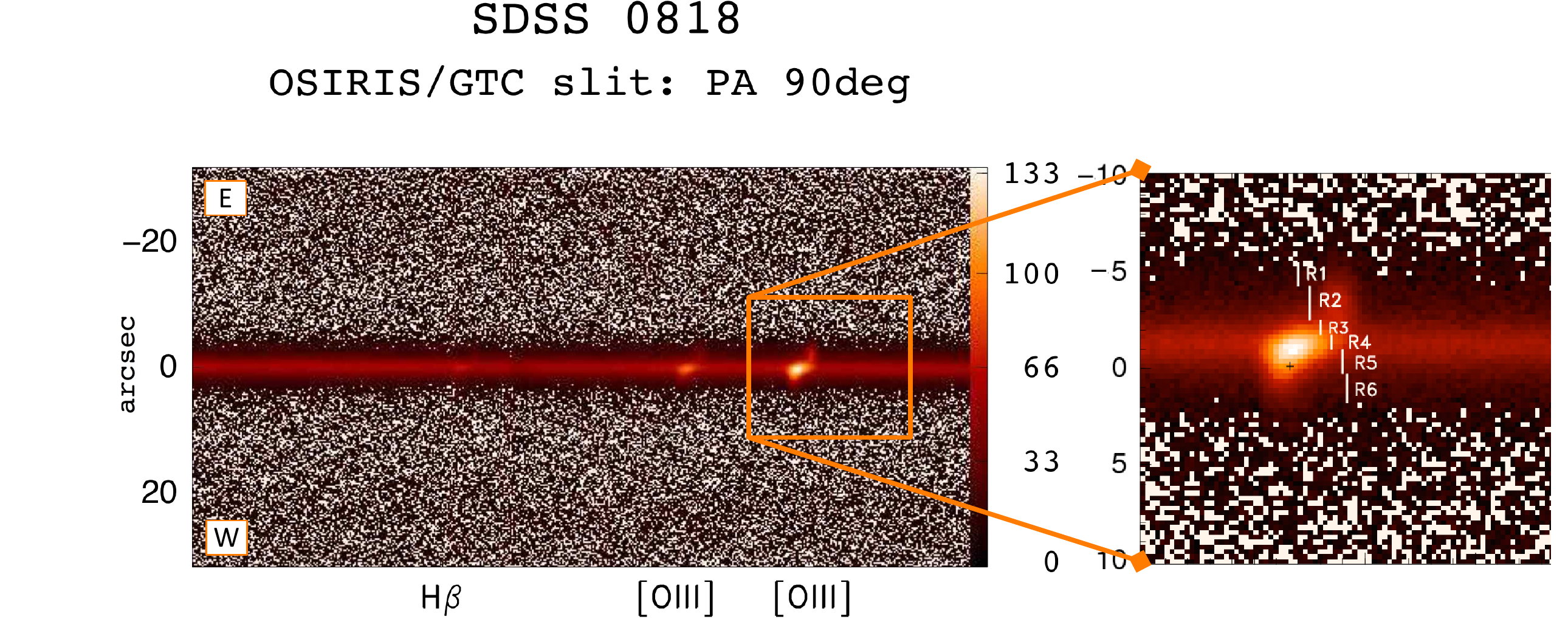}
\vskip15mm
\hskip0cm\includegraphics[scale=0.65]{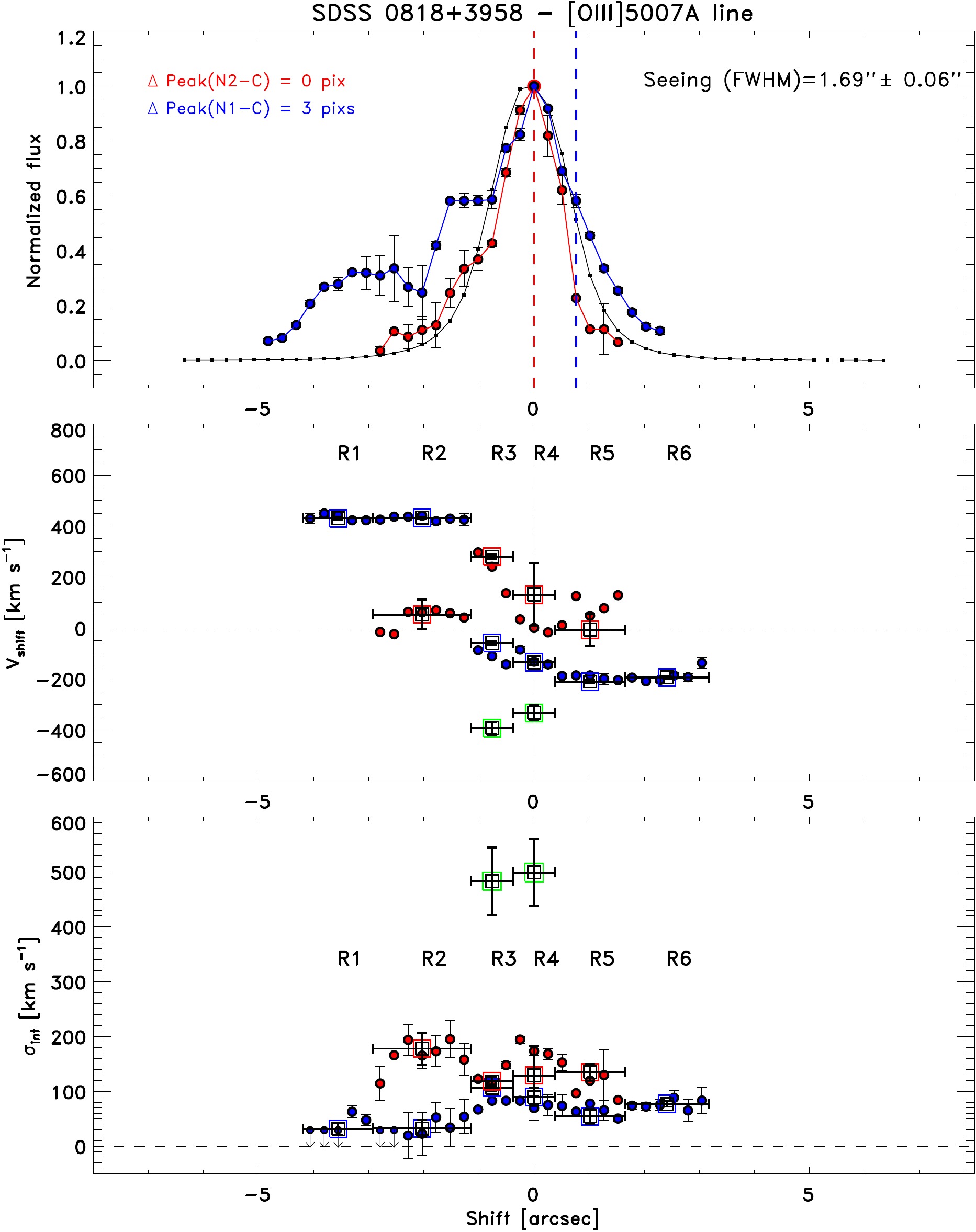}
\caption{SDSS 0818+3958: Same figure caption as in Fig.~\ref{panel_0741}. The fact that N2 shows a more compact spatial profile than the seeing shows that this improved during the spectroscopic observations in comparison with the acquisition image.}
\label{panel_0818}
\end{center}
\end{figure*}

\begin{figure*}
\begin{center}
\includegraphics[scale=0.6]{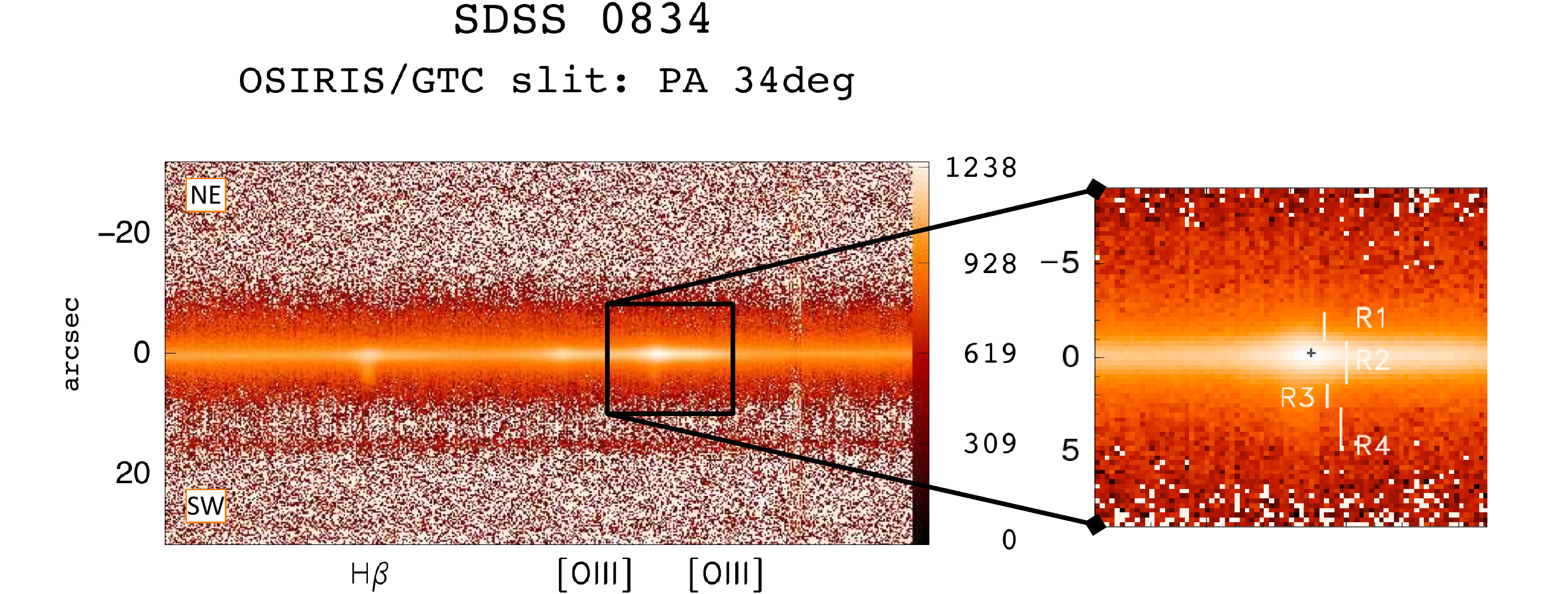}
\vskip15mm
\hskip0cm\includegraphics[scale=0.65]{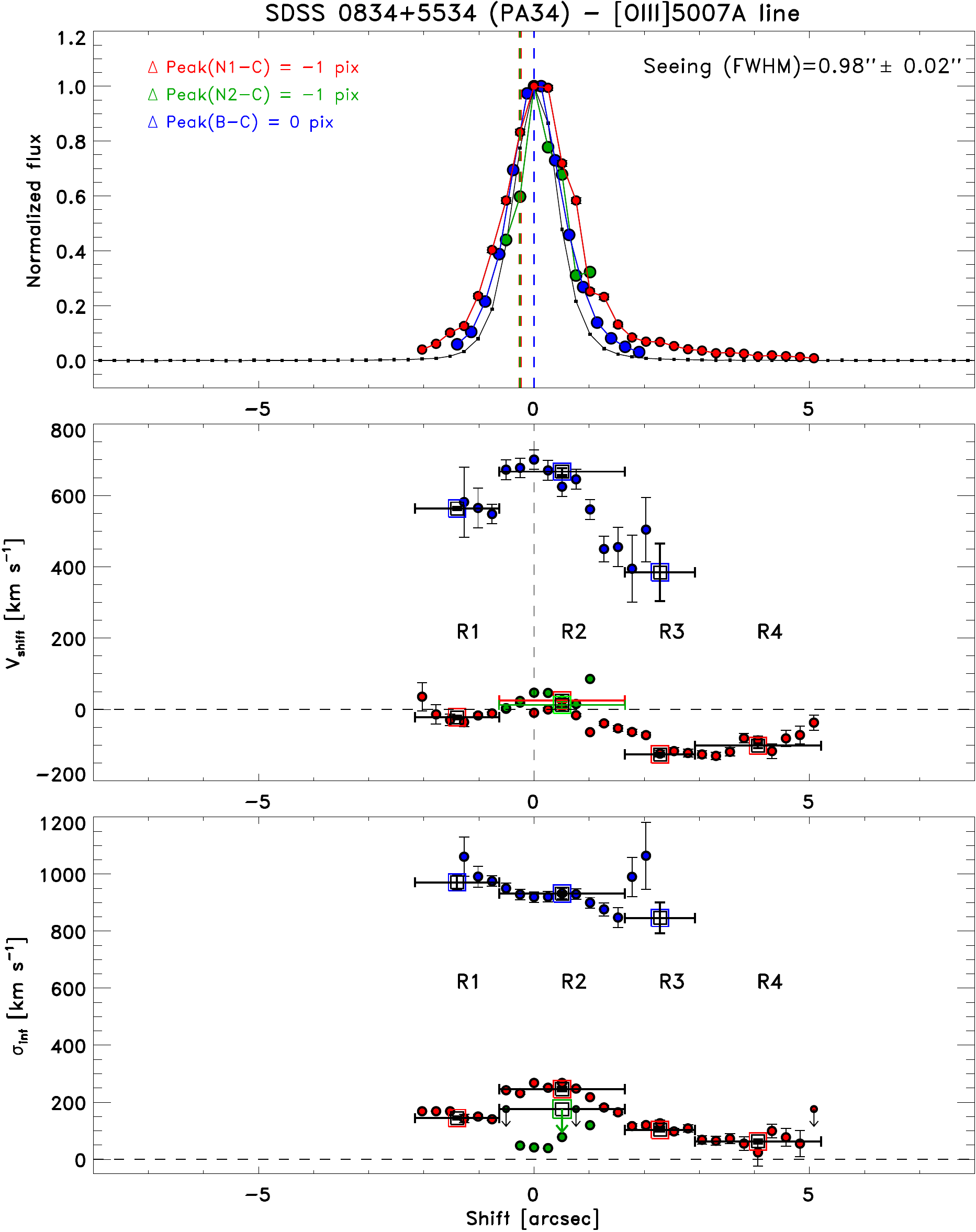}
\caption{SDSS 0834+5534: Same figure caption as in Fig.~\ref{panel_0741}.}
\label{panel_0834}
\end{center}
\end{figure*}

\begin{figure*}
\begin{center}
\includegraphics[scale=0.6]{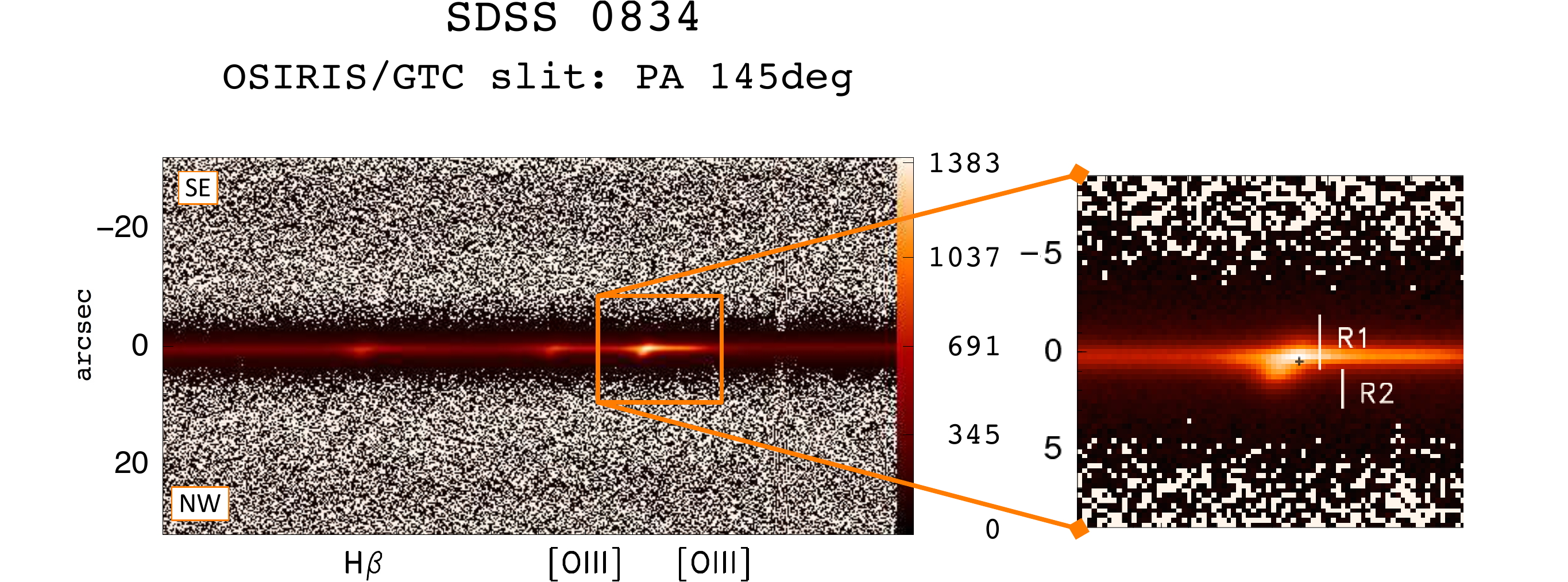}
\vskip15mm
\hskip0cm\includegraphics[scale=0.65]{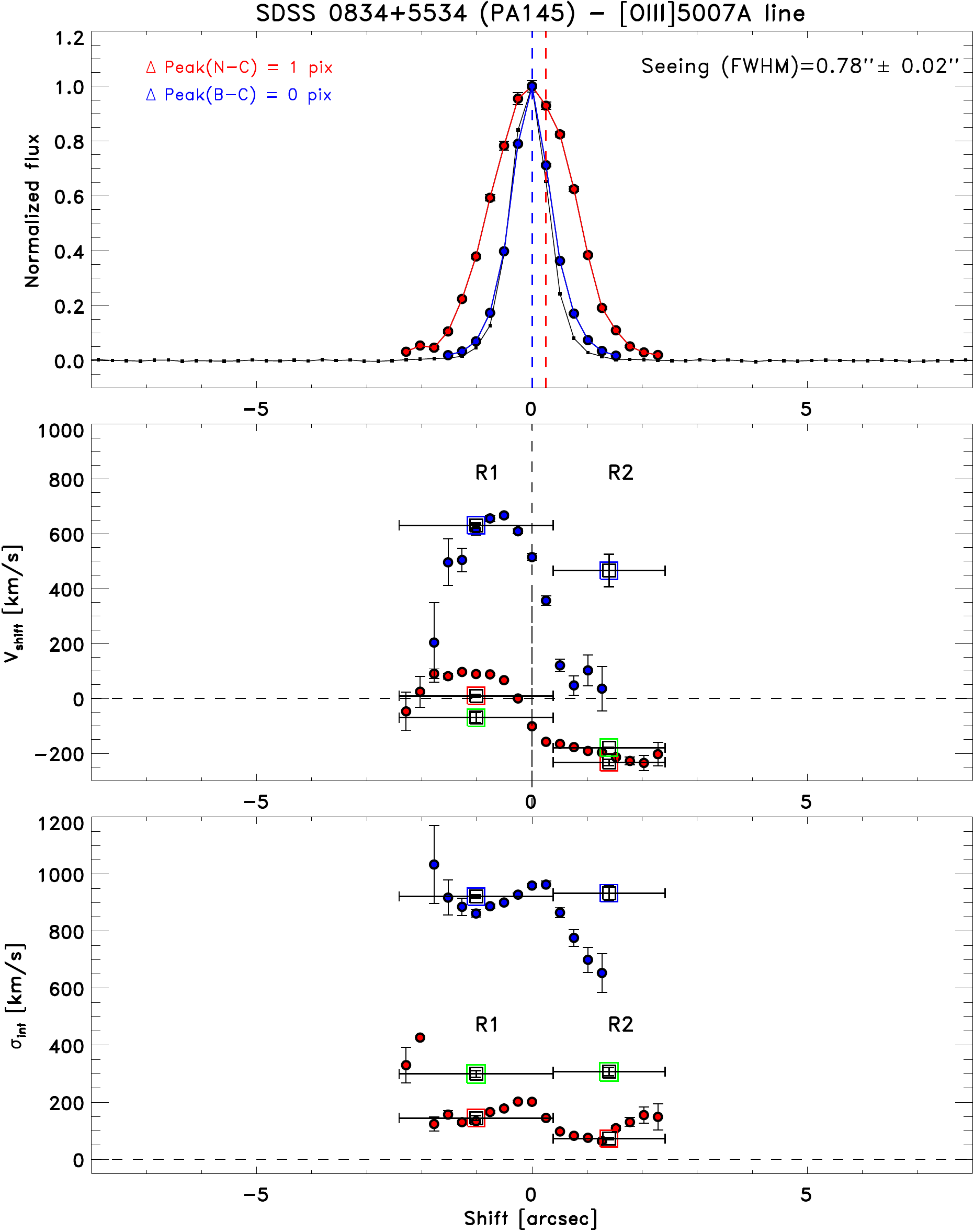}
\caption{SDSS 0834+5534: Same figure caption as in Fig.~\ref{panel_0741}. }
\label{panel_0834_2}
\end{center}
\end{figure*}

\begin{figure*}
\begin{center}
\includegraphics[scale=0.6]{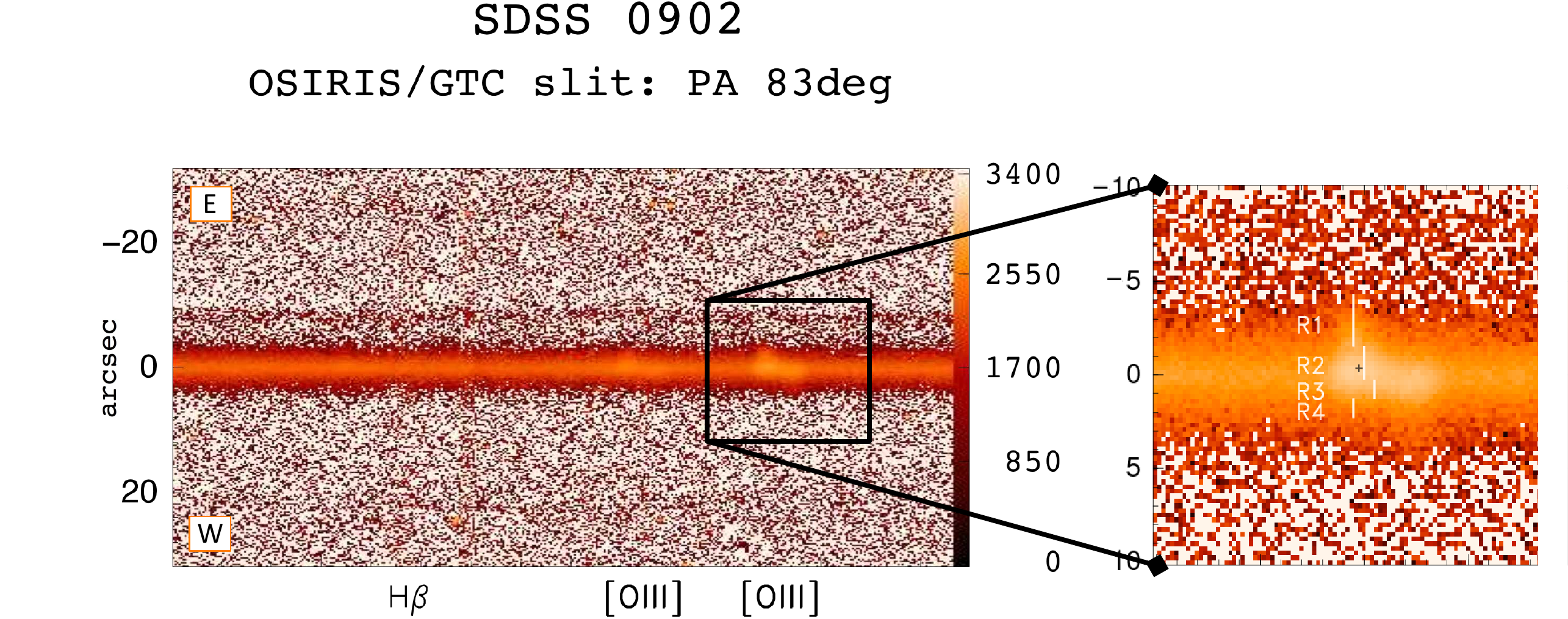}
\vskip15mm
\hskip0cm\includegraphics[scale=0.65]{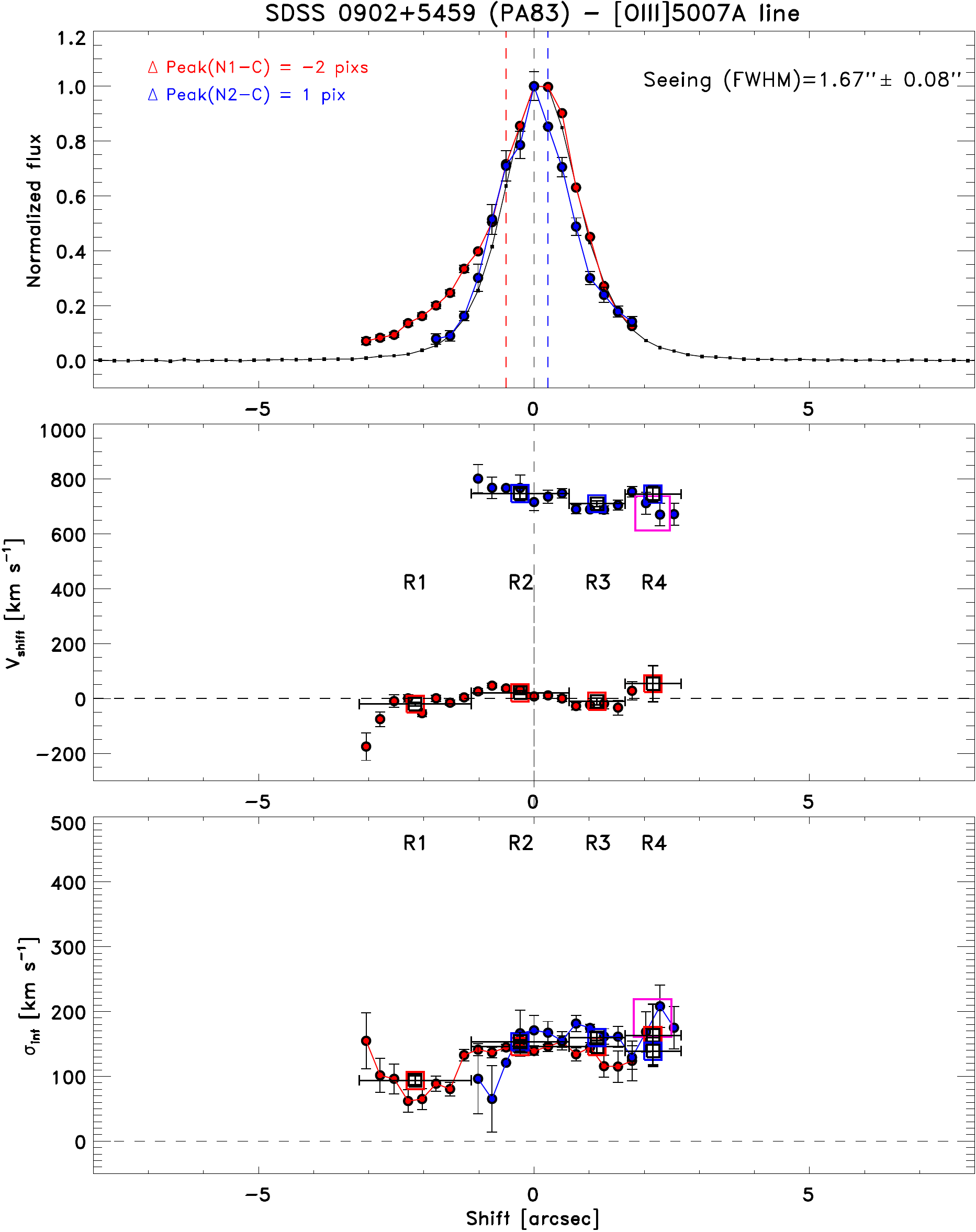}
\caption{SDSS 0902+5459 (PA 83): Same figure caption as in Fig.~\ref{panel_0741}. The magenta squares in the velocity field and velocity dispersion plots are used to highlight the pixels that are characterized by low S/N, which are well fit by using only one component instead of two. }
\label{panel_0902}
\end{center}
\end{figure*}

\begin{figure*}
\begin{center}
\includegraphics[scale=0.6]{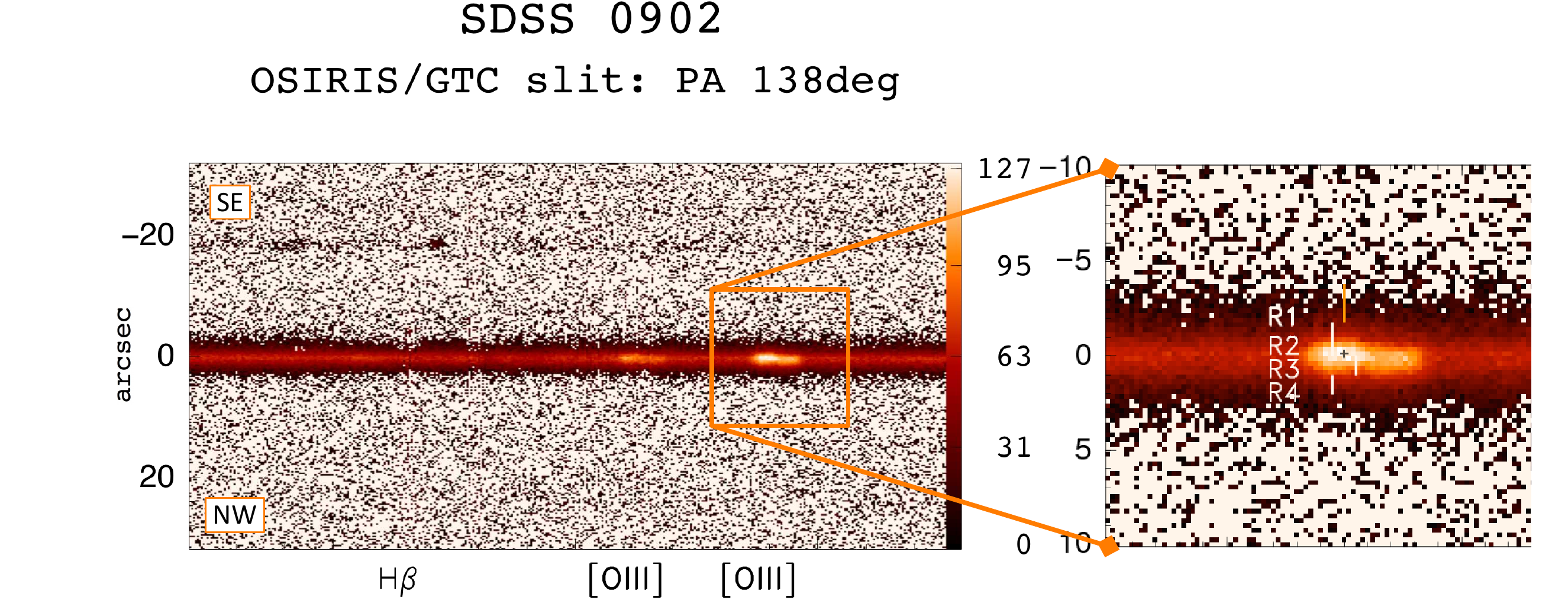}
\vskip15mm
\hskip0cm\includegraphics[scale=0.65]{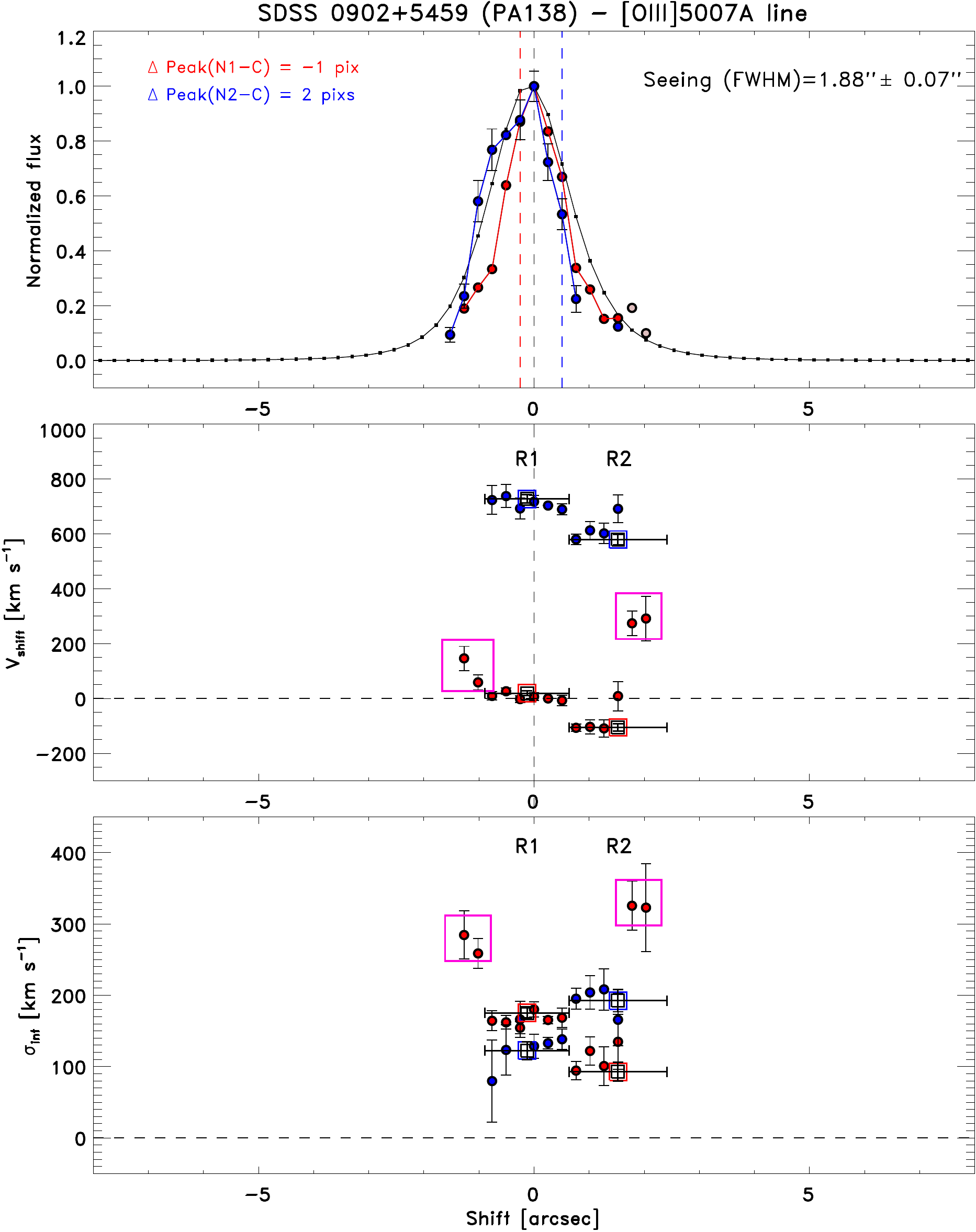}
\caption{SDSS 0902+5459 (PA 138): Same figure caption as in Fig.~\ref{panel_0741}. The fact that N1 has a more compact spatial profile than the seeing shows that this improved during the observations in comparison with the acquisition image. Magenta squares are the same as in Fig.~\ref{panel_0902}. The kinematic parameters are probably a blend of the two narrow components isolated in inner pixels with a higher S/N.}
\label{panel_0902_2}
\end{center}
\end{figure*}

\begin{figure*}
\begin{center}
\includegraphics[scale=0.6]{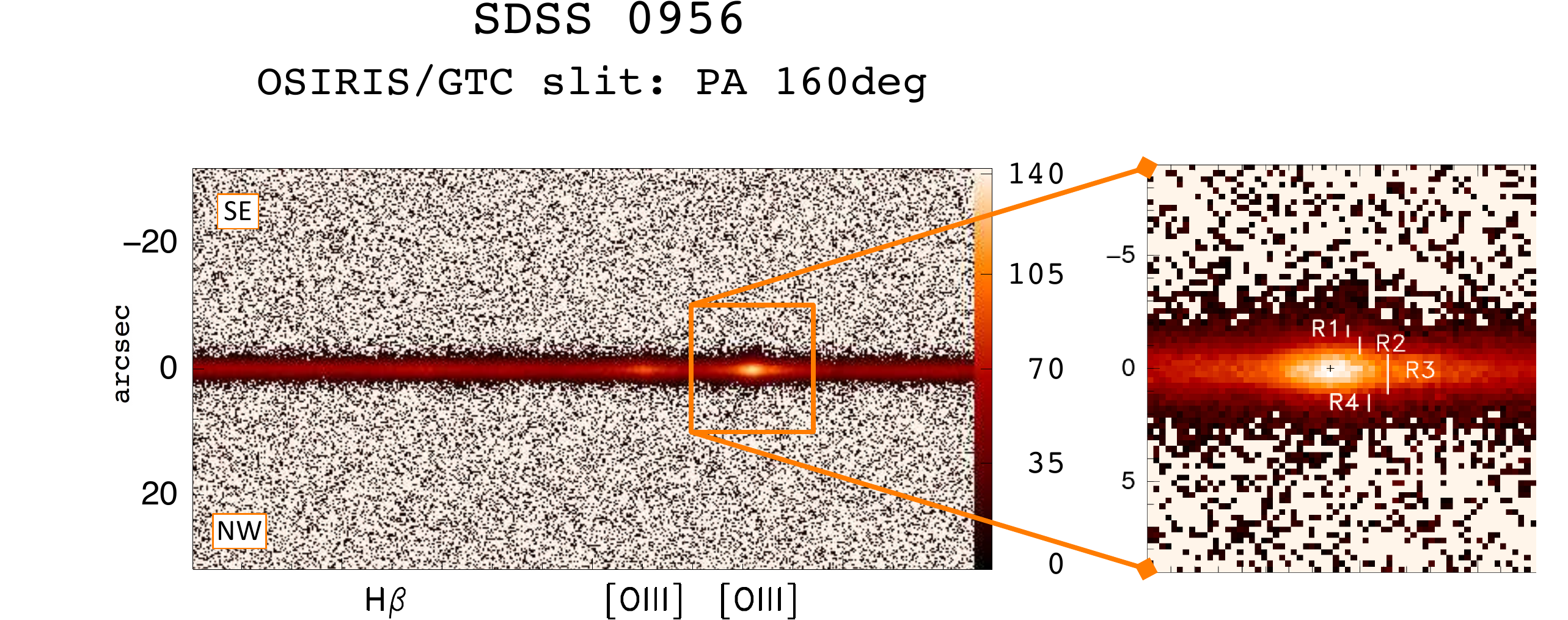}
\vskip15mm
\hskip0cm\includegraphics[scale=0.65]{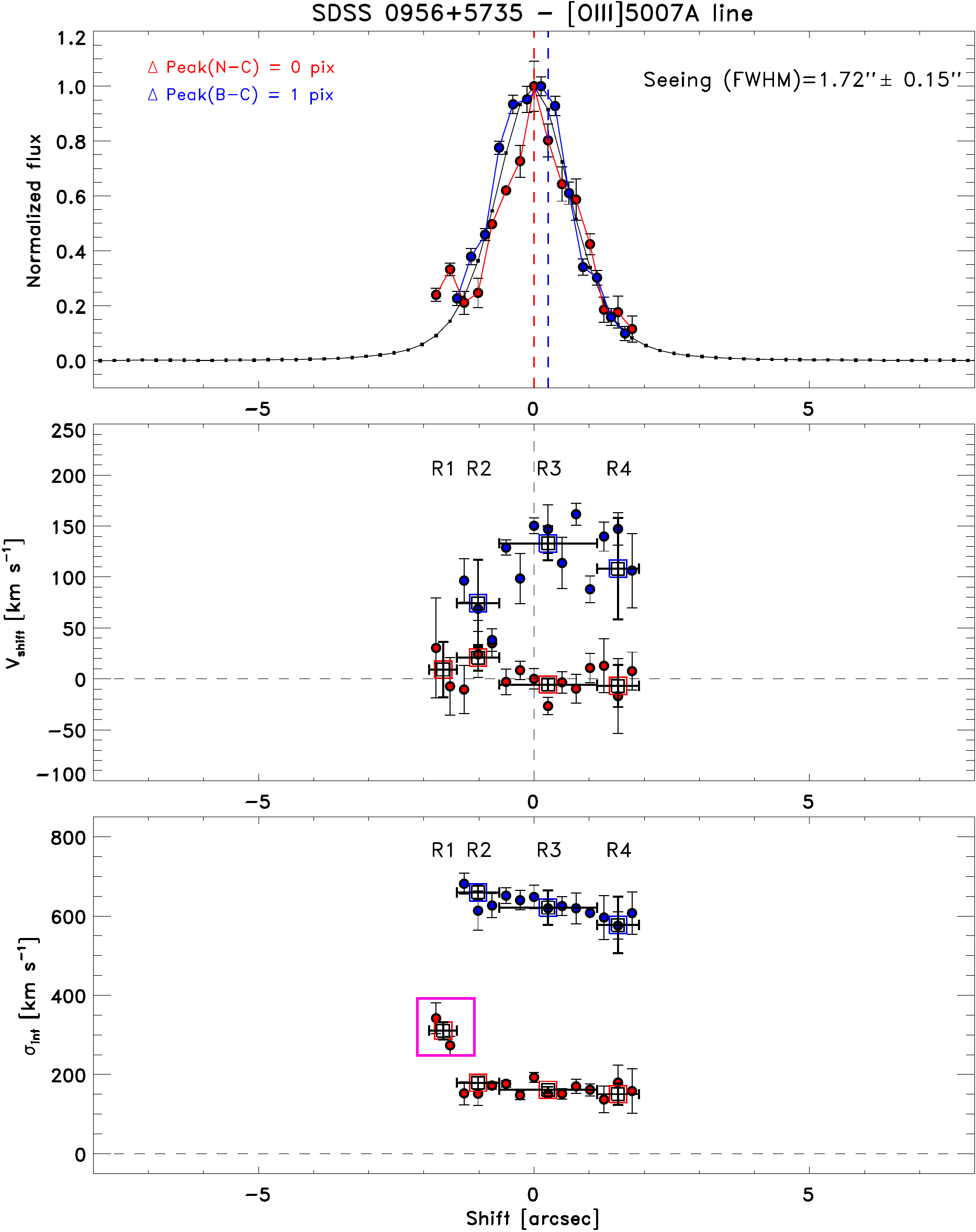}
\caption{SDSS 0956+5735: Same figure caption as in Fig.~\ref{panel_0741}. Magenta squares are the same as in Fig.~\ref{panel_0902}. }
\label{panel_0956}
\end{center}
\end{figure*}

\begin{figure*}
\begin{center}
\includegraphics[scale=0.6]{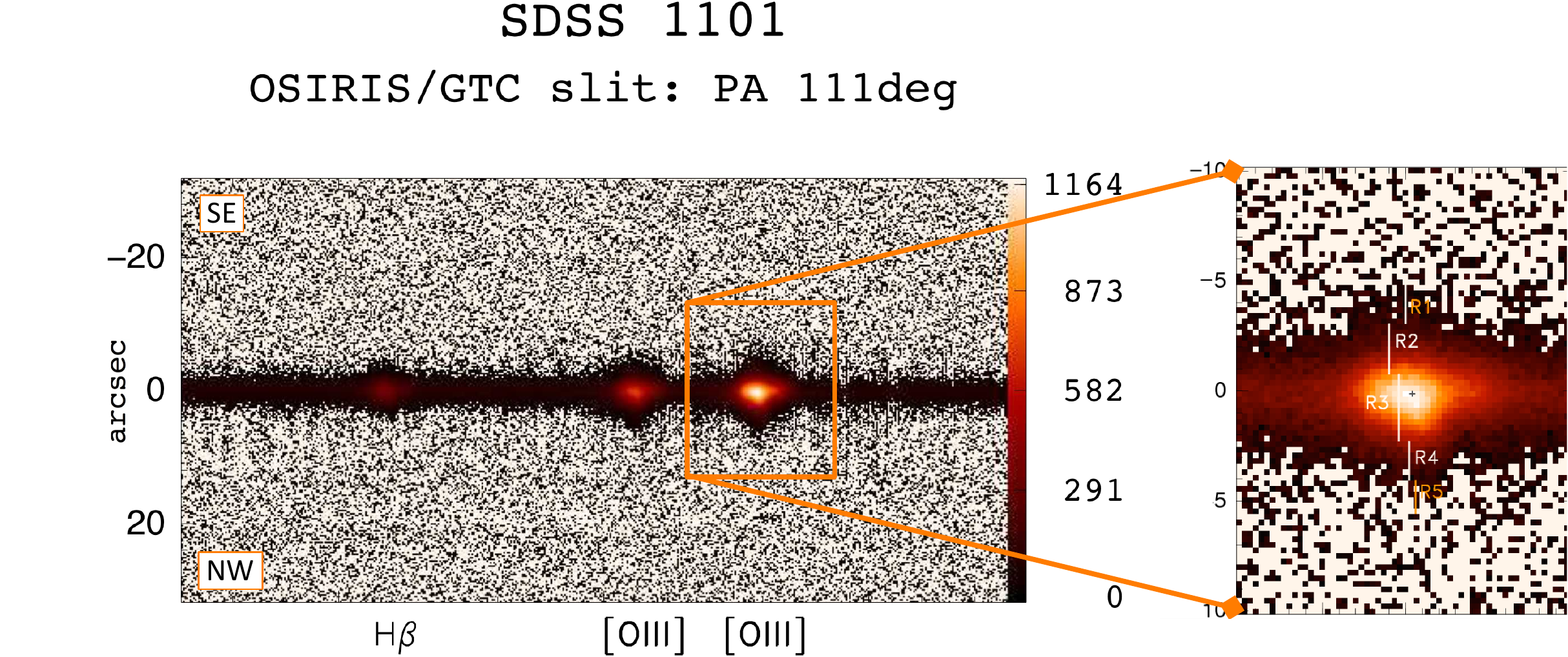}
\vskip15mm
\hskip0cm\includegraphics[scale=0.65]{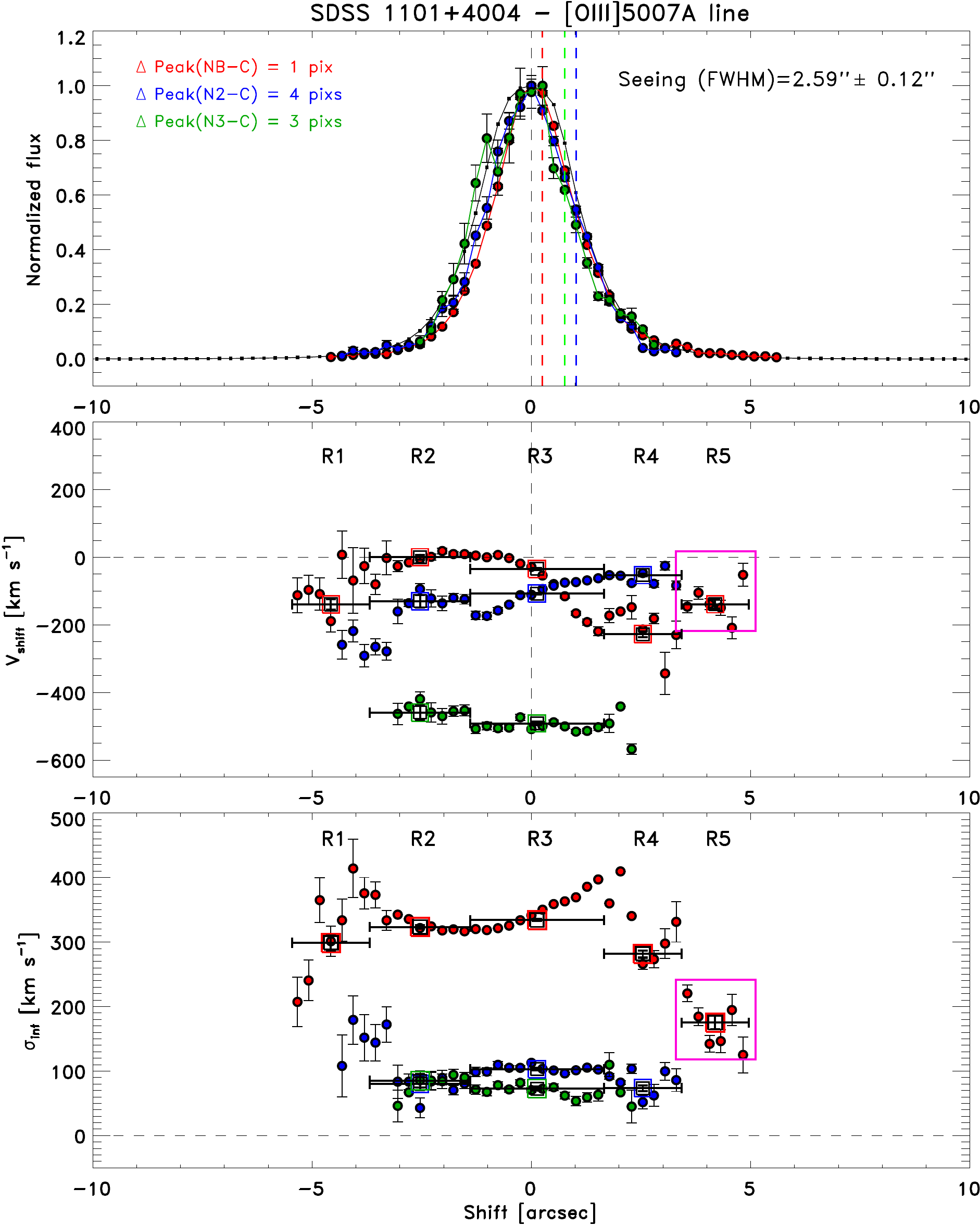}
\caption{SDSS 1101+4004: Same figure caption as in Fig.~\ref{panel_0741}. The fact that NB, N2, and N3 show  more compact spatial profiles than the seeing disk shows that this improved during the spectroscopic observations. Magenta squares are the same as in Fig.~\ref{panel_0902}. }
\label{panel_1101}
\end{center}
\end{figure*}

\begin{acknowledgements}

We thank the anonymous referee for useful comments and suggestions that helped us to improve the quality and presentation of the manuscript.

This work is based on observations made with OSIRIS on the Gran Telescopio Canarias (GTC), installed in the Spanish Observatorio del Roque de los Muchachos of the Instituto de Astrof\'isica de Canarias on the island of La Palma (programme GTC13-16B).

EB, MVM, AC acknowledge support from the Spanish Ministerio de Econom\'ia y Competitividad through the grants AYA2015-64346-C2-2-P. 

This research made use of the NASA/IPAC Extragalactic Database (NED), which is operated by the Jet Propulsion Laboratory, California Institute of Technology, under contract with the National Aeronautic and Space Administration and also the data from Sloan Digital Sky Survey (\url{http://www.sdss.org/}).

The National Radio Astronomy Observatory is a facility of the National Science Foundation operated under cooperative agreement by Associated Universities, Inc.

\end{acknowledgements}


\begin{thebibliography}{73}
\expandafter\ifx\csname natexlab\endcsname\relax\def\natexlab#1{#1}\fi

\bibitem[{{Arribas} {et~al.}(2014){Arribas}, {Colina}, {Bellocchi}, {Maiolino},
  \& {Villar-Mart{\'{\i}}n}}]{Arribas14}
{Arribas}, S., {Colina}, L., {Bellocchi}, E., {Maiolino}, R., \&
  {Villar-Mart{\'{\i}}n}, M. 2014, \aap, 568, A14

\bibitem[{{Bae} \& {Woo}(2014)}]{Bae14}
{Bae}, H.-J. \& {Woo}, J.-H. 2014, \apj, 795, 30

\bibitem[{{Baldwin} {et~al.}(2003){Baldwin}, {Ferland}, {Korista}, {Hamann}, \&
  {Dietrich}}]{Baldwin03}
{Baldwin}, J.~A., {Ferland}, G.~J., {Korista}, K.~T., {Hamann}, F., \&
  {Dietrich}, M. 2003, \apj, 582, 590

\bibitem[{{Bellocchi} {et~al.}(2013){Bellocchi}, {Arribas}, {Colina}, \&
  {Miralles-Caballero}}]{Bellocchi13}
{Bellocchi}, E., {Arribas}, S., {Colina}, L., \& {Miralles-Caballero}, D. 2013,
  \aap, 557, A59

\bibitem[{{Bessiere} {et~al.}(2012){Bessiere}, {Tadhunter}, {Ramos Almeida}, \&
  {Villar Mart{\'{\i}}n}}]{Bessiere12}
{Bessiere}, P.~S., {Tadhunter}, C.~N., {Ramos Almeida}, C., \& {Villar
  Mart{\'{\i}}n}, M. 2012, \mnras, 426, 276

\bibitem[{{Best} {et~al.}(2005){Best}, {Kauffmann}, {Heckman}, {Brinchmann},
  {Charlot}, {Ivezi{\'c}}, \& {White}}]{Best05}
{Best}, P.~N., {Kauffmann}, G., {Heckman}, T.~M., {et~al.} 2005, \mnras, 362,
  25

\bibitem[{{Bieri} {et~al.}(2017){Bieri}, {Dubois}, {Rosdahl}, {Wagner}, {Silk},
  \& {Mamon}}]{Bieri17}
{Bieri}, R., {Dubois}, Y., {Rosdahl}, J., {et~al.} 2017, \mnras, 464, 1854

\bibitem[{{Cappellari} {et~al.}(2007){Cappellari}, {Emsellem}, {Bacon},
  {Bureau}, {Davies}, {de Zeeuw}, {Falc{\'o}n-Barroso}, {Krajnovi{\'c}},
  {Kuntschner}, {McDermid}, {Peletier}, {Sarzi}, {van den Bosch}, \& {van de
  Ven}}]{Cap07}
{Cappellari}, M., {Emsellem}, E., {Bacon}, R., {et~al.} 2007, \mnras, 379, 418

\bibitem[{{Cielo} {et~al.}(2018){Cielo}, {Bieri}, {Volonteri}, {Wagner}, \&
  {Dubois}}]{Cielo18}
{Cielo}, S., {Bieri}, R., {Volonteri}, M., {Wagner}, A.~Y., \& {Dubois}, Y.
  2018, \mnras, 477, 1336

\bibitem[{{Crenshaw} {et~al.}(2010){Crenshaw}, {Schmitt}, {Kraemer},
  {Mushotzky}, \& {Dunn}}]{Crenshaw10}
{Crenshaw}, D.~M., {Schmitt}, H.~R., {Kraemer}, S.~B., {Mushotzky}, R.~F., \&
  {Dunn}, J.~P. 2010, \apj, 708, 419

\bibitem[{{Croton}(2006)}]{Croton06}
{Croton}, D.~J. 2006, \mnras, 369, 1808

\bibitem[{{Dasyra} {et~al.}(2006){Dasyra}, {Tacconi}, {Davies}, {Naab},
  {Genzel}, {Lutz}, {Sturm}, {Baker}, {Veilleux}, {Sanders}, \&
  {Burkert}}]{Dasyra06b}
{Dasyra}, K.~M., {Tacconi}, L.~J., {Davies}, R.~I., {et~al.} 2006, \apj, 651,
  835

\bibitem[{{De Robertis} \& {Osterbrock}(1984)}]{DeRob84}
{De Robertis}, M.~M. \& {Osterbrock}, D.~E. 1984, \apj, 286, 171

\bibitem[{{Di Matteo} {et~al.}(2005){Di Matteo}, {Springel}, \&
  {Hernquist}}]{DiMatteo05}
{Di Matteo}, T., {Springel}, V., \& {Hernquist}, L. 2005, in Growing Black
  Holes: Accretion in a Cosmological Context, ed. A.~{Merloni}, S.~{Nayakshin},
  \& R.~A. {Sunyaev}, 340--345

\bibitem[{{Epinat} {et~al.}(2010){Epinat}, {Amram}, {Balkowski}, \&
  {Marcelin}}]{Epi10}
{Epinat}, B., {Amram}, P., {Balkowski}, C., \& {Marcelin}, M. 2010, \mnras,
  401, 2113

\bibitem[{{Epinat} {et~al.}(2012){Epinat}, {Tasca}, {Amram}, {Contini}, {Le
  F{\`e}vre}, {Queyrel}, {Vergani}, {Garilli}, {Kissler-Patig}, {Moultaka},
  {Paioro}, {Tresse}, {Bournaud}, {L{\'o}pez-Sanjuan}, \& {Perret}}]{Epi12}
{Epinat}, B., {Tasca}, L., {Amram}, P., {et~al.} 2012, \aap, 539, A92

\bibitem[{{Espey} {et~al.}(1994){Espey}, {Turnshek}, {Lee}, {Bergeron},
  {Boksenberg}, {Hartig}, {Jannuzi}, {Sargent}, {Savage}, {Schneider},
  {Weymann}, \& {Wolfe}}]{Espey94}
{Espey}, B.~R., {Turnshek}, D.~A., {Lee}, L., {et~al.} 1994, \apj, 434, 484

\bibitem[{{Fabian}(2012)}]{Fabian12}
{Fabian}, A.~C. 2012, \araa, 50, 455

\bibitem[{{Ferrarese} \& {Merritt}(2000)}]{Ferrarese00}
{Ferrarese}, L. \& {Merritt}, D. 2000, \apjl, 539, L9

\bibitem[{{Fischer} {et~al.}(2018){Fischer}, {Kraemer}, {Schmitt}, {Longo
  Micchi}, {Crenshaw}, {Revalski}, {Vestergaard}, {Elvis}, {Gaskell}, {Hamann},
  {Ho}, {Hutchings}, {Mushotzky}, {Netzer}, {Storchi-Bergmann}, {Straughn},
  {Turner}, \& {Ward}}]{Fischer18}
{Fischer}, T.~C., {Kraemer}, S.~B., {Schmitt}, H.~R., {et~al.} 2018, \apj, 856,
  102

\bibitem[{{Genzel} {et~al.}(2001){Genzel}, {Tacconi}, {Rigopoulou}, {Lutz}, \&
  {Tecza}}]{Genzel01}
{Genzel}, R., {Tacconi}, L.~J., {Rigopoulou}, D., {Lutz}, D., \& {Tecza}, M.
  2001, \apj, 563, 527

\bibitem[{{Gon{\c c}alves} {et~al.}(2010){Gon{\c c}alves}, {Basu-Zych},
  {Overzier}, {Martin}, {Law}, {Schiminovich}, {Wyder}, {Mallery}, {Rich}, \&
  {Heckman}}]{Gon10}
{Gon{\c c}alves}, T.~S., {Basu-Zych}, A., {Overzier}, R., {et~al.} 2010, \apj,
  724, 1373

\bibitem[{{Granato} {et~al.}(2001){Granato}, {Silva}, {Monaco}, {Panuzzo},
  {Salucci}, {De Zotti}, \& {Danese}}]{Granato01}
{Granato}, G.~L., {Silva}, L., {Monaco}, P., {et~al.} 2001, \mnras, 324, 757

\bibitem[{{Greene} \& {Ho}(2005)}]{Greene05}
{Greene}, J.~E. \& {Ho}, L.~C. 2005, \apj, 627, 721

\bibitem[{{Greene} {et~al.}(2011){Greene}, {Zakamska}, {Ho}, \&
  {Barth}}]{Greene11}
{Greene}, J.~E., {Zakamska}, N.~L., {Ho}, L.~C., \& {Barth}, A.~J. 2011, \apj,
  732, 9

\bibitem[{{Greene} {et~al.}(2009){Greene}, {Zakamska}, {Liu}, {Barth}, \&
  {Ho}}]{Greene09}
{Greene}, J.~E., {Zakamska}, N.~L., {Liu}, X., {Barth}, A.~J., \& {Ho}, L.~C.
  2009, \apj, 702, 441

\bibitem[{{G{\"u}rkan} {et~al.}(2015){G{\"u}rkan}, {Hardcastle}, {Jarvis},
  {Smith}, {Bourne}, {Dunne}, {Maddox}, {Ivison}, \& {Fritz}}]{Gurkan15}
{G{\"u}rkan}, G., {Hardcastle}, M.~J., {Jarvis}, M.~J., {et~al.} 2015, \mnras,
  452, 3776

\bibitem[{{Harrison} {et~al.}(2014){Harrison}, {Alexander}, {Mullaney}, \&
  {Swinbank}}]{Harrison14}
{Harrison}, C.~M., {Alexander}, D.~M., {Mullaney}, J.~R., \& {Swinbank}, A.~M.
  2014, \mnras, 441, 3306

\bibitem[{{Harrison} {et~al.}(2015){Harrison}, {Thomson}, {Alexander}, {Bauer},
  {Edge}, {Hogan}, {Mullaney}, \& {Swinbank}}]{Harrison15}
{Harrison}, C.~M., {Thomson}, A.~P., {Alexander}, D.~M., {et~al.} 2015, \apj,
  800, 45

\bibitem[{{Holt} {et~al.}(2011){Holt}, {Tadhunter}, {Morganti}, \&
  {Emonts}}]{Holt11}
{Holt}, J., {Tadhunter}, C.~N., {Morganti}, R., \& {Emonts}, B.~H.~C. 2011,
  \mnras, 410, 1527

\bibitem[{{Hopkins}(2010)}]{Hopkins10}
{Hopkins}, P.~F. 2010, in IAU Symposium, Vol. 267, Co-Evolution of Central
  Black Holes and Galaxies, ed. B.~M. {Peterson}, R.~S. {Somerville}, \&
  T.~{Storchi-Bergmann}, 421--428

\bibitem[{{Husemann} {et~al.}(2016){Husemann}, {Scharw{\"a}chter}, {Bennert},
  {Mainieri}, {Woo}, \& {Kakkad}}]{Husemann16}
{Husemann}, B., {Scharw{\"a}chter}, J., {Bennert}, V.~N., {et~al.} 2016, \aap,
  594, A44

\bibitem[{{Husemann} {et~al.}(2013){Husemann}, {Wisotzki}, {S{\'a}nchez}, \&
  {Jahnke}}]{Husemann13}
{Husemann}, B., {Wisotzki}, L., {S{\'a}nchez}, S.~F., \& {Jahnke}, K. 2013,
  \aap, 549, A43

\bibitem[{{Karouzos} {et~al.}(2013){Karouzos}, {Trichas}, {Im}, {Malkan}, \&
  {the AKARI-NEP team}}]{Karouzos13}
{Karouzos}, M., {Trichas}, M., {Im}, M., {Malkan}, M., \& {the AKARI-NEP team}.
  2013, ArXiv e-prints

\bibitem[{{Karouzos} {et~al.}(2016){Karouzos}, {Woo}, \& {Bae}}]{Karouzos16}
{Karouzos}, M., {Woo}, J.-H., \& {Bae}, H.-J. 2016, \apj, 828, 64

\bibitem[{{Kellermann} {et~al.}(1994){Kellermann}, {Sramek}, {Schmidt},
  {Green}, \& {Shaffer}}]{Kellermann94}
{Kellermann}, K.~I., {Sramek}, R.~A., {Schmidt}, M., {Green}, R.~F., \&
  {Shaffer}, D.~B. 1994, \aj, 108, 1163

\bibitem[{{Lal} \& {Ho}(2010)}]{Lal10}
{Lal}, D.~V. \& {Ho}, L.~C. 2010, \aj, 139, 1089

\bibitem[{{Liu} {et~al.}(2013{\natexlab{a}}){Liu}, {Zakamska}, {Greene},
  {Nesvadba}, \& {Liu}}]{Liu13}
{Liu}, G., {Zakamska}, N.~L., {Greene}, J.~E., {Nesvadba}, N.~P.~H., \& {Liu},
  X. 2013{\natexlab{a}}, \mnras, 430, 2327

\bibitem[{{Liu} {et~al.}(2013{\natexlab{b}}){Liu}, {Zakamska}, {Greene},
  {Nesvadba}, \& {Liu}}]{Liu13_2}
{Liu}, G., {Zakamska}, N.~L., {Greene}, J.~E., {Nesvadba}, N.~P.~H., \& {Liu},
  X. 2013{\natexlab{b}}, \mnras, 436, 2576

\bibitem[{{McConnell} \& {Ma}(2013)}]{McConnell13}
{McConnell}, N.~J. \& {Ma}, C.-P. 2013, \apj, 764, 184

\bibitem[{{McElroy} {et~al.}(2015){McElroy}, {Croom}, {Pracy}, {Sharp}, {Ho},
  \& {Medling}}]{McElroy15}
{McElroy}, R., {Croom}, S.~M., {Pracy}, M., {et~al.} 2015, \mnras, 446, 2186

\bibitem[{{Mullaney} {et~al.}(2013){Mullaney}, {Alexander}, {Fine}, {Goulding},
  {Harrison}, \& {Hickox}}]{Mullaney13}
{Mullaney}, J.~R., {Alexander}, D.~M., {Fine}, S., {et~al.} 2013, \mnras, 433,
  622

\bibitem[{{Nesvadba} {et~al.}(2011){Nesvadba}, {Bryant}, {de Breuck},
  {Hunstead}, {Johnston}, {Lehnert}, \& {Collet}}]{Nesvadba11_0}
{Nesvadba}, N., {Bryant}, J., {de Breuck}, C., {et~al.} 2011, {Jets and AGN
  feedback at high-z: The role of radio power}, ATNF Proposal

\bibitem[{{Nesvadba} {et~al.}(2010){Nesvadba}, {Boulanger}, {Salom{\'e}},
  {Guillard}, {Lehnert}, {Ogle}, {Appleton}, {Falgarone}, \& {Pineau Des
  Forets}}]{Nesvadba10}
{Nesvadba}, N.~P.~H., {Boulanger}, F., {Salom{\'e}}, P., {et~al.} 2010, \aap,
  521, A65

\bibitem[{{Nesvadba} {et~al.}(2006){Nesvadba}, {Lehnert}, {Eisenhauer},
  {Gilbert}, {Tecza}, \& {Abuter}}]{Nesvadba06}
{Nesvadba}, N.~P.~H., {Lehnert}, M.~D., {Eisenhauer}, F., {et~al.} 2006, \apj,
  650, 693

\bibitem[{Osterbrock(1989)}]{Oster89}
Osterbrock, D.~E. 1989, Physics Today, 42, 123

\bibitem[{{Page} {et~al.}(2012){Page}, {Symeonidis}, {Vieira}, {Altieri},
  {Amblard}, {Arumugam}, {Aussel}, {Babbedge}, {Blain}, {Bock}, {Boselli},
  {Buat}, {Castro-Rodr{\'{\i}}guez}, {Cava}, {Chanial}, {Clements}, {Conley},
  {Conversi}, {Cooray}, {Dowell}, {Dubois}, {Dunlop}, {Dwek}, {Dye}, {Eales},
  {Elbaz}, {Farrah}, {Fox}, {Franceschini}, {Gear}, {Glenn}, {Griffin},
  {Halpern}, {Hatziminaoglou}, {Ibar}, {Isaak}, {Ivison}, {Lagache},
  {Levenson}, {Lu}, {Madden}, {Maffei}, {Mainetti}, {Marchetti}, {Nguyen},
  {O'Halloran}, {Oliver}, {Omont}, {Panuzzo}, {Papageorgiou}, {Pearson},
  {P{\'e}rez-Fournon}, {Pohlen}, {Rawlings}, {Rigopoulou}, {Riguccini},
  {Rizzo}, {Rodighiero}, {Roseboom}, {Rowan-Robinson}, {Portal}, {Schulz},
  {Scott}, {Seymour}, {Shupe}, {Smith}, {Stevens}, {Trichas}, {Tugwell},
  {Vaccari}, {Valtchanov}, {Viero}, {Vigroux}, {Wang}, {Ward}, {Wright}, {Xu},
  \& {Zemcov}}]{Page12}
{Page}, M.~J., {Symeonidis}, M., {Vieira}, J.~D., {et~al.} 2012, \nat, 485, 213

\bibitem[{{Perna} {et~al.}(2015){Perna}, {Brusa}, {Cresci}, {Comastri},
  {Lanzuisi}, {Lusso}, {Marconi}, {Salvato}, {Zamorani}, {Bongiorno},
  {Mainieri}, {Maiolino}, \& {Mignoli}}]{Perna15}
{Perna}, M., {Brusa}, M., {Cresci}, G., {et~al.} 2015, \aap, 574, A82

\bibitem[{{Prieto} {et~al.}(2016){Prieto}, {Fern{\'a}ndez-Ontiveros},
  {Markoff}, {Espada}, \& {Gonz{\'a}lez-Mart{\'{\i}}n}}]{Prieto16}
{Prieto}, M.~A., {Fern{\'a}ndez-Ontiveros}, J.~A., {Markoff}, S., {Espada}, D.,
  \& {Gonz{\'a}lez-Mart{\'{\i}}n}, O. 2016, \mnras, 457, 3801

\bibitem[{{Ramos Almeida} {et~al.}(2017){Ramos Almeida}, {Piqueras L{\'o}pez},
  {Villar-Mart{\'{\i}}n}, \& {Bessiere}}]{RAlmeida17}
{Ramos Almeida}, C., {Piqueras L{\'o}pez}, J., {Villar-Mart{\'{\i}}n}, M., \&
  {Bessiere}, P.~S. 2017, \mnras, 470, 964

\bibitem[{{Reyes} {et~al.}(2008){Reyes}, {Zakamska}, {Strauss}, {Green},
  {Krolik}, {Shen}, {Richards}, {Anderson}, \& {Schneider}}]{Reyes08}
{Reyes}, R., {Zakamska}, N.~L., {Strauss}, M.~A., {et~al.} 2008, \aj, 136, 2373

\bibitem[{{Rodr{\'{\i}}guez Zaur{\'{\i}}n} {et~al.}(2013){Rodr{\'{\i}}guez
  Zaur{\'{\i}}n}, {Tadhunter}, {Rose}, \& {Holt}}]{RZ13}
{Rodr{\'{\i}}guez Zaur{\'{\i}}n}, J., {Tadhunter}, C.~N., {Rose}, M., \&
  {Holt}, J. 2013, \mnras, 432, 138

\bibitem[{{Rose} {et~al.}(2015){Rose}, {Elvis}, \& {Tadhunter}}]{Rose15}
{Rose}, M., {Elvis}, M., \& {Tadhunter}, C.~N. 2015, \mnras, 448, 2900

\bibitem[{{Rose} {et~al.}(2018){Rose}, {Tadhunter}, {Ramos Almeida},
  {Rodr{\'{\i}}guez Zaur{\'{\i}}n}, {Santoro}, \& {Spence}}]{Rose18}
{Rose}, M., {Tadhunter}, C., {Ramos Almeida}, C., {et~al.} 2018, \mnras, 474,
  128

\bibitem[{{Rupke} {et~al.}(2005){Rupke}, {Veilleux}, \& {Sanders}}]{Rupke05}
{Rupke}, D.~S., {Veilleux}, S., \& {Sanders}, D.~B. 2005, \apjs, 160, 115

\bibitem[{{Rupke} {et~al.}(2017){Rupke}, {G{\"u}ltekin}, \&
  {Veilleux}}]{Rupke17}
{Rupke}, D.~S.~N., {G{\"u}ltekin}, K., \& {Veilleux}, S. 2017, \apj, 850, 40

\bibitem[{{Rupke} \& {Veilleux}(2011)}]{Rupke11}
{Rupke}, D.~S.~N. \& {Veilleux}, S. 2011, \apjl, 729, L27

\bibitem[{{Spence} {et~al.}(2018){Spence}, {Tadhunter}, {Rose}, \&
  {Rodr{\'{\i}}guez Zaur{\'{\i}}n}}]{Spence18}
{Spence}, R.~A.~W., {Tadhunter}, C.~N., {Rose}, M., \& {Rodr{\'{\i}}guez
  Zaur{\'{\i}}n}, J. 2018, \mnras, 478, 2438

\bibitem[{{Storchi-Bergmann} {et~al.}(2018){Storchi-Bergmann}, {Dall' Agnol de
  Oliveira}, {Longo Micchi}, {Schmitt}, {Fischer}, {Kraemer}, {Crenshaw},
  {Maksym}, {Elvis}, {Fabbiano}, \& {Colina}}]{StorchiB18}
{Storchi-Bergmann}, T., {Dall' Agnol de Oliveira}, B., {Longo Micchi}, L.~F.,
  {et~al.} 2018, \apj, 868, 14

\bibitem[{{Tadhunter} {et~al.}(2014){Tadhunter}, {Morganti}, {Rose}, {Oonk}, \&
  {Oosterloo}}]{Tadhunter14}
{Tadhunter}, C., {Morganti}, R., {Rose}, M., {Oonk}, J.~B.~R., \& {Oosterloo},
  T. 2014, \nat, 511, 440

\bibitem[{{Tadhunter} {et~al.}(2018){Tadhunter}, {Rodr{\'{\i}}guez
  Zaur{\'{\i}}n}, {Rose}, {Spence}, {Batcheldor}, {Berg}, {Ramos Almeida},
  {Spoon}, {Sparks}, \& {Chiaberge}}]{Tadhunter18}
{Tadhunter}, C., {Rodr{\'{\i}}guez Zaur{\'{\i}}n}, J., {Rose}, M., {et~al.}
  2018, \mnras, 478, 1558

\bibitem[{{Tecza} {et~al.}(2000){Tecza}, {Genzel}, {Tacconi}, {Anders},
  {Tacconi-Garman}, \& {Thatte}}]{Tecza00}
{Tecza}, M., {Genzel}, R., {Tacconi}, L.~J., {et~al.} 2000, \apj, 537, 178

\bibitem[{{Urbano-Mayorgas} {et~al.}(2019){Urbano-Mayorgas}, {Villar
  Mart{\'{\i}}n}, {Buitrago}, {Piqueras L{\'o}pez}, {Rodr{\'{\i}}guez del
  Pino}, {Koekemoer}, {Huertas-Company}, {Dom{\'{\i}}nguez-Tenreiro},
  {Carrera}, \& {Tadhunter}}]{UM19}
{Urbano-Mayorgas}, J.~J., {Villar Mart{\'{\i}}n}, M., {Buitrago}, F., {et~al.}
  2019, \mnras, 483, 1829

\bibitem[{{Veale} {et~al.}(2017){Veale}, {Ma}, {Greene}, {Thomas}, {Blakeslee},
  {McConnell}, {Walsh}, \& {Ito}}]{Veale17}
{Veale}, M., {Ma}, C.-P., {Greene}, J.~E., {et~al.} 2017, \mnras, 471, 1428

\bibitem[{{Villar-Mart{\'{\i}}n} {et~al.}(2016){Villar-Mart{\'{\i}}n},
  {Arribas}, {Emonts}, {Humphrey}, {Tadhunter}, {Bessiere}, {Cabrera Lavers},
  \& {Ramos Almeida}}]{VM16}
{Villar-Mart{\'{\i}}n}, M., {Arribas}, S., {Emonts}, B., {et~al.} 2016, \mnras,
  460, 130

\bibitem[{{Villar Mart{\'{\i}}n} {et~al.}(2015){Villar Mart{\'{\i}}n},
  {Bellocchi}, {Stern}, {Ramos Almeida}, {Tadhunter}, \& {Gonz{\'a}lez
  Delgado}}]{VM15}
{Villar Mart{\'{\i}}n}, M., {Bellocchi}, E., {Stern}, J., {et~al.} 2015,
  \mnras, 454, 439

\bibitem[{{Villar-Mart{\'{\i}}n} {et~al.}(2012){Villar-Mart{\'{\i}}n}, {Cabrera
  Lavers}, {Bessiere}, {Tadhunter}, {Rose}, \& {de Breuck}}]{VM12}
{Villar-Mart{\'{\i}}n}, M., {Cabrera Lavers}, A., {Bessiere}, P., {et~al.}
  2012, \mnras, 423, 80

\bibitem[{{Villar-Mart{\'{\i}}n} {et~al.}(2018){Villar-Mart{\'{\i}}n},
  {Cabrera-Lavers}, {Humphrey}, {Silva}, {Ramos Almeida}, {Piqueras-L{\'o}pez},
  \& {Emonts}}]{VM18}
{Villar-Mart{\'{\i}}n}, M., {Cabrera-Lavers}, A., {Humphrey}, A., {et~al.}
  2018, \mnras, 474, 2302

\bibitem[{{Villar-Mart{\'{\i}}n} {et~al.}(2017){Villar-Mart{\'{\i}}n},
  {Emonts}, {Cabrera Lavers}, {Tadhunter}, {Mukherjee}, {Humphrey},
  {Rodr{\'{\i}}guez Zaur{\'{\i}}n}, {Ramos Almeida}, {P{\'e}rez Torres}, \&
  {Bessiere}}]{VM17}
{Villar-Mart{\'{\i}}n}, M., {Emonts}, B., {Cabrera Lavers}, A., {et~al.} 2017,
  \mnras, 472, 4659

\bibitem[{{Villar Mart{\'{\i}}n} {et~al.}(2014){Villar Mart{\'{\i}}n},
  {Emonts}, {Humphrey}, {Cabrera Lavers}, \& {Binette}}]{VM14}
{Villar Mart{\'{\i}}n}, M., {Emonts}, B., {Humphrey}, A., {Cabrera Lavers}, A.,
  \& {Binette}, L. 2014, \mnras, 440, 3202

\bibitem[{{Villar-Mart{\'{\i}}n} {et~al.}(2011){Villar-Mart{\'{\i}}n},
  {Tadhunter}, {Humphrey}, {Encina}, {Delgado}, {Torres}, \&
  {Mart{\'{\i}}nez-Sansigre}}]{VM11}
{Villar-Mart{\'{\i}}n}, M., {Tadhunter}, C., {Humphrey}, A., {et~al.} 2011,
  \mnras, 416, 262

\bibitem[{{Xu} {et~al.}(1999){Xu}, {Livio}, \& {Baum}}]{Xu99}
{Xu}, C., {Livio}, M., \& {Baum}, S. 1999, \aj, 118, 1169

\bibitem[{{Zakamska} \& {Greene}(2014)}]{Zakamska14}
{Zakamska}, N.~L. \& {Greene}, J.~E. 2014, \mnras, 442, 784

\end{thebibliography}

\end{document}